\documentclass[11pt]{article}

\usepackage[a4paper,includeheadfoot,top=1cm,bottom=2cm,left=2cm,right=2cm]{geometry}
\usepackage{here}
\usepackage{graphicx}
\usepackage[colorlinks=true,urlcolor=blue]{hyperref}
\usepackage{authblk}
\usepackage{caption}
\usepackage{xcolor}
\usepackage{soul}
\usepackage{amsmath}
\usepackage{float}
\usepackage[section]{placeins}

\graphicspath{{Figs/}}

\title{Role of Shafranov shift, zonal structures on the behavior of TAEs, AAEs and microinstabilities in the presence of energetic particles}

\author[1]{B.~Rofman}
\author[1]{G.~Di~Giannatale}
\author[2]{A.~Mishchenko}
\author[1]{E.~Lanti}
\author[3]{A.~Bottino}
\author[3]{T.~Hayward-Schneider}
\author[6]{J.N.~Sama}
\author[5]{A.~Biancalani}
\author[4]{B.F.~McMillan}
\author[1]{S.~Brunner}
\author[1]{L.~Villard}

\affil[1]{Ecole Polytechnique F\'ed\'erale de Lausanne, Swiss Plasma Center, CH-1015 Lausanne, Switzerland}
\affil[2] {Max-Planck-Institut f\"ur Plasmaphysik, Greifswald, Germany}
\affil[3]{Max-Planck-Institut f\"ur Plasmaphysik, Garching, Germany}
\affil[4]{University of Warwick, Department of Physics, UK}
\affil[5]{De Vinci Higher Education, De Vinci Research Center, 92916 Paris, France}
\affil[6]{Universit\'e de Lorraine, CNRS, IJL, Nancy, France}
\date{}             

\begin{document}

\maketitle

\begin{abstract}
In future nuclear fusion reactors, even a small fraction of fusion-born energetic particles (EP) about 100 times hotter than the thermal bulk species, contributes substantially to the kinetic pressure and therefore affect the MHD equilibrium, mainly via the Shafranov shift.
In this work, we perform first-principles numerical simulations using the gyrokinetic, electromagnetic, global code ORB5 to study the effect of a self-consistent finite $\beta$ equilibrium on the arising Alfv\'en Eigenmodes (destabilized by EPs), Ion Temperature Gradient (ITG), and Kinetic Ballooning Modes (KBM) microturbulence (destabilized by thermal species). 
Linearly, we explore the complex interplay between EP fraction, bulk gradients and a self-consistent Shafranov shift on the plasma stability.
We choose single toroidal mode numbers to represent the system's instabilities and study the characteristic nonlinear evolutions of TAEs, KBMs and ITGs separately and including the axisymmetric field response to each mode separately. This study focuses on the impact of Shafranov shift equilibrium consistency, as well as the self-generated zonal ${E \times B}$ flows, the saturation levels and resulting heat and particle fluxes.
In the ITG cases including the $n=0$ perturbations reduces turbulent fluxes, as expected, however, for the TAE cases including the $n=0$ perturbations is shown to enhance the fluxes. We show for the first time that Axisymmetric Alfv\'en Eigenmodes (AAEs) play a role in this mechanism.
\end{abstract}

\section{Introduction} 

A burning plasma is a complex dynamical system where turbulence and instabilities arise from sources of free energy like kinetic gradients and resonant energetic particles. Linear and nonlinear coupling across a wide range of spatio-temporal frequencies and wavelengths, gives rise to global ($low-n$) Toroiadal Alfv\'en Eigenmodes (TAEs), and Zonal Structures(ZS), which become the meso-scale in a self-organizing system. The magnetic geometry plays a role in determining the particle drifts and resonances, thus changing the nonlinear saturation level (of the fluctuating fields), profile relaxation, and turbulent transport of heat and particles \cite{Chen_RevModPhys2016}. In this work we aim to systematically explore the effects of Shafranov shift on the stability and self-organization of the plasma, emphasizing the importance of considering ideal MHD equilibria consistent with the plasma kinetic profiles.

In the core of a tokamak, the hot collisionless plasma is confined in a twisted magnetic field, forming nested magnetic surfaces. The finite plasma pressure pushes against the outer mid-plane, i.e. the low field side, and compresses the flux surfaces, causing the magnetic axis to shift radially outwards by $\Delta (s) = R_{mag}-R_{geom}(s)$. Where $R_{mag}$ is the location of the magnetic axis, and $R_{geom}(s) = (R_{max}(s)+R_{min}(s))/2$ \cite{Shafranov_RPP1966}.

Increasing the Shafranov shift has a stabilizing effect on the ideal MHD ballooning modes (in the $n \rightarrow \infty $ limit) due to reduction of the connection length or the effective bad curvature, and opening of a second stability region beyond Troyon's $\beta$ limit \cite{Coppi_NF1979, Troyon_PPCF1984, Ramos_PoF1991}. When kinetic effects are considered a stability $\beta$ threshold is found slightly below the ideal one, for the pressure-driven Kinetic Ballooning Mode (KBM) \cite{Tang_NF1980}. 

In many fusion-relevant cases, that are ideal MHD stable, other instabilities may exisit, such as Ion Temperature Gradient (ITG) mode for $\eta_i = \frac{R/\langle {L_T} \rangle_{i}}{R/\langle {L_n} \rangle_{i}} > 1$, i.e. the flux-surface-averaged (fsa) temperature gradient is stronger than the flux-surface-averaged density gradient. \cite{Rudakov_DAN1961, Coppi_PoF1967, Guzdar_PoF1983}. The ITG instability is a low frequency mode which is primarily electrostatic in nature, i.e. found in the $\beta = 0$ limit, where $\beta =  \frac{P}{B_0^2/2\mu_0}$ is the ratio between kinetic and magnetic pressure, and $B_0 = B_{mag}$ is the magnetic field on axis. Nonetheless, even without accounting for Shafranov shift, including finite $\beta$ (electromagnetic) effects acts to stabilize the ITG mode \cite{Weiland_NF1992}, up to the KBM limit \cite{Pueschel_PoP2010}.

In addition to various drift wave instabilities which lead to turbulence and anomalous transport \cite{Drummond_PoF1962}, highly magnetized plasmas confined in a toroidal magnetic geometry are subject to a wide range of Alfv\'en Eigenmodes (AE). In general both compressional and shear Alfv\'en waves can propagate through the magnetized plasma but in this work we shall ignore the former. Noticeable are the low-$n$ ($n$ being the toroidal mode number) toroidicity induced Alfv\'en Eigenmodes (TAE), which resonate in the toroidicity induced gaps in the SAW continuum, a.k.a the Alfv\'en continuum \cite{Cheng_PoF1986}. These modes are excited by resonant energetic particles ubiquitous in burning plasmas and in turn lead to significant EP redistribution and loss \cite{Heidbrink_NF1994}. These are global modes, i.e. scale with system size, peak at radial locations where $nq = (m + 1/2)$, with a typical frequency of $\omega_0 \cong \omega_A/2q$ (mid gap). Where $m$ is the poloidal mode number, $\omega_A = v_A/R$ is the Alfv\'en frequency, $v_A = B_0/\sqrt{\mu_0\rho_m}$ is the Alfv\'en velocity, and $\rho_m$ is the mass density. The plasma pressure combined with the geodesic curvature couples the Alfv\'en and sound wave continuum, thus modifying the lower order gaps and opening the so called $0^{th}$ frequency gap where the Beta Alfv\'en Eigenmodes (BAEs) can exist \cite{Chu_PoF1992, Heidbrink_PoP1999}. Another, usually less discussed, mode is the $ n = 0$ Axisymmetric Alfv\'en Eigenmode (AAE), first seen during NBI experiments on TFTR \cite{Chang_NF1995}. This mode does not require geometrically induced gaps and exists even in cylindrical geometry. However, with a frequency just below a minimum of the Alfv\'en continuum it still depends on it. 

Energetic Particles (EPs) are ubiquitous in burning plasmas as fusion-born $\alpha$-particles ($3.5 \ MeV$ for D-T fusion). They can also be found in current tokamaks at lower energies as part of the heating schemes, either injected by a neutral beam (NBI), or produced by ICRH \cite{Fasoli_NF2007}. As a result, the EPs are approximately mono-energetic when they enter the plasma and are assumed to slowdown on the electrons, resulting in either isotropic or anisotropic slowing-down distributions \cite{Gaffey_PoP1976}. These hot ions contribute to the plasma kinetic pressure and therefore to the Shafranov shift. In addition to their effect on the magnetic geometry, the EPs precessional frequencies can stabilize ballooning modes \cite{Rosenbluth_PRL1983}, or destabilize internal kink modes \cite{Porcelli_PPCF1996}. The parallel velocity of EPs can become comparable to the Alfv\'en speed. Radial gradients in EP density can excite various AEs such as TAEs, via linear wave - particle interactions \cite{Rosenbluth_PRL1975, Fu_PoF1989}. Moreover, even when intersecting the continuum, EPs can excite energetic particles modes (EPMs) such as "fishbones" \cite{Chen_PRL1984, Zonca_PoP2000}.  

Unstable modes cause fluctuations in the electrostatic potential $\phi$ leading to the formation of zonal bands of $E \times B$ shear flows known as Zonal Flows (ZF) \cite{Hasegawa_PoF1979, Diamond_PPCF2005}. Similarly, fluctuations in the parallel magnetic vector potential $A_{\parallel}$, lead to the formation of zonal currents which together with the ZF are known as zonal structures (ZS) and phase space zonal structures (PSZS) \cite{Zonca_NJP2015}. These axisymmetric $(n=0)$ modes are self-consistently generated through nonlinear wave-wave coupling, remain undamped in collisionless plasmas, and as a result play a crucial role in mitigating core turbulence. \cite{Rosenbluth_PRL1998, Hinton_PPCF1999}. Their shear reduces the eddy correlation length, ushering them down the turbulent energy cascade towards smaller spatio-temporal scales. The shearing rate grows with the unstable mode until it becomes comparable to the instability's growth rate $\gamma$, leading to the nonlinear saturation of the system. Additionally, PSZS can nonlinearly modify the resonances in phase space, resulting in the suppression Alfv\'enic instabilities \cite{Qiu_PoP2016} or, surprisingly, to an enhancement of turbulent fluxes \cite{Chen_NF2024}. These zonal structures (ZS) act as a storage for the free energy in the system, and a nonlinear coupling channel accessible by both the turbulence and the Alfv\'en eigenmodes, the combined effect of which will be further explored in future works. Our study shows that including the $n = 0$ response allows for the excitation of a non-zonal i.e. $m \neq 0$, finite frequency Alfv\'enic global mode during the nonlinear saturation phase. This Axisymmetric Alfv\'en Eigenmode (AAE) might be linked to enhanced turbulent fluxes observed in the TAE cases, in contrast to the ITG cases which experience a reduction in turbulent fluxes due to (the shearing rate produced by the) ZF. 

In this work, we use ORB5, a first-principles, global, electromagnetic, gyrokinetic Particle-In-Cell (PIC) code \cite{Lanti_CPC2020}, to study the linear and nonlinear behavior of AEs, EPs and DWs in the presence of self-consistently generated ZS. This is done in a configuration inspired by the Cyclone Based Case \cite{Dimits_PoP2000}, where a finite $\beta$, circular MHD equilibrium is generated by the ideal MHD code CHEASE \cite{Lutjens_CPC1996}. A recently introduced 'pullback' scheme with a mixed - variables representation \cite{Mishchenko_CPC2019} which deals well with the so called "cancellation problem", allows for electromagnetic linear and nonlinear simulations with fully kinetic species to be performed at a reasonable computational effort \cite{Mishchenko_PPCF2023}. This allowed us to successfully capture a ITG-KBM transition \cite{Cole_PoP2021}, investigate the interactions between EPs, Alfv\'enic modes and microinstabilities \cite{Biancalani_PPCF2021, HaywardSchneider_NF2022, Sama_PoP2024, Ivanov_arXiv2025}, and carry out an extensive numerical linear and nonlinear validation campaign \cite{Vlad_NF2021, Vlad_RevModPhys2025}. 

The following work is divided between the linear and nonlinear parts. In the linear part we map the dispersion relation for several different configurations to study the effects of bulk gradients, EP fraction, and Shafranov shift. At the end of the linear part, we select from this matrix of scans several representative modes with a single non zero toroidal mode number. A TAE with a low toroidal mode number $n$, a KBM in the mid-$n$ range, and an ITG in the high-$n$ range. 
In the nonlinear part, we study the saturation dynamics of the system and the importance of including the self consistent Shafranov shift and axisymmetric response . We analyze the self generated zonal structures, and find an Axisymmetric Alfvn\'en Eigenmode (AAE) is generated by both TAEs and ITGs. We further study the temperature, density and q-profile relaxations and the associated fluxes. We find that including the axisymmetric response  strongly enhances the heat and particle fluxes generated by the TAE while reducing as expected the fluxes in the case of the ITG. We find a similar strong AAE activity in the TAE generated fluxes. 

\begin{figure}
\begin{center}
\includegraphics[width=0.495\textwidth]{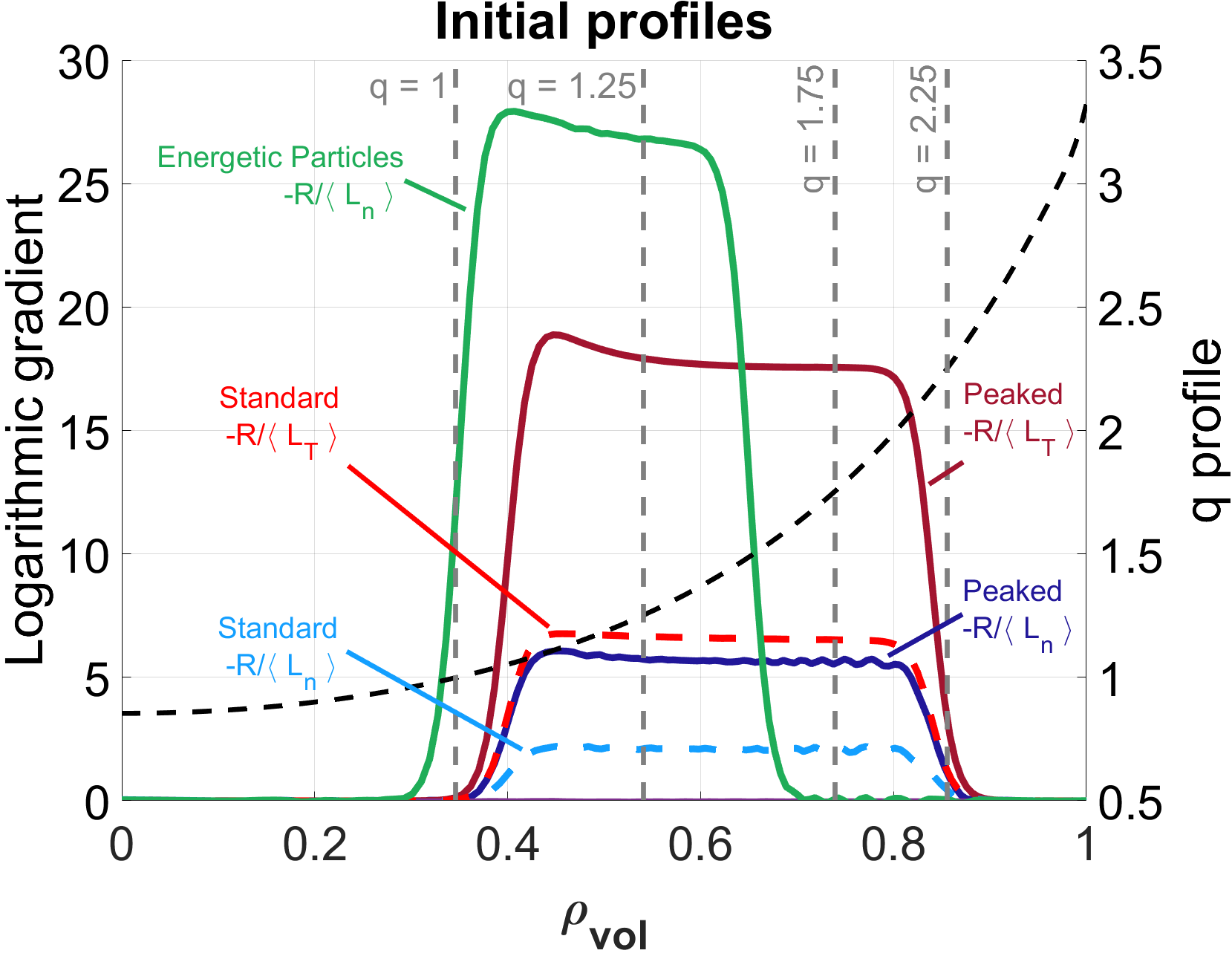}
\includegraphics[width=0.495\textwidth]{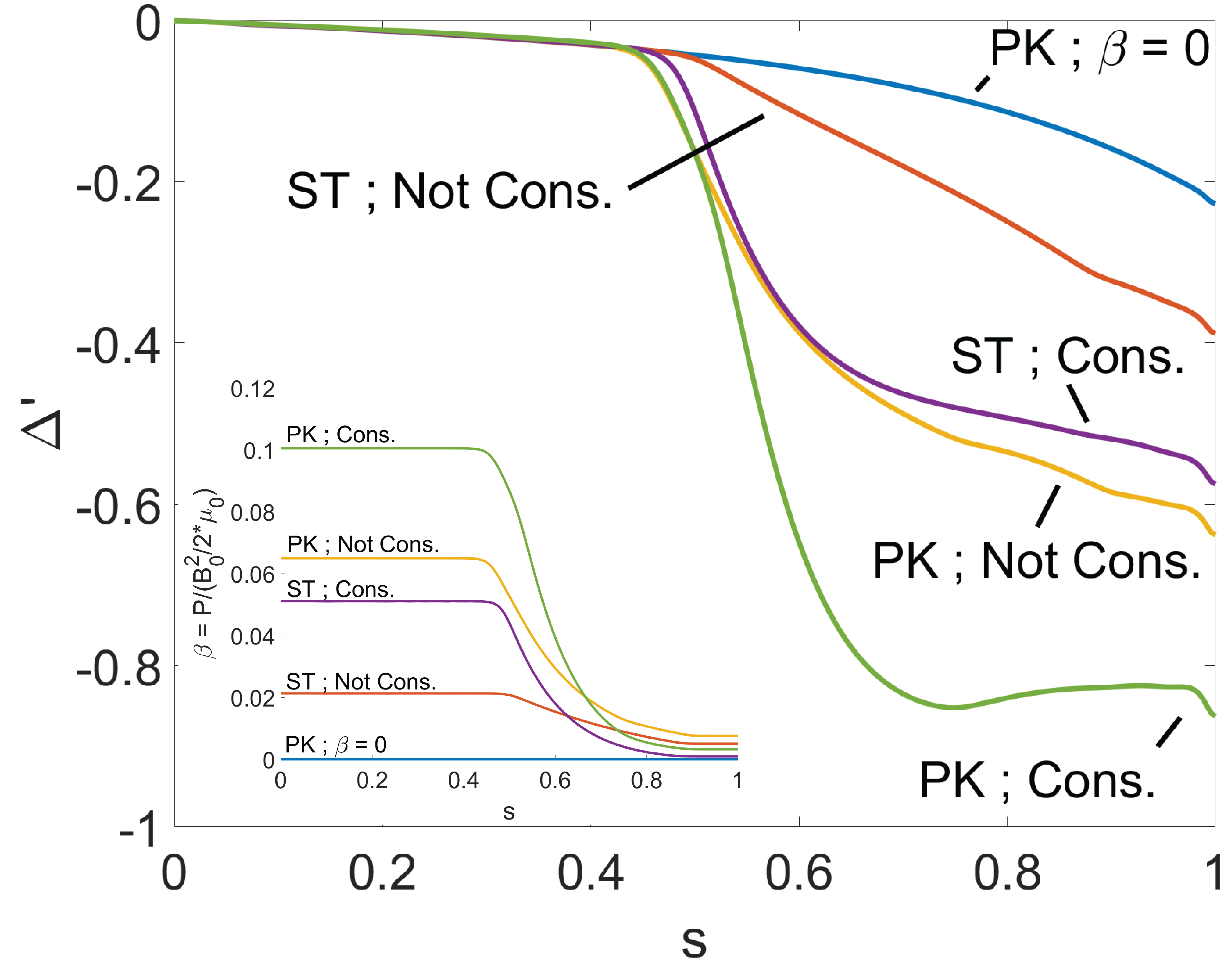}

\caption{\label{FIG:Combined_profiles_particles} \it On the left: initial temperature and density (flux-surface-averaged) logarithmic gradients, for ions, electrons, and EPs. The profiles of the bulk species (ions and electrons) change between the standard (ST) and peaked (PK) cases, while the EP profiles remain the same. The q profile is indicated by the black dashed line. The vertical gray dashed lines mark the mode rational surfaces $(nq = m + 1/2)$ of the $n = 2$ TAE.
On the right: $\Delta^\prime$ and the $\beta$ profiles of the MHD equilibrium for the main 5 cases in this work: $\beta = 0$, not consistent ST and PK equilibria based on pressure arising solely from the bulk profiles, and consistent ST and PK equilibria which account for $1\%$ EP pressure as well.}
\end{center}
\end{figure}

\section{Simulation parameters}
The global multi-species, PIC code, ORB5 \cite{Lanti_CPC2020}, solves the nonlinear, electromagnetic, gyrokinetic Vlasov–Maxwell system where the fields are represented with two potentials $(\phi,A_\parallel)$ using B-splines on a 3D finite element grid, and the distribution function $f$ is discretized by numerical markers representing the particles. To deal with the Monte Carlo sampling noise, ORB5 splits the full-$f$ into a fast changing perturbation $\delta f/f \ll 1$ and a fixed or slowly evolving background $f_0$ used as a control variate \cite{Murugappan_PoP2024}. As most physical modes align with the magnetic field lines, an additional noise reduction scheme is implemented using a field-aligned Fourier filter in a straight-field-line representation, which allows us to keep only the $m \in [nq - \Delta m \ ,\ nq + \Delta m]$ modes. where $n$ and $m$ are the toroidal and poloidal mode numbers, with $\Delta m$ is set to 5 in our simulations, and $q$ is the safety factor.

We performed our simulations with a $dt = 0.5$ for numerical stability, and used $40 M$ numerical particles $n_p$ per species, and per toroidal mode number $n$. For specific linear and nonlinear cases we tested the convergence with ten times number of particles, $n_p$, We found overall good convergance with the linear growth rate changing by less than $4\%$ and the nonlinear saturation dynamics captured in essence. More details found in figure \ref{FIG:nonlinear_TAE_Sat_dynemics}.

The fields in ORB5 are represented on a 3D finite element grid of $n_s\times n_{\theta^{*}}\times n_{\varphi}$, were $(s,\theta^*,\varphi)$ are the straight-field-line coordinates. In our simulations we used two grids depending on the typical mode range of the instability. Specifically, for the $n = \{[2],[0,2]\}$ TAE cases we used a grid size $128 \times 128 \times 64$, and for the $n = \{[25],[0,25]\}$ ITG cases we used a grid size $512\times512\times256$.  

In this work we take further advantage of the Fourier representation and perform a toroidal decomposition which allows us to simulate only a restricted set of toroidal mode numbers, $n$, while including a large range of poloidal mode numbers $m$. In the nonlinear simulations we use a modified Krook operator both as noise control and as a source with a damping rate/drive of about $5\%$ of the linear growth rate. The Krook operator is designed to conserve flux-surface-averaged moments such as density, energy, parallel flows and residual zonal flows. Considering the drive for microturbulence, we set the Krook operator to conserve all moments except for the energy of the bulk species, i.e. ions and electrons. In simulations with energetic particles, they are introduced as a third species with a separate Krook operator set to conserve all moments except for density \cite{McMillan_PoP2008}. Thus, besides their noise reduction action, the Krook operators act as a heat source for the bulk species and a particle source for the EPs. 

As a starting point for our study, we use the DIII-D toy-model known as the Cyclone Base Case (CBC) \cite{Dimits_PoP2000}, which was based on an experimental study that investigated the stabilizing effects of plasma shaping \cite{Greenfield_NF1997}. Here we account for the profile effects in a finite $\beta$ equilibrium. We use CHEASE \cite{Lutjens_CPC1996}, a fixed boundary Grad - Shafranov solver to generate a set of circular finite $\beta$ equilibria to isolate the effect of $\Delta'$.

CHEASE accepts $q$ and pressure profiles as inputs, which allowed us to keep the q profile constant while varying the pressure between the equilibria. Specifically, we looked at 5 different cases: a $\beta = 0$ case and four cases based on two sets of bulk species profiles: standard (ST) and peaked (PK), with or without EPs.

The modeled plasma has a major radius $R_0 = 1.7\ m$, and a minor radius $a = 0.61\ m$, with a magnetic field on axis $B_0 = 1.9\ T$.
Assuming collisionless, hot plasma with $T_i = T_e$, we keep the machine size constant with $1/\rho^*(s=0.5) = 180$, where s is $s = \sqrt{\psi/\psi_{edge}}$, and $\psi$ is the poloidal magnetic flux. Here $\rho^* = \rho_s/a$, where $\rho_s = \sqrt{T_e/m_i}$ is the ion sound Larmor radius. Notably, the peaked profiles with a higher core temperature will have a larger local $\rho^\ast$ value in the core - and a lower $\rho^\ast$ near the edge - than the standard profiles. All species (ions, electrons, EPs) are taken to be kinetic and described by local Maxwellian distribution functions. Specifically, we perform a gyro-averaging for the ions and EPs, while electrons are drift kinetic. We used slightly heavy electrons with a mass ratio (to ions) of $m_i/m_e = 1000$ (vs. the physical ratio of $m_i/m_e = 3670$ for deuterium).

\FloatBarrier
Microturbulence is induced by bulk gradients in temperature and density for three cases of 
\begin{itemize} 
    \item Flat (FL) bulk gradients: $R/\langle {L_T} \rangle_{i,e} = R/\langle {L_n} \rangle_{i,e} = 2.22$, such that $\eta_{i,e} = 1$
    \item Standard (ST) bulk gradients: $-R/\langle {L_T} \rangle_{i,e} = 6.716$ $\&$ $-R/\langle {L_n} \rangle_{i,e} = -2.195$, such that $\eta_{i,e} = 3.06 $
    \item Peaked (PK) bulk gradients: $-R/\langle {L_T} \rangle_{i,e} = 19.25$ $\&$ $R/\langle {L_n} \rangle_{i,e} = -6.23$, such that $\eta_{i,e} = 3.09 $. 
\end{itemize}
\FloatBarrier

The values above are defined at the radial location $s = 0.54$ where $q = 1.25$. Here due to the flux surface shaping we adopt a coordinate-independent definition for the logarithmic gradient \cite{Villard_PPCF2013}

\begin{equation}
R/\langle {L_T} \rangle = R_0 \frac{{dT}/{d\psi}}{{dV}/{d\psi}}S
\end{equation}

where $S$ and $V$ are the flux surface area and enclosed volume, respectively. The same definition is used for $R/L_n$, replacing the temperature with the species density. 

The Alfv\'en Eigenmodes are excited by energetic particles with either $1\%$ or $3\%$ dilution from the bulk ions. Both are considered deuterium, and the effect on bulk density profiles is considered negligible. The EPs have 120 times the energy of the bulk ions such that $\tau_{EP}(s=0.5) = T_{EP}/T_e = 120 $, no temperature gradient $R/\langle {L_T} \rangle_{EP} = 0$, and a logarithmic density gradient of $R/\langle {L_n} \rangle_{EP} = 28.2$.

We choose the destabilizing kinetic profiles with several goals in mind. First, to induce the desired instability while avoiding other instabilities that might exist in the system, e.g. an internal kink due to the $q < 1$ profile at the core $(s < 0.35)$. Second, to induce the microinstabilities and the AEs at a similar radial location, and third, to obtain a system where both the low-$n$ AE and the mid-$n$ ITG have comparable growth rates. Figure \ref{FIG:Combined_profiles_particles} presents the temperature, density and $q$ profiles for the standard and peaked cases as well as for the EPs. On the right in Figure \ref{FIG:Combined_profiles_particles} we present the $(\Delta')$, and the $(\beta)$ profiles of the 5 different MHD configurations accounting for bulk plasma and EP profiles.

\section{Results and Discussion}
\subsection{Linear simulations - dispersion relation and the mode structure}

We begin by performing a set of linear simulations with two objectives in mind. First, to gain a deeper understanding of the Alfv\'en eigenmodes (AE) and drift waves in our system with an emphasis placed on the effects of Shafranov shift. Second, obtaining a relevant case study for our future nonlinear investigations. Specifically, we are interested in a case where, for the same initial conditions, both an AE and a microinstability (ITG in our case) are excited at a similar radial location, and with comparable growth rates. In the linear regime, the toroidal modes in Fourier space are decoupled from each other, allowing us to study them separately.

We examine several cases of EP concentrations i.e. EP$_{fraction} = [0\% ; 1\% ; 3\% ;5\%]$, with the same density gradient. Even in such small fractions the energetic particles have a $\beta$ comparable to the bulk plasma, thus a self-consistent MHD equilibrium will depend significantly on the EP fraction. Therefore, throughout this work we denote 3 types of MHD equilibria. Those that account for EP pressure in addition to the bulk profiles as "Consistent" (Cons.), which also includes cases with $0\%$ EPs. Those that account only for the bulk pressure in cases where the EP $ [\cdot]\% > 0$ as "Not Consistent" (Not Cons.), and those that account for no pressure at all as $\beta = 0$ MHD.

\begin{figure} [H]
\begin{center}
\includegraphics[width=1\textwidth]{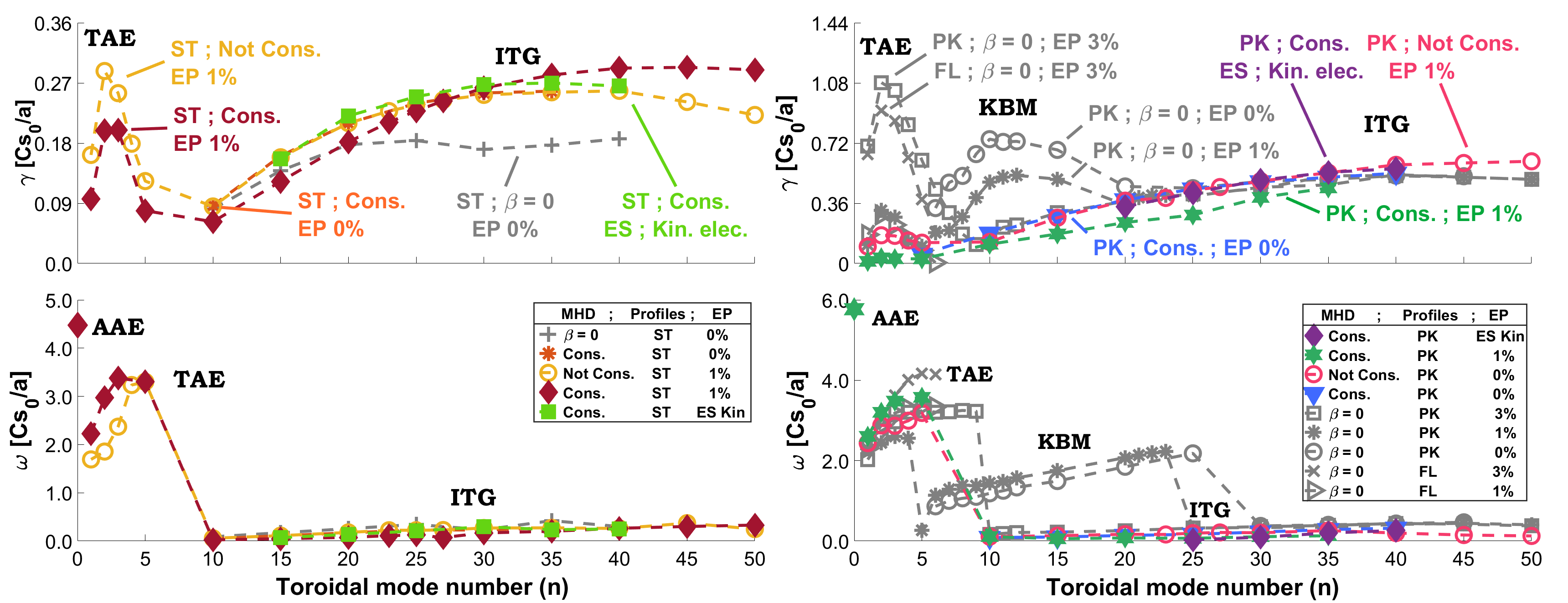}
\caption{\label{FIG:Linear_disperssion} \it
Linear growth rate $\gamma$ (top row) and real frequency $\omega$ (bottom row) vs. toroidal mode number $n$. 
On the left are ST cases, on the right cases with peaked (PK) and flat (FL) bulk gradients. 
Bottom plot left: frequency of the $n = 0$ Axisymmetric Alfv\'en eigenmodes (AAE).
Color coding: Filled markers - Consistent (Cons.), Empty markers - Not Consistent (Not Cons.), and Gray - $\beta = 0$ MHD equilibria.}
\end{center}
\end{figure}

Figure \ref{FIG:Linear_disperssion} presents the linear dispersion relations for the various cases, with mainly standard (ST) profiles on the left and mainly peaked (PK) profiles on the right. In the low-$n = [1-6]$ range we identify TAEs excited by the EPs, and absent without them. As anticipated by design, the $n = 2$ TAE has the highest growth rate. In the mid-$n$ range ($6 - 26$) for the peaked gradients we identify a KBM-like mode by its frequency in the bottom plot. It is worth pointing out, that ORB5 was used to successfully capture a ITG-KBM transition \cite{Cole_PoP2021}. And in the mid- to high-$n$ range $(10 - 50)$ depending on the scenario, we find the electromagnetic ITG branch.

\begin{figure}
\begin{center}
\includegraphics[width=0.325\textwidth]
{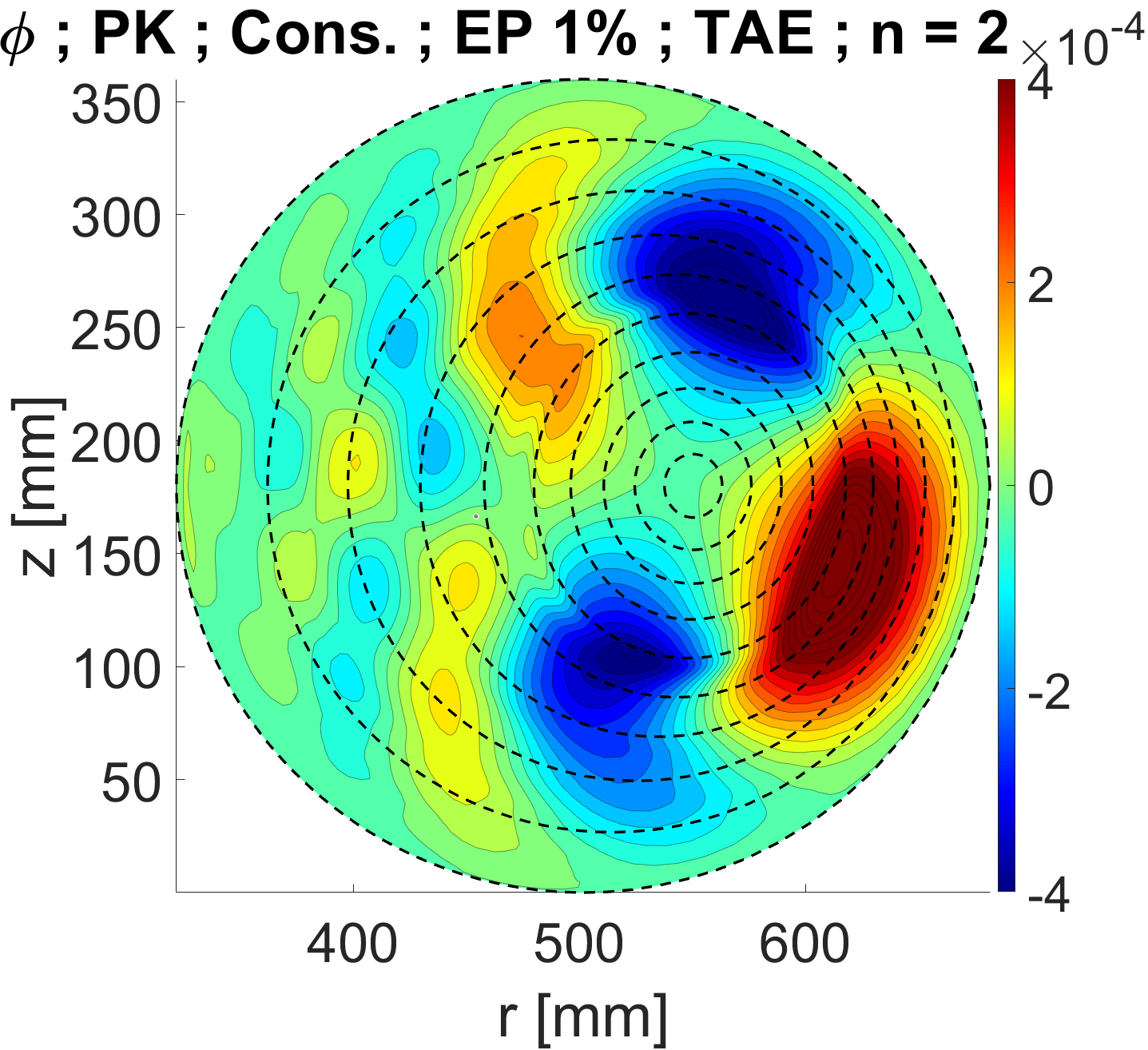}
\includegraphics[width=0.325\textwidth]{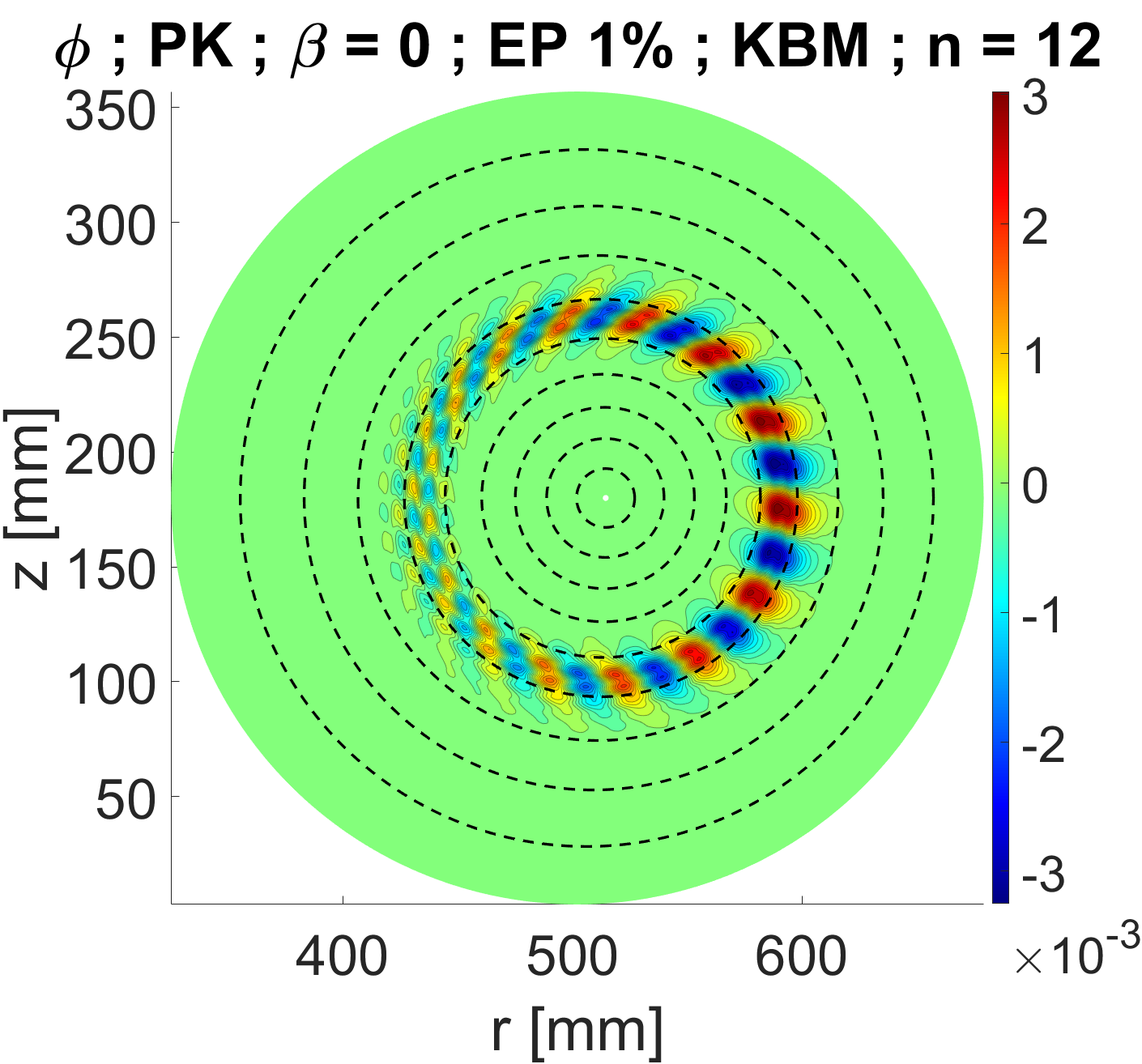}
\includegraphics[width=0.325\textwidth]
{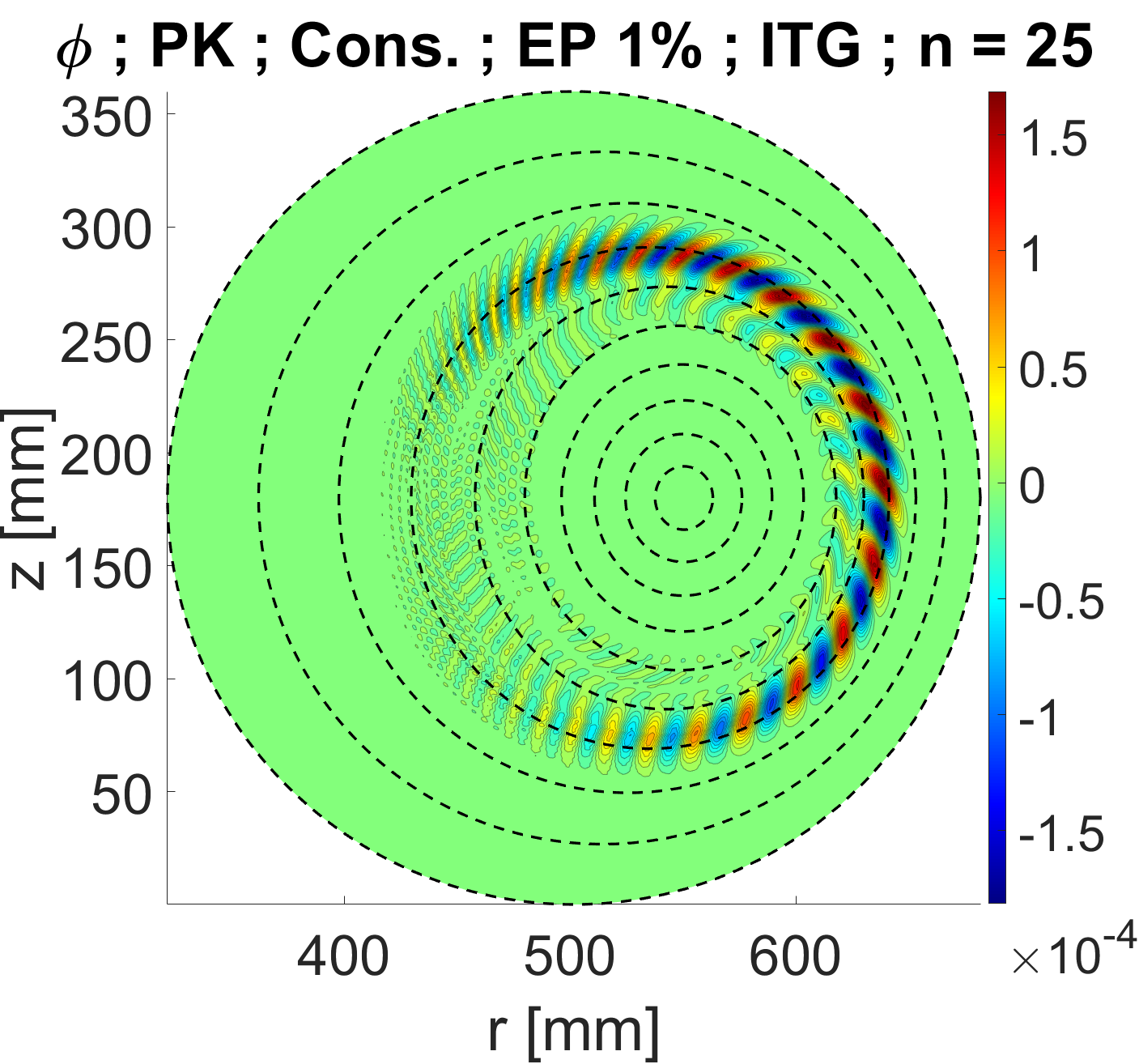}
\includegraphics[width=0.325\textwidth]{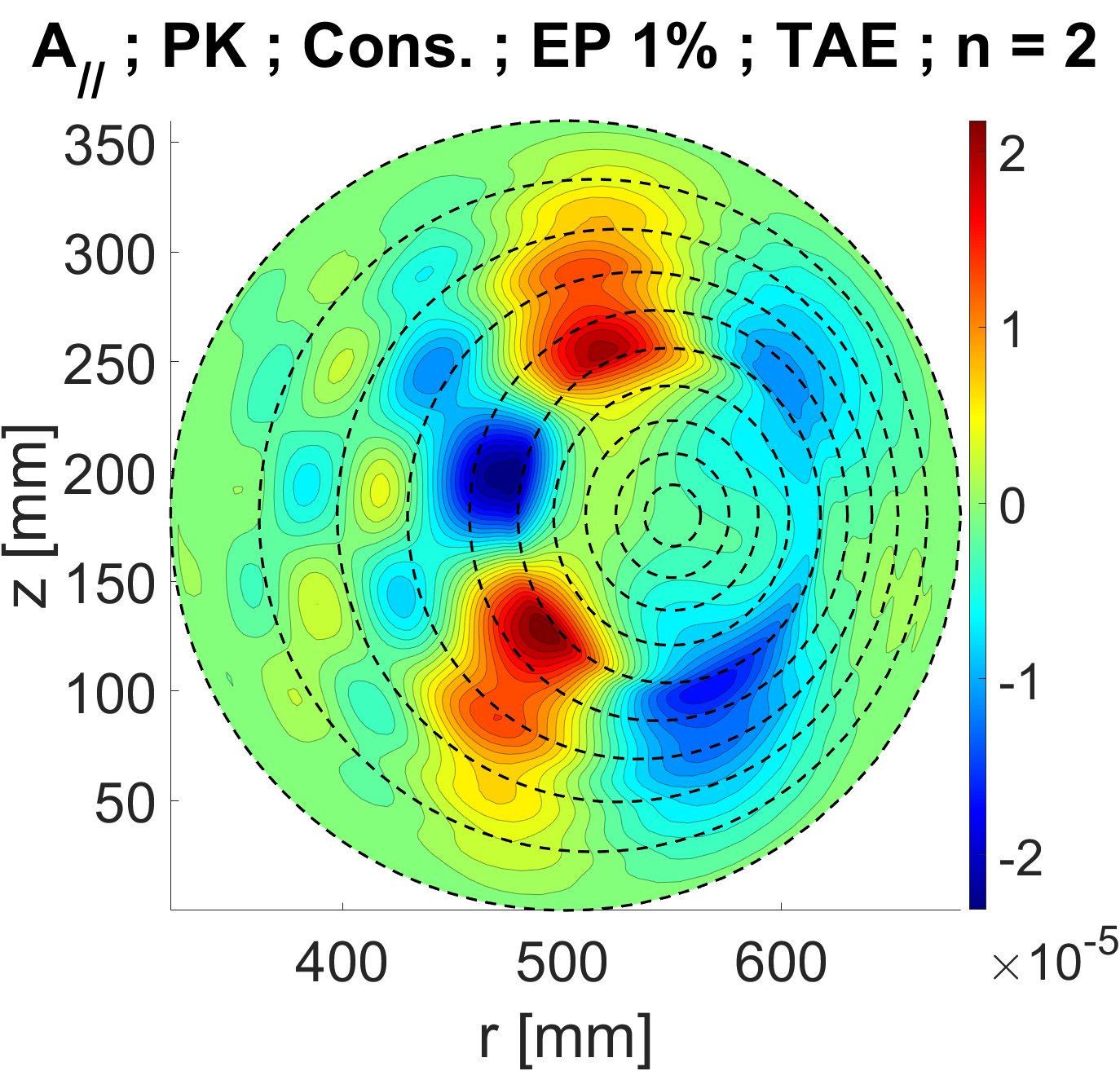}
\includegraphics[width=0.325\textwidth]{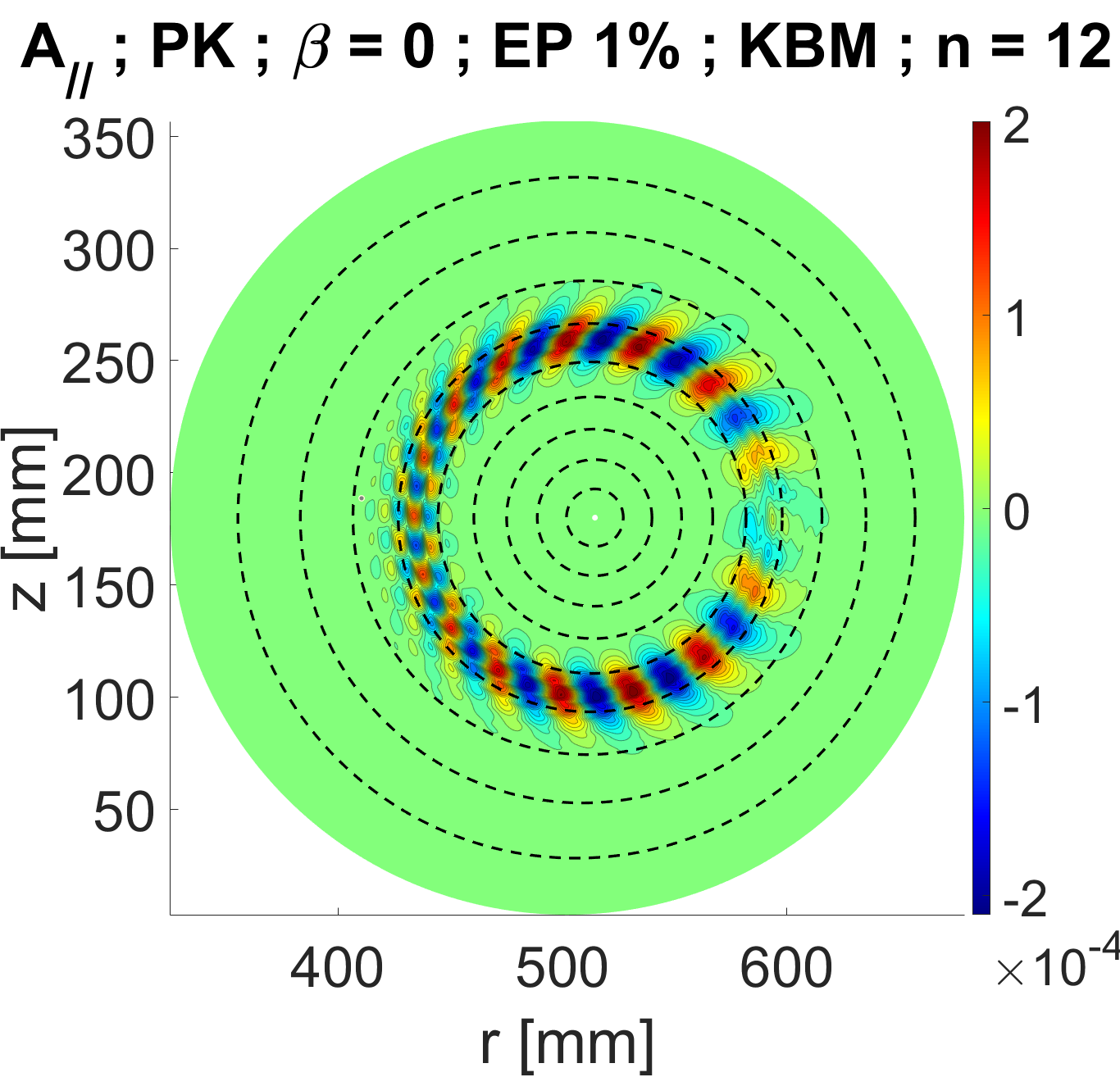}
\includegraphics[width=0.325\textwidth]{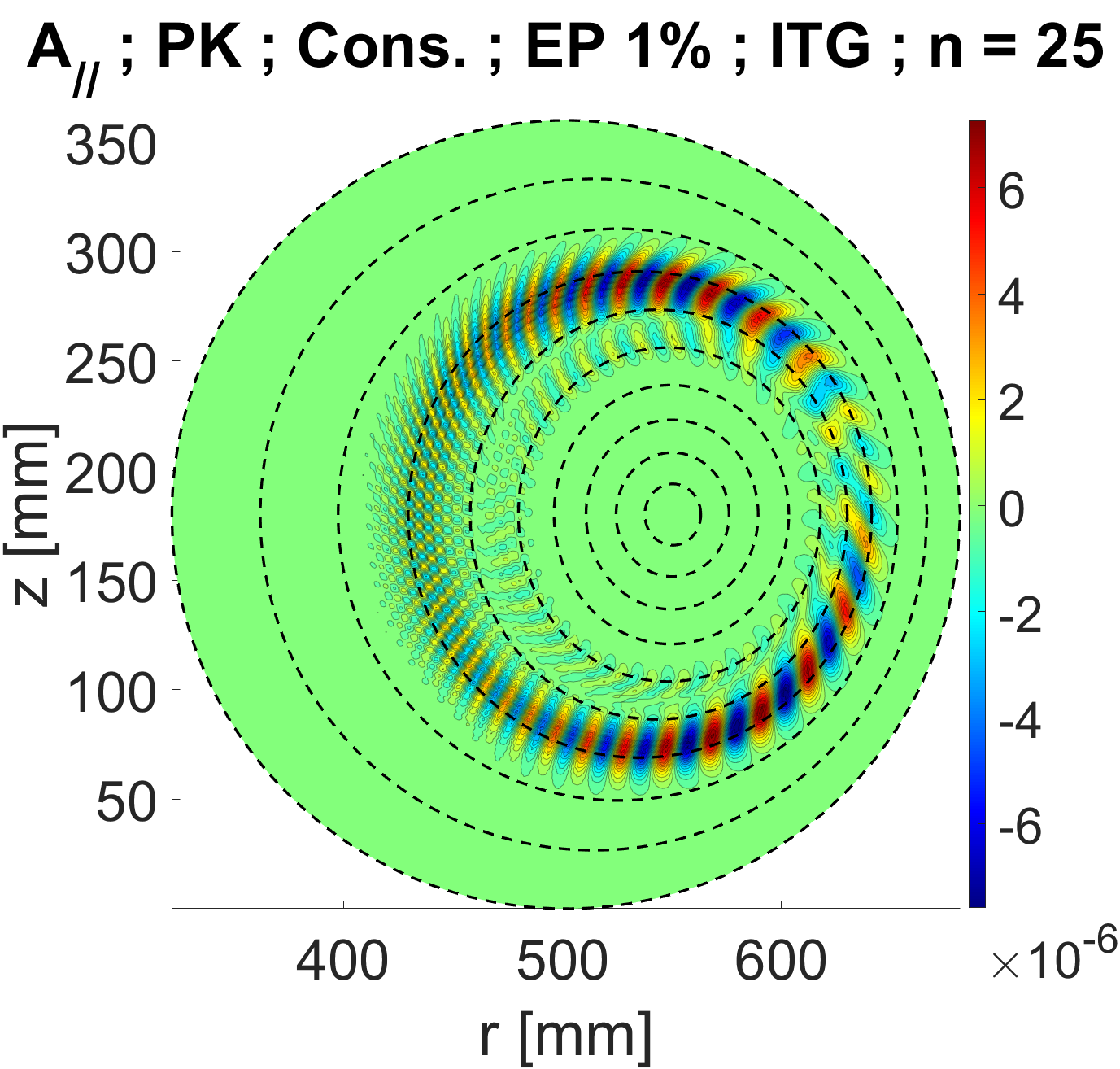}
\includegraphics[width=0.325\textwidth]
{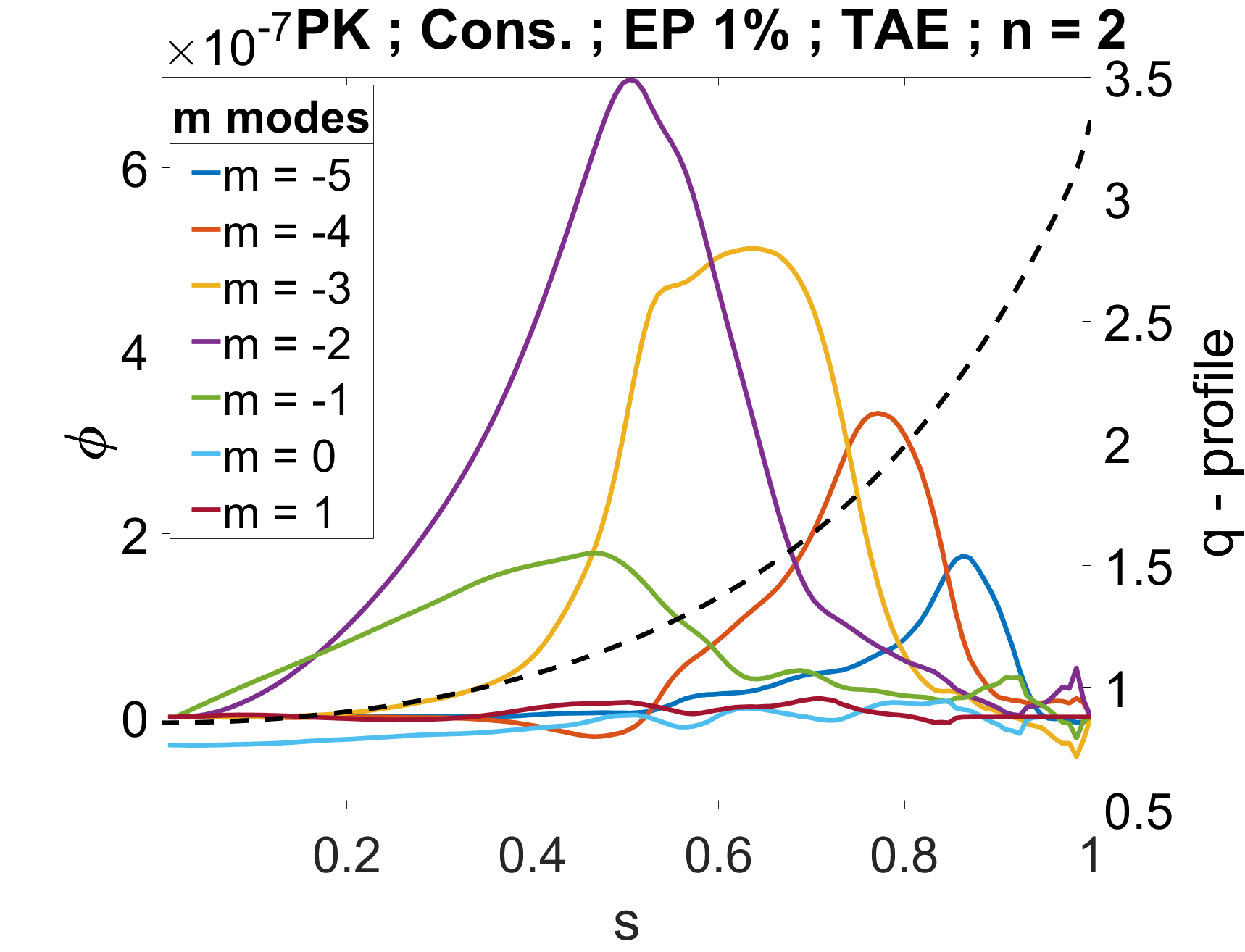}
\includegraphics[width=0.325\textwidth]{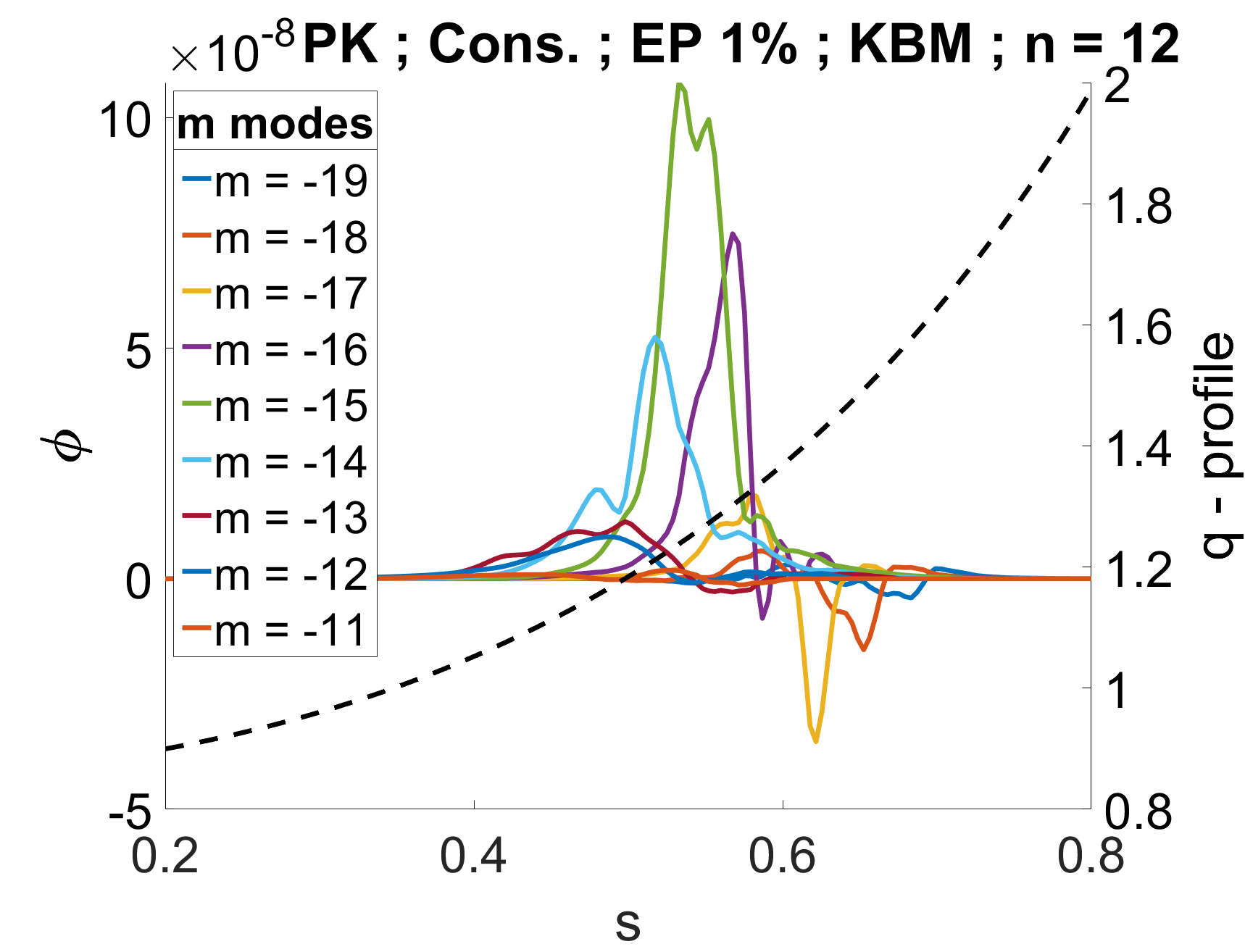}
\includegraphics[width=0.325\textwidth]{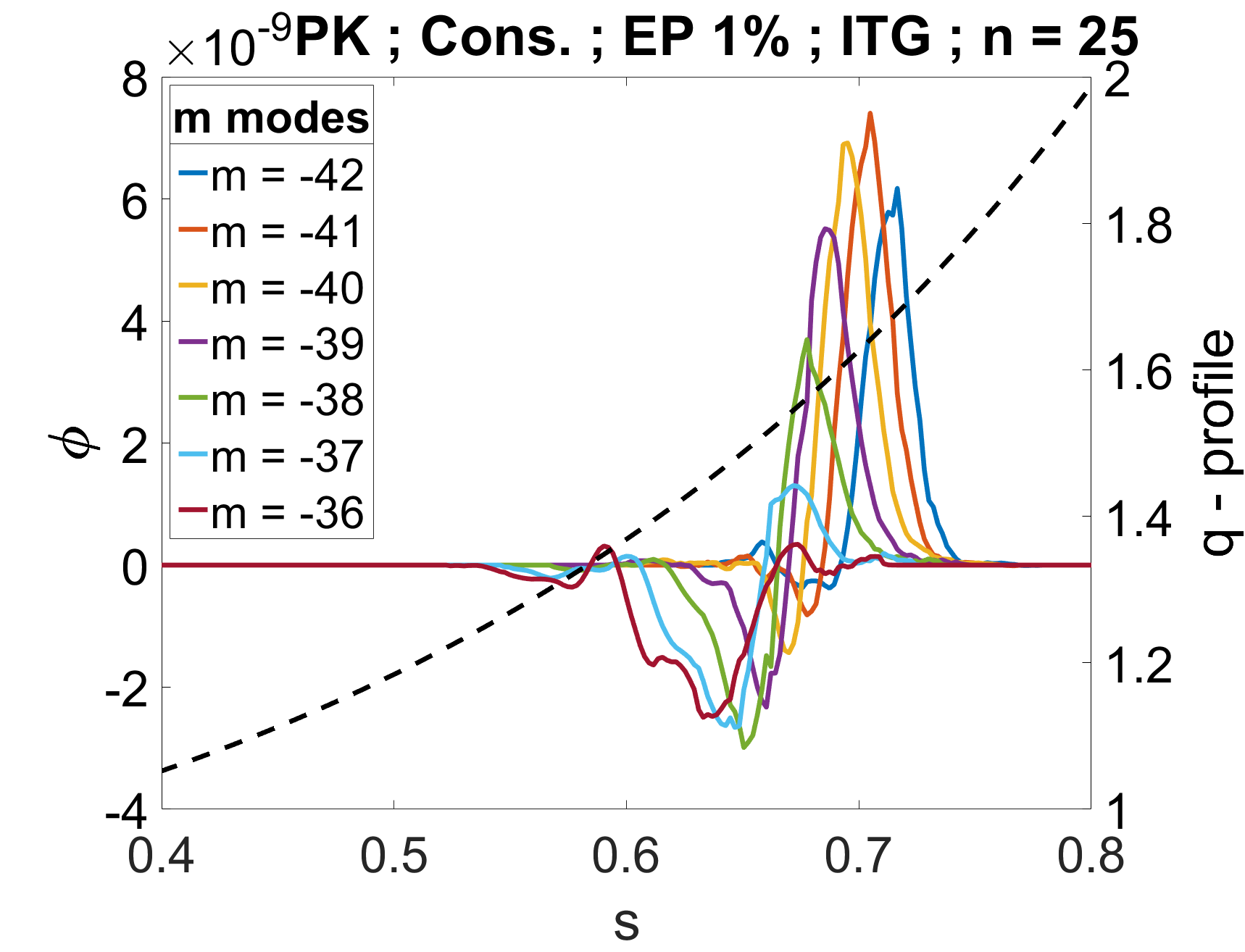}
\includegraphics[width=0.325\textwidth]{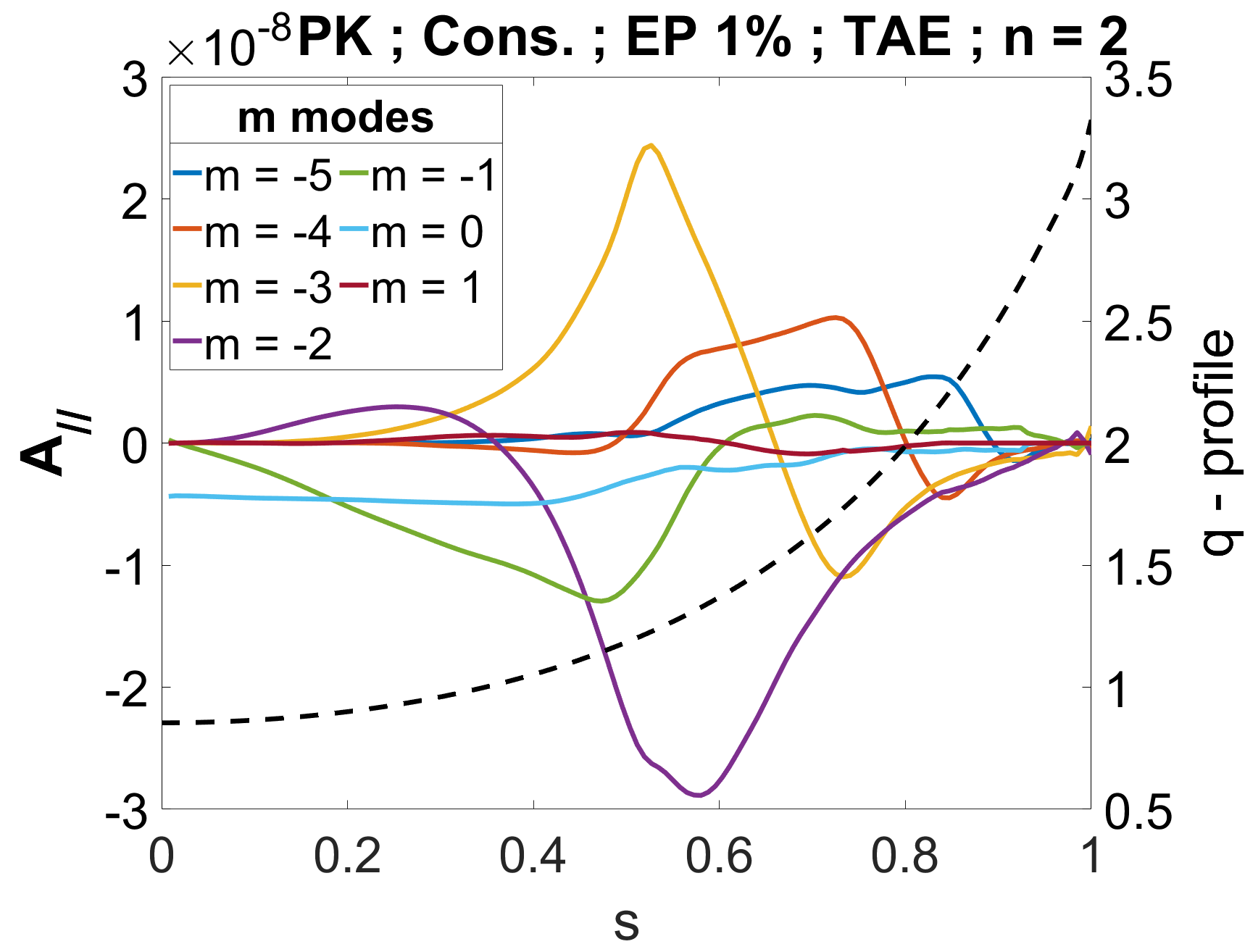}
\includegraphics[width=0.325\textwidth]{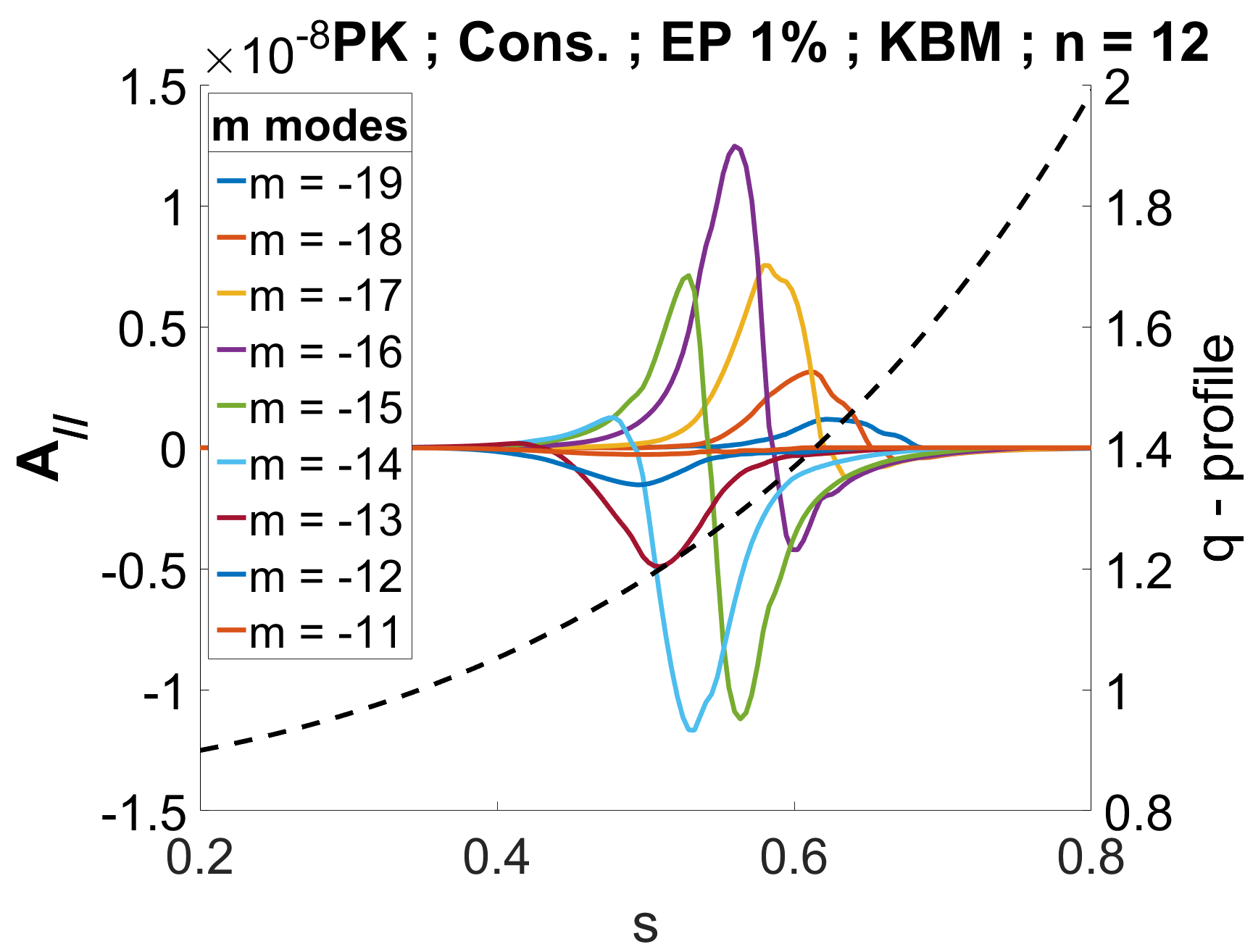}
\includegraphics[width=0.325\textwidth]{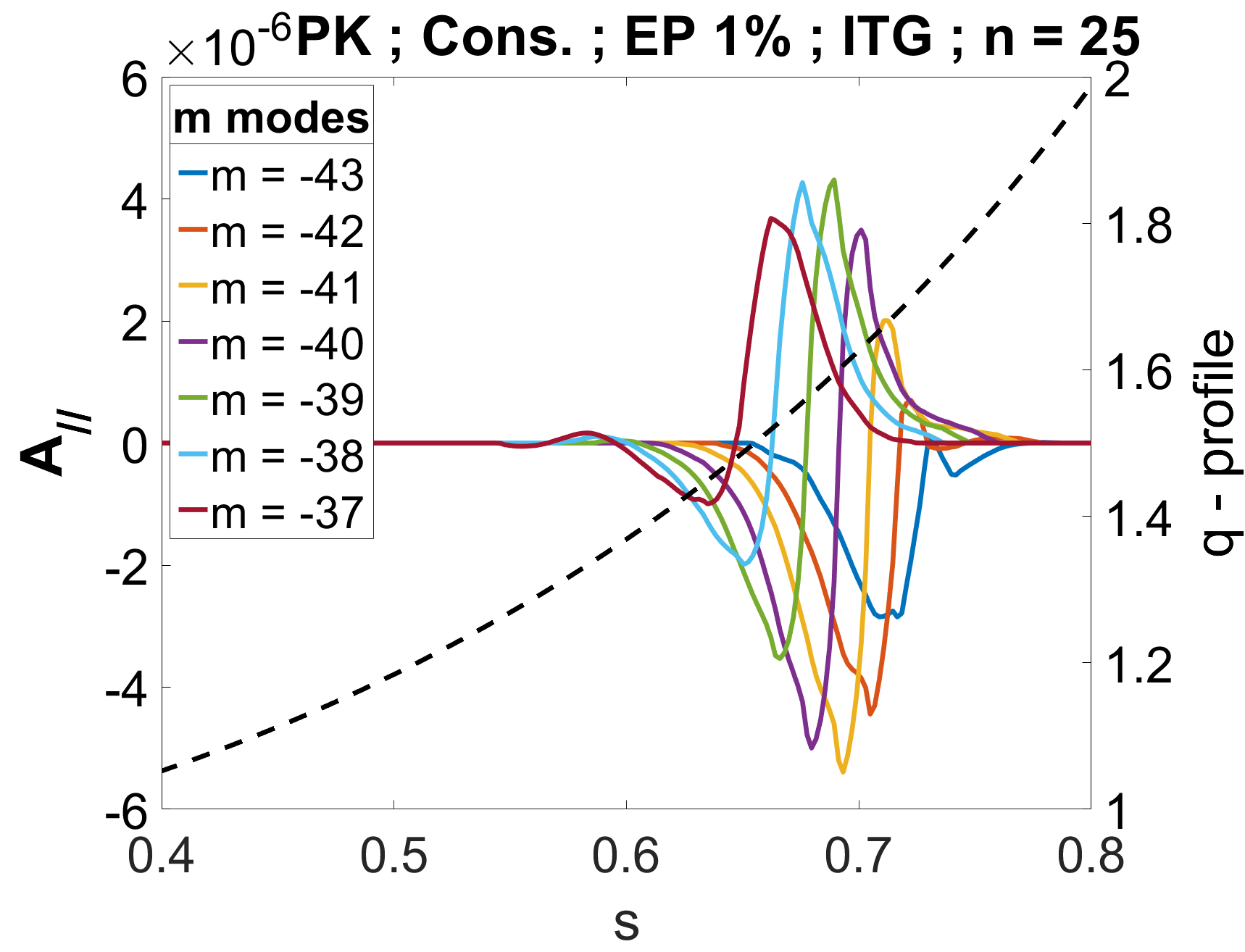}

\caption{\label{FIG:lin_mode_structure} \it 
Characteristic linear mode structure in $\phi$ ($1^{st}$ and $3^{rd}$ rows), and $A_{\parallel}$ ($2^{nd}$ and $4^{th}$ rows) of the $n = 2$ TAE (left column) and $n = 25$ ITG (right column) for cases with a self-consistent MHD equilibrium, and $n = 12$ KBM (middle column) in a case with a $\beta = 0$ MHD equilibrium. All cases are with peaked bulk gradients and include $1\%$ EPs.
The dashed black curves are magnetic surfaces $s=[0;0.1;1]$. In the bottom two rows we see the radial structure of the poloidal harmonics. Here the black dashed line is the q profile referring to the y-axis on the right.}
\end{center}
\end{figure}

Figure \ref{FIG:lin_mode_structure} presents, for scenarios with peaked gradients (PK) and $1\%$ EPs, the linear mode structure for a [Cons. ; $n = 2$] TAE, [$\beta = 0$ ; $n = 12$] KBM, and [Cons. ; $n = 25$] ITG. The low-$n$ TAE has a global structure, spanning the entire width of the plasma in a set of coupled poloidal harmonics, while the KBM and ITG have a radially localized structure. The Shafranov shift is clearly evident in the two self-consistent cases (TAE and ITG), in contrast to the $\beta = 0$ MHD of the KBM mode.

\subsection{Linear simulations: Alfv\'en Eigenmodes}
Toroidal Alfv\'en Eigenmodes (TAEs) are gap modes that are driven by unstable by EPs, through their real space (density) or velocity space (bump-on-tail) gradients. And peak at radial locations where $nq = (m + 1/2)$ with a typical frequency of $\omega_0 \cong \omega_A/2q$. Where $\omega_A = v_A/R$ is the Alfv\'en frequency, $v_A = B_0/\sqrt{\mu_0\rho_m}$ is the Alfv\'en velocity, and $\rho_m$ is the mass density.

\begin{figure}
\begin{center}
\includegraphics[width=0.395\textwidth] {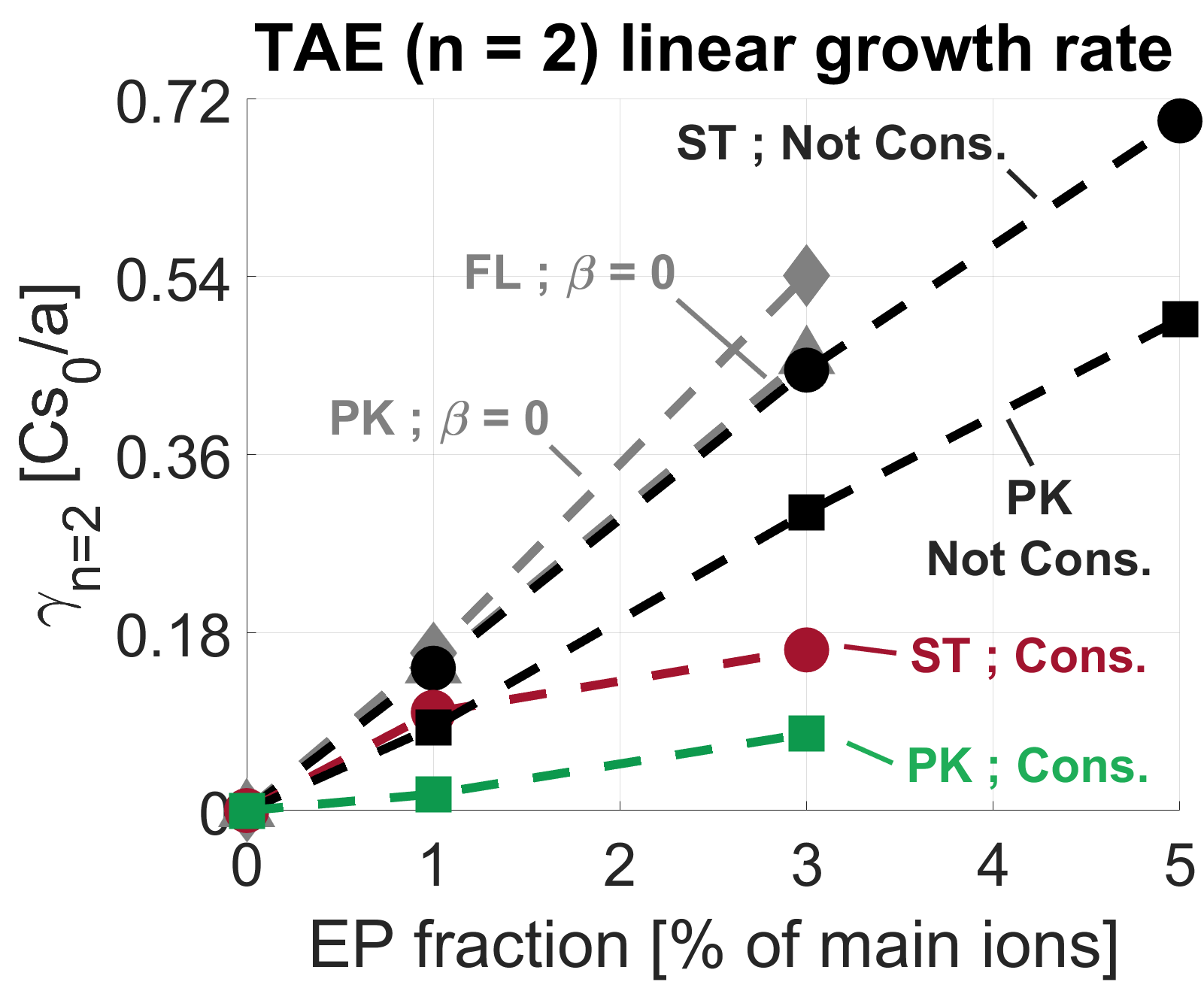}
\includegraphics[width=0.595\textwidth]{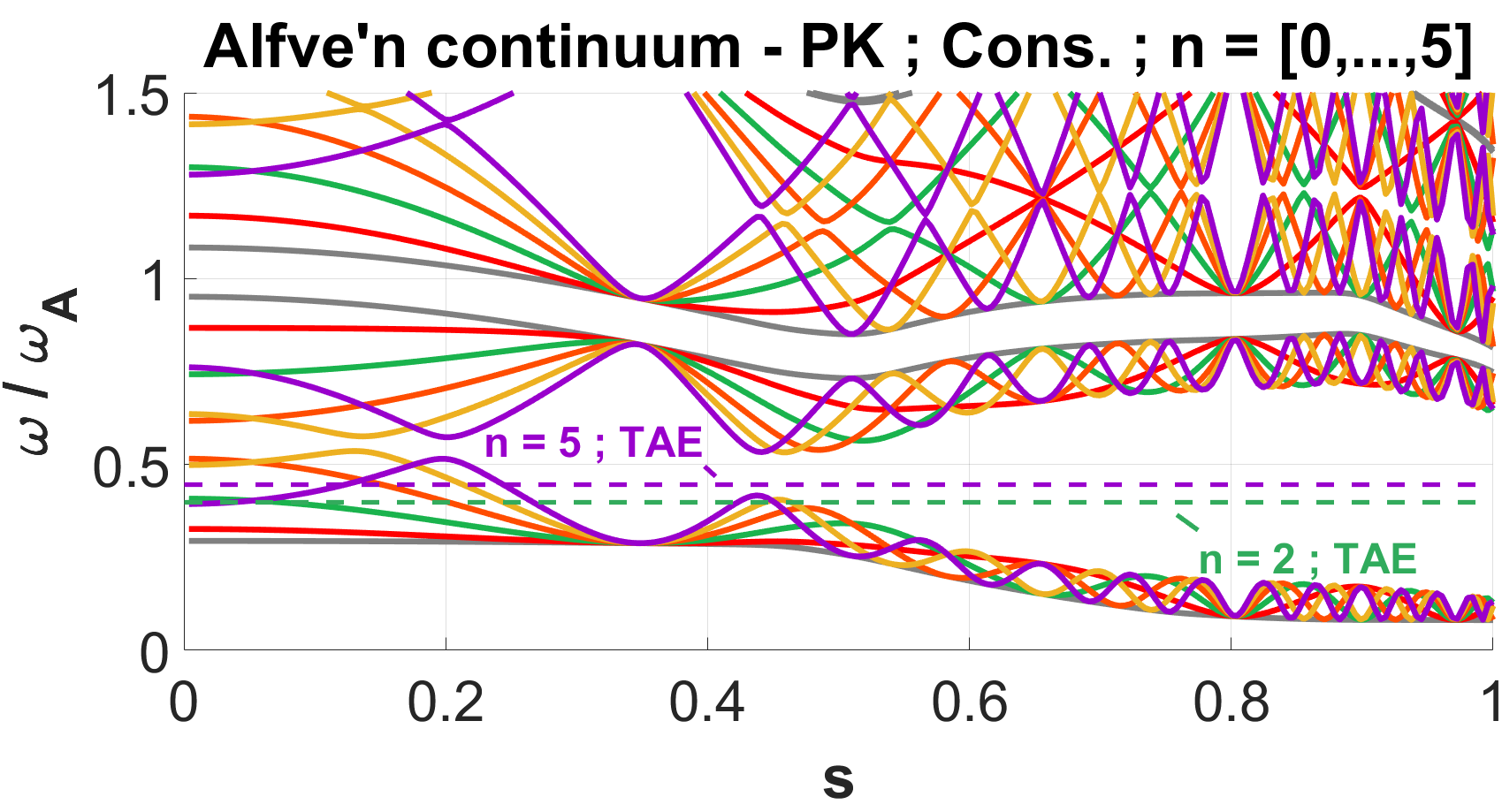}
\caption{\label{FIG:lin_TAE_GR_EP_scan} \it 
On the left is the growth rate of an $n=2$ TAE as a function of the EP fraction. 
Color coding: Gray - $\beta=0$, Black - Not Consistent (Not Cons.), Colored - Consistent (Cons.), MHD equilibria. The colors match Figure \ref{FIG:Linear_disperssion}.
The two plots on the right show the TAE's saturation level to growth rate ratio as a function of EP fraction.}
\end{center}
\end{figure}

From figure \ref{FIG:Linear_disperssion} we learn that the TAE is much more sensitive to Shafranov shift stabilization than the ITG mode. 
Thus, increasing the EP fraction increases both the drive and the Shafranov shift stabilization. In figure \ref{FIG:lin_TAE_GR_EP_scan} (left), we examine these competing effects on the growth rate of an $n = 2$ TAE by performing a scan in EP fraction for several scenarios, and notice several trends.

\begin{itemize}
    \item The TAE growth rate increases with EP fraction.
    \item The Shafranov shift due to bulk gradients stabilizes the TAE. Thus the TAE growth rate is higher in plasmas with standard gradients (ST) than with peaked gradients (PK), i.e. $ \gamma_{[\text{ST, Not Cons.}]} > \gamma_{[\text {PK, Not Cons.}]}$ and $ \gamma_{[\text{ST, Cons.}]} > \gamma_{[\text {PK, Cons.}]}$ for all EP fractions.
    \item Additional Shafranov shift due to EP pressure (on top of the bulk) stabilizes the TAE as well, leading to a saturation in mode growth rate with EP fraction, i.e.
    \begin{equation*}
    (\gamma_{[\text{ST, Cons.}]} / \gamma_{[\text {ST, Not Cons.}]})_{ \text{EP } 1\%} > (\gamma_{[\text{ST, Cons.}]} / \gamma_{[\text {ST, Not Cons.}]})_{ \text{EP } 3\%}
    \end{equation*}  
    \item When the Shafranov shift effect is neglected (the $\beta = 0$ case) the growth rate is the highest and slightly increases with bulk gradients, an opposite trend to the cases with Shafranov shift.  
\end{itemize}

\FloatBarrier
Furthermore, from the bottom plots in figure \ref{FIG:Linear_disperssion} we see that the TAE frequency increases with both the EP fraction,

\begin{equation*}
    (\omega^{\small TAE}_{\small[\text{$\beta = 0$, $ \text{EP } 3\%$}]} / \omega^{\small TAE}_{\small[\text{$\beta =0$, $ \text{EP } 1\%$}]})_{[FL;PK]}>1 
\end{equation*}

and Shafranov shift, e.g. for EP $1\%$

\begin{equation*}
    \omega^{\small TAE}_{\small[\text{Cons., $ \text{EP } 1\%$}]} >\omega^{\small TAE}_{\small[\text{Not Cons., $ \text{EP } 1\%$}]} >\omega^{\small TAE}_{\small[\text{$\beta = 0$, $ \text{EP } 1\%$}]}
\end{equation*}.

Figure \ref{FIG:lin_TAE_GR_EP_scan} shows on the right the Alfv\'en continua frequencies under the slow-sound approximation for the $n = [0 ... 5]$ toroidal modes, with the $n = [2,5]$ TAEs indicated by the dashed lines.  
\FloatBarrier

\begin{figure}
\begin{center}
\includegraphics[width=0.495\textwidth] {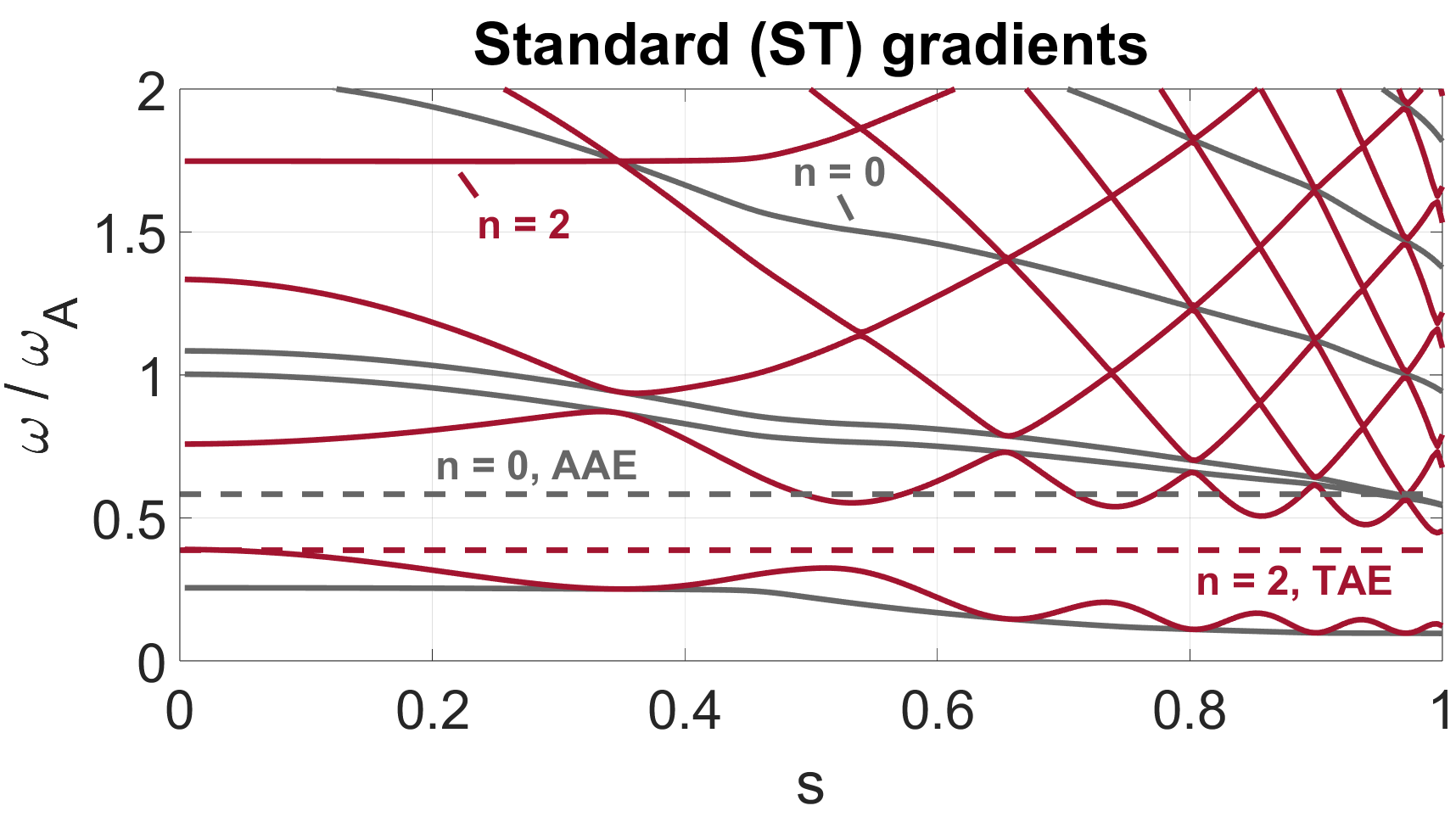}
\includegraphics[width=0.495\textwidth] {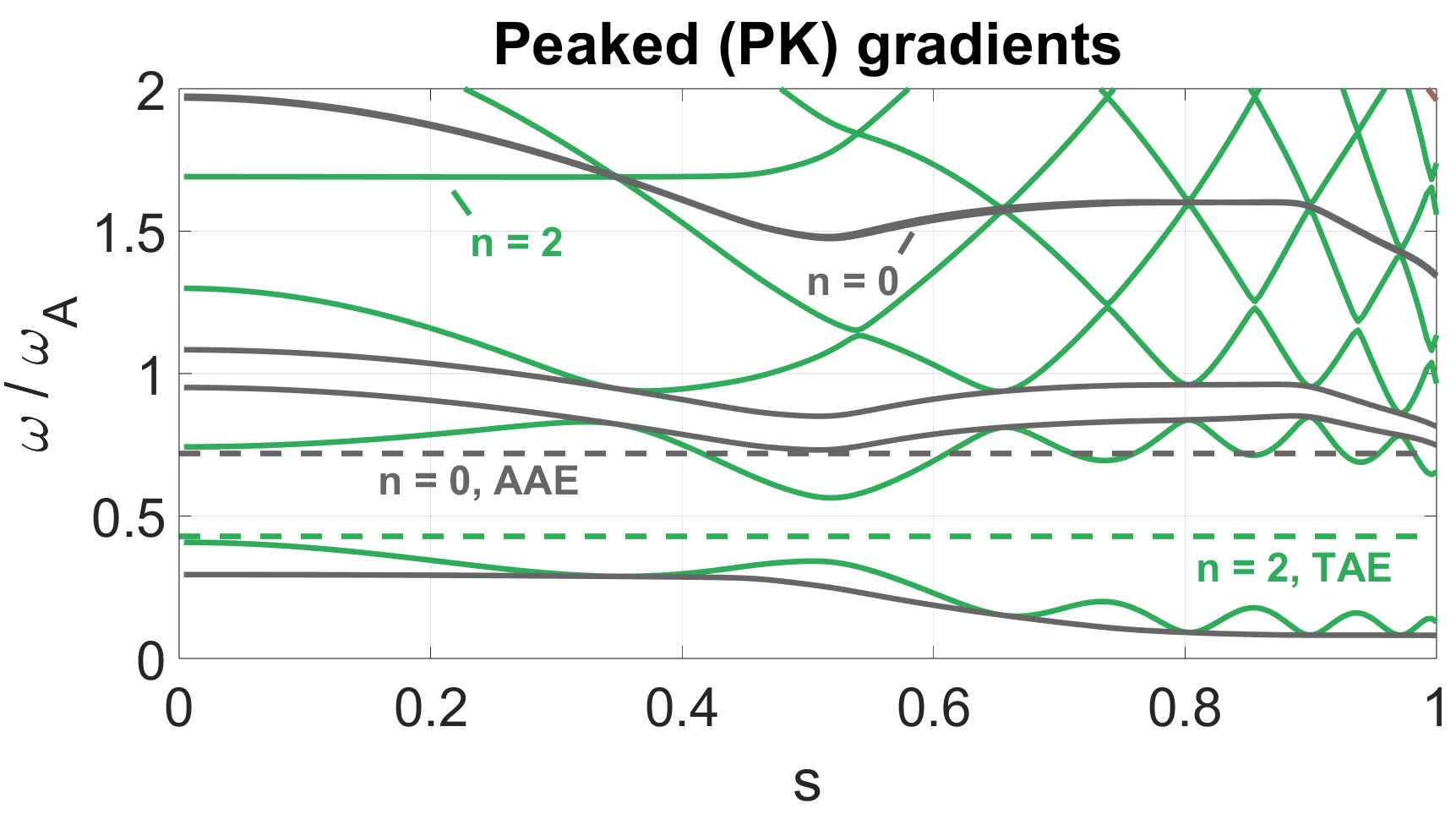}
\caption{\label{FIG:TAE_conttinuum} \it 
The Alfv\'en continuum with the slow-sound approximation for the $n = 0$ $\&$ $n = 2$ toroidal mode numbers, for standard and peaked profiles with consistent MHD equilibria. The labeled dashed lines indicate the frequency (in Alfv\'enic units), of either the $n = 2$ TAE or the $n = 0$ Axisymmetric Alfv\'en Eigenmode (AAE). 
Color coding: Dark gray - $n = 0$, Colored - $n = 2$ modes. The colors match Figure \ref{FIG:Linear_disperssion}.}
\end{center}
\end{figure}

To further investigate the TAE behavior, we plot in Figure \ref{FIG:TAE_conttinuum} the $n = [0 , 2] $ Alfv\'en continua. And, because the density profiles of the consistent and not consistent cases are very similar, we only plot the consistent cases. Especially for the Peaked bulk density profiles, the gaps happen to align well along the radial axis, allowing for a TAE to spread across a wide radial range without much continuum damping. To further illustrate this point, we mark the $n = 2$ TAE frequency in Alfv\'enic units with horizontal colored dashed lines. Due to finite $\beta$ effects, the BAE gap opens on the bottom of the frequency range due to coupling with the ion sound continuum. The continuum structure does not offer a clear explanation for the strong Shafranov shift stabilization of the TAE modes. Therefore, further investigation into the effect of Shafranov shift on phase space resonances is needed and left for future work.

Under the conditions explored in this work, the $n=0$ AAE mode is linearly stable and nonlinearly unstable. The frequency of the mode is obtained from the data presented in figure \ref{FIG:nonlinear_Omega_ExB_2D} and will be discussed further in the nonlinear section. The mode frequency (gray dashed lines in Figure \ref{FIG:TAE_conttinuum}) is located near the minima of the $n = 0$ Alfv\'en continuum. Notably, for the PK cases, the strong Shafranov shift leads to a new minimum in the $n = 0$ continuum found around $s=0.53$. While for the ST cases it is the edge of the continuum which dictated the AAE's frequency (minmum). This is in good agreement with the experimental observations on TFTR \cite{Chang_NF1995}, and the theoretical explanation offered by Villard $\&$ Vaclavik \cite{Villard_NF1997}.

\subsection{Linear simulations: Electromagnetic KBM and ITG}
%
In Figure \ref{FIG:Linear_disperssion}, we first turn our attention to the gray plots representing the $\beta = 0$ MHD equilibrium on the right (PK). In the mid-$n$ range ($6 - 26$) we identify a KBM mode by its frequency found in the bottom right plot. The KBM is unstable for peaked kinetic gradients (PK), while the mode remains stable for the ST cases on the left. We can identify two stabilizing mechanisms in our simulations. First is the EP stabilization \cite{Rosenbluth_PRL1983}, which shows as the decrease in growth rate with the increase in EP fraction from $0\%$ to $3\%$. At $3\%$ EPs, the KBM is completely suppressed. The second mechanism is Shafranov shift stabilization; already for the case with no EPs the KBM is fully stabilized. Qualitatively similar results were obtained by Aleynikova et. al. \cite{Aleynikova_PoP2018}, (see Figs. 10, 11 of that paper) where it was concluded it is essential to consider self-consistent MHD equilibria.    

The electromagnetic ITG branch is found in Figure \ref{FIG:Linear_disperssion} in the mid- to high-$n$ range ($n = 10$ or $20$, depending on the scenario to $n = 50$). Unlike the KBM, the ITG instability is much less sensitive to EP stabilization for both standard (ST) and peaked (PK) bulk gradients. This is evident in the similar dispersion relation of the [ST, Cons., EP $0\%$] and [ST, Not Cons., EP $1\%$] cases, which have the same MHD equilibrium and differ only by the EP fraction.

\begin{figure}
\begin{center}
\includegraphics[width=0.495\textwidth]{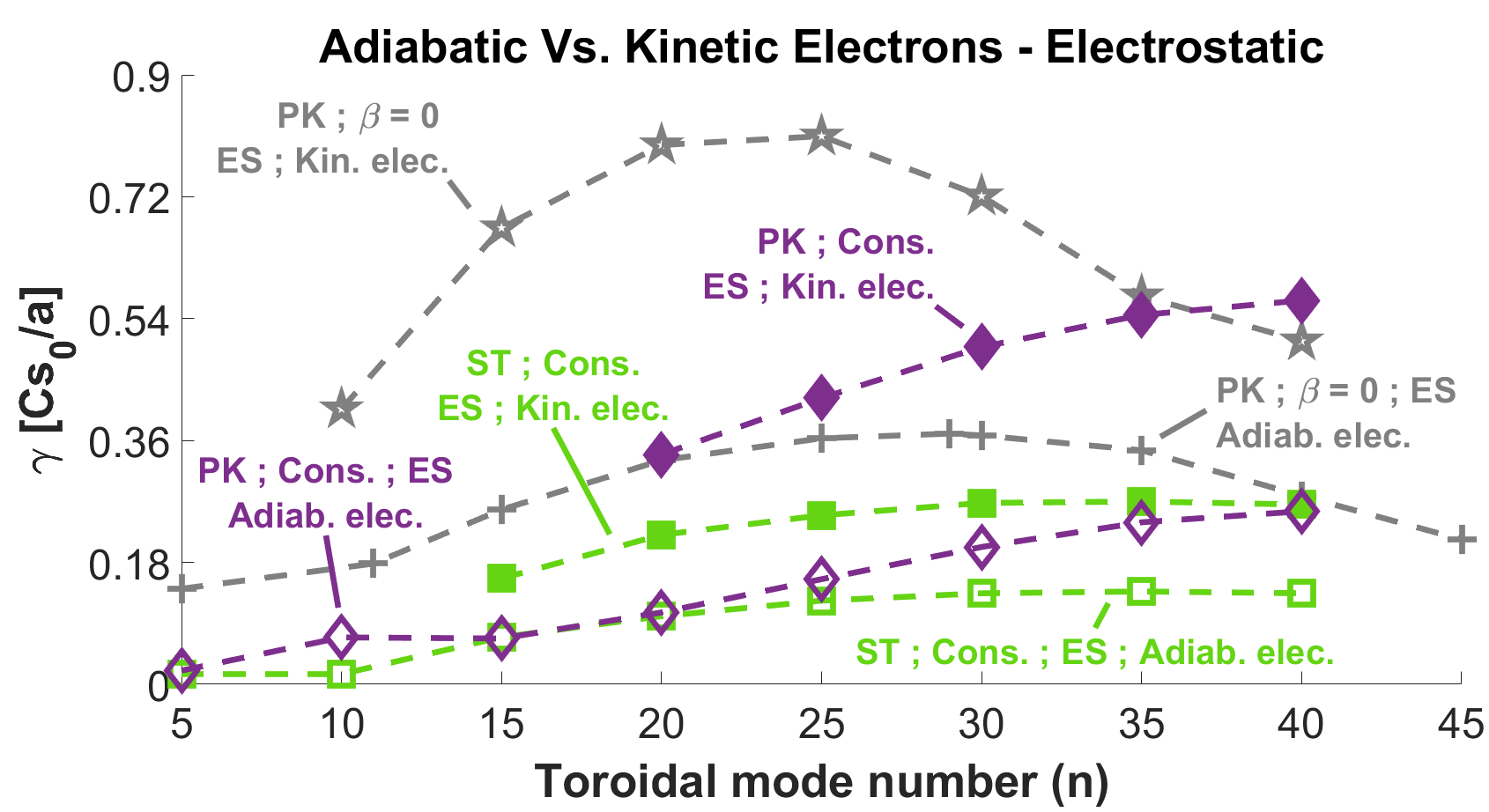}
\includegraphics[width=0.495\textwidth]{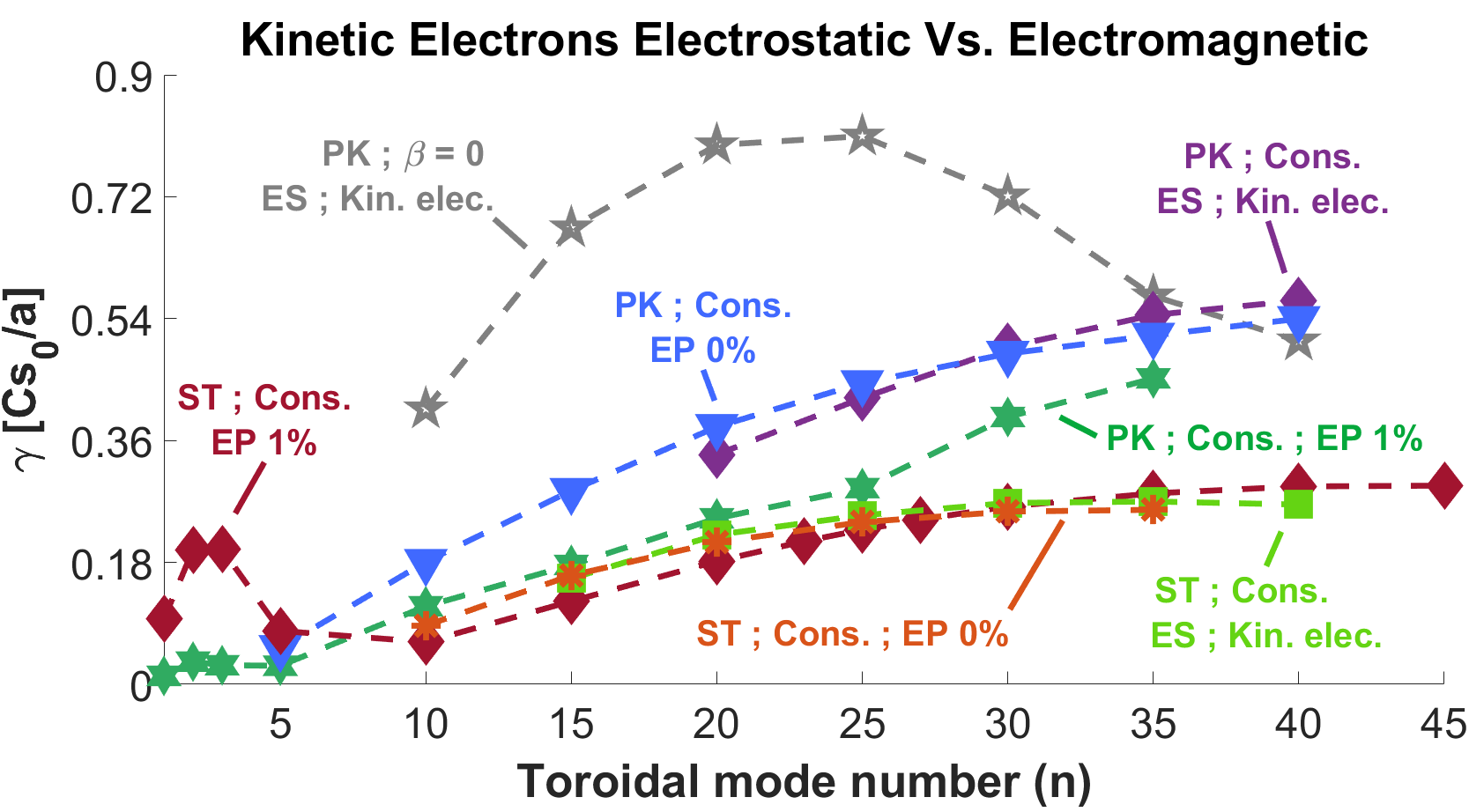}
\caption{\label{FIG:lin_ITG} \it 
Growth rate $\gamma$ per toroidal mode number, comparing between adiabatic and kinetic electrons for electrostatic cases on the left and between electrostatic and electromagnetic cases with kinetic electrons on the right.
Color coding indicates cases with: Gray - $\beta=0$ MHD equilibrium, light green - standard gradients (ST) and purple - peaked gradients (PK) consistent MHD equilibria. Full markers are for cases with kinetic electrons, and hollow markers are for cases with adiabatic electrons.The color coding of the curves in the figure on the right is identical to one used in \ref{FIG:Linear_disperssion}.}
\end{center}
\end{figure}

 For comparison, we specifically use the results of an electrostatic simulation with kinetic electrons (ES Kin.). A strong Shafranov shift stabilization is evident in the difference between the [PK ; $\beta = 0$ ; ES Kin.] and the [PK ; Cons. ; ES Kin.] cases in Figure \ref{FIG:lin_ITG} on the left. 

The Shafranov shift stabilization appears to be toroidal mode dependent; with a stronger effect on the lower-$n$ modes. A good example of this is found in the cases just mentioned and in the [PK ; Cons. ; EP 1$\%$] case presented in Figure \ref{FIG:Linear_disperssion} on the right. A strange outlier is found in the [ST ; $\beta = 0$ ; EP $0\%$] case (Fig. \ref{FIG:Linear_disperssion} on the left) when compared to its consistent counterpart [ST ; Cons. ; EP $0\%$] which has both a higher growth rate and a stronger Shafranov shift.

In Figure \ref{FIG:lin_ITG} on the left we recover the result of Dominski et. al. \cite{Dominski_PoP2015}, by showing that cases with adiabatic electrons have about half the growth rate of the cases with kinetic electrons, i.e. $ \gamma_{adiab. \ elec.} \approx 0.5 \ \gamma_{kin. \ elec.} $. And in Figure \ref{FIG:lin_ITG} on the right we find that in our case, for the ITG instability, there is a reasonably good match (in the dispersion relation) between the electrostatic cases with kinetic electrons and the electromagnetic cases under the same conditions.

The combined effects of bulk gradients on the drive, and Shafranov shift stabilization of the ITG instability, results in the PK cases being more unstable than ST cases, in contrast to the trend found for TAEs (Fig. \ref{FIG:lin_TAE_GR_EP_scan}) which are mainly stabilized by increased bulk gradients. 

\section{Nonlinear simulations}
Following the linear results we choose several representative cases for the nonlinear study. We perform the simulations on a single toroidal mode number with or without an additional axisymmetric component ($n = 0$). For the Alfv\'en physics (TAE), we choose $n = 2$, and for the ITG microinstability we chose $n = 25$. All the simulations, unless stated otherwise, include $1\%$ EP. 
Our nonlinear study focuses on exploring the emerging ZS and their feedback on the saturation mechanism of the TAE and microturbulence. We further highlight the changes in plasma behavior arising from taking into account the EP pressure in the MHD equilibrium. For the peaked background profiles we include the KBM-like mode which is present in the $\beta = 0$ limit of the MHD equilibrium, while keeping in-mind that the KBM mode instability is the result of an inconsistent MHD equilibrium as shown in the previous section. In this section we explore the system's time-evolution, from the linear phase to the early nonlinear phase. This can be generally divided into two stages:
\begin{itemize}
    \item Stage 0: The linear phase which was discussed in the previous section. During this stage the growth rate of the mode is dominated by linear physics, and the zonal component grows due to numerical noise.  
    \item Stage 1: The saturation phase begins when the amplitude of the unstable mode becomes sufficiently high for nonlinear effects to become significant. At this stage, the zonal response growth rate deviates from that of the driving mode, indicating the existence of a nonlinear physical drive. At the end of this phase the zonal structures are already established and strong bursts of flux occur. In this work we refer to the beginning of this stage as "quasi-linear" 

    \item Stage 2: The early nonlinear phase, when the system settles down due to a newly achieved balance between nonlinear damping and drive in the presence of fluxes and zonal structures. Here we can discuss about time and space average quantities.
\end{itemize}

In the following sections we analyze the system behavior in these stages. We study both the transient dynamical response and the long lasting structures of the fields, the arising $E \times  B$ and parallel flows, and heat and particle fluxes.

\begin{figure}
\begin{center}
\includegraphics[width=1\textwidth]{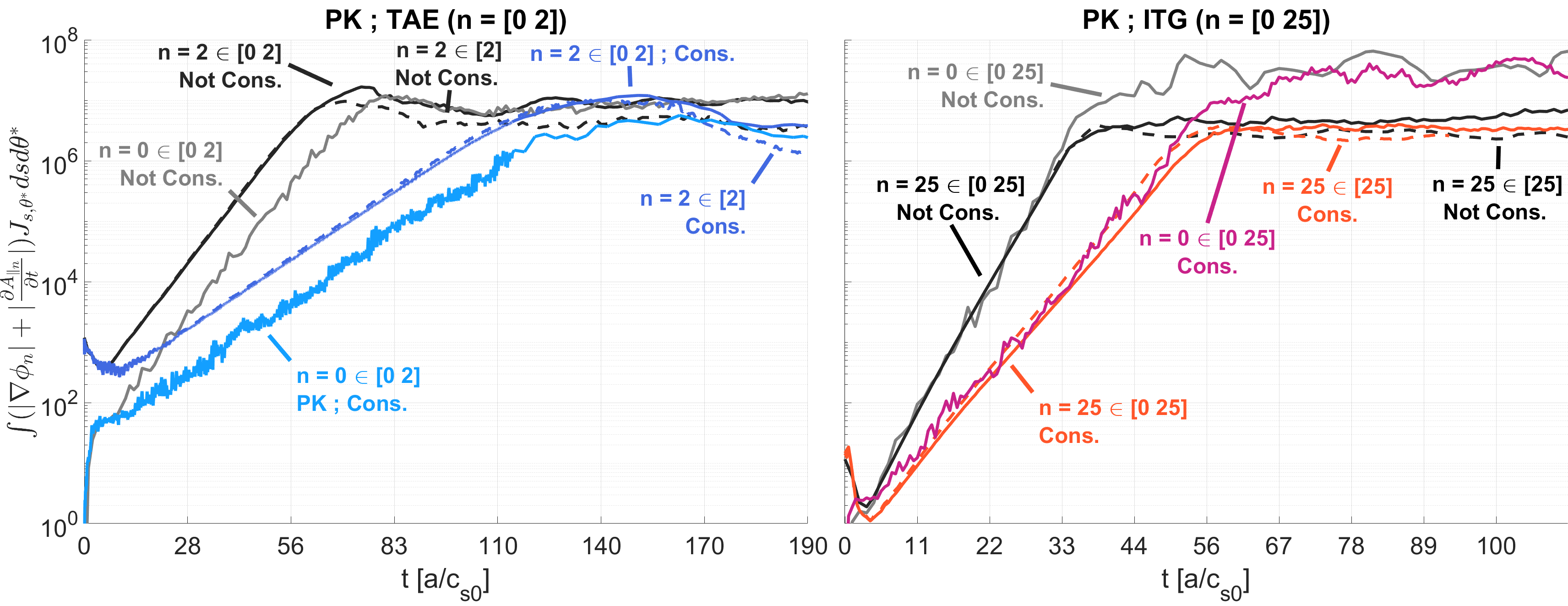}
\caption{\label{FIG:nonlinear_timetrace} \it
Time evolution of the electric field $ |E_n(t)| = \int (| \nabla \phi_n| + |\frac{\partial A_{\parallel_n}}{\partial t}|)J_{s,\theta^*} ds d\theta^* $, up to the early nonlinear phase, after the initial saturation. Nonlinear simulations contain either a single toroidal mode or that mode and $n=0$ mode (allowing for self generated ZS).} 
\end{center}
\end{figure}

The nonlinear results are either time-averaged or presented in the middle of a time window of $ \sim 55 \ [a/c_{s_0}] = 10^3 \ [\Omega^{-1}_{c_i}]$ situated in the early nonlinear saturation phase. Large variations in the linear growth rates result in similar variations in the time of the early nonlinear phase. 

\begin{itemize}
    \item ST ; Cons ; TAE $n \in{[2],[0,2]}$ : $\Delta$t =$(111 - 172) \ [a/c_{s_0}] \approx (2 - 3.1)\cdot10^4 \ [\Omega^{-1}_{c_i}]$
    \item ST ; Cons. ; ITG $n \in {[25],[0,25]}$ : $\Delta$t =$(89 - 144) \ [a/c_{s_0}] \approx (1.6 - 2.6)\cdot10^4 \ [\Omega^{-1}_{c_i}]$
    \item PK ; Cons. ; TAE $n \in{[2],[0,2]}$ : $\Delta$t =$(194 - 278) \ [a/c_{s_0}] \approx (3.5 - 5)\cdot10^4 \ [\Omega^{-1}_{c_i}]$
    \item PK ; Cons. ; ITG $n \in {[25],[0,25]}$ : $\Delta$t =$(66 - 139) \ [a/c_{s_0}] \approx (1.2 - 2.5)\cdot10^4 \ [\Omega^{-1}_{c_i}]$
    \item PK ; $\beta = 0$ MHD ; TAE {[2],[0,2]} : $\Delta$t =$(66 - 110) \ [a/c_{s_0}] \approx (1.2 - 2.0)\cdot10^4 \ [\Omega^{-1}_{c_i}]$
    \item PK ; $\beta = 0$ MHD ; KBM {[12],[0,12]} : $\Delta$t =$(66 - 128) \ [a/c_{s_0}] \approx (1.2 - 2.3)\cdot10^4 \ [\Omega^{-1}_{c_i}]$
\end{itemize}

\begin{figure}
\begin{center}
\includegraphics[width=0.33\textwidth]{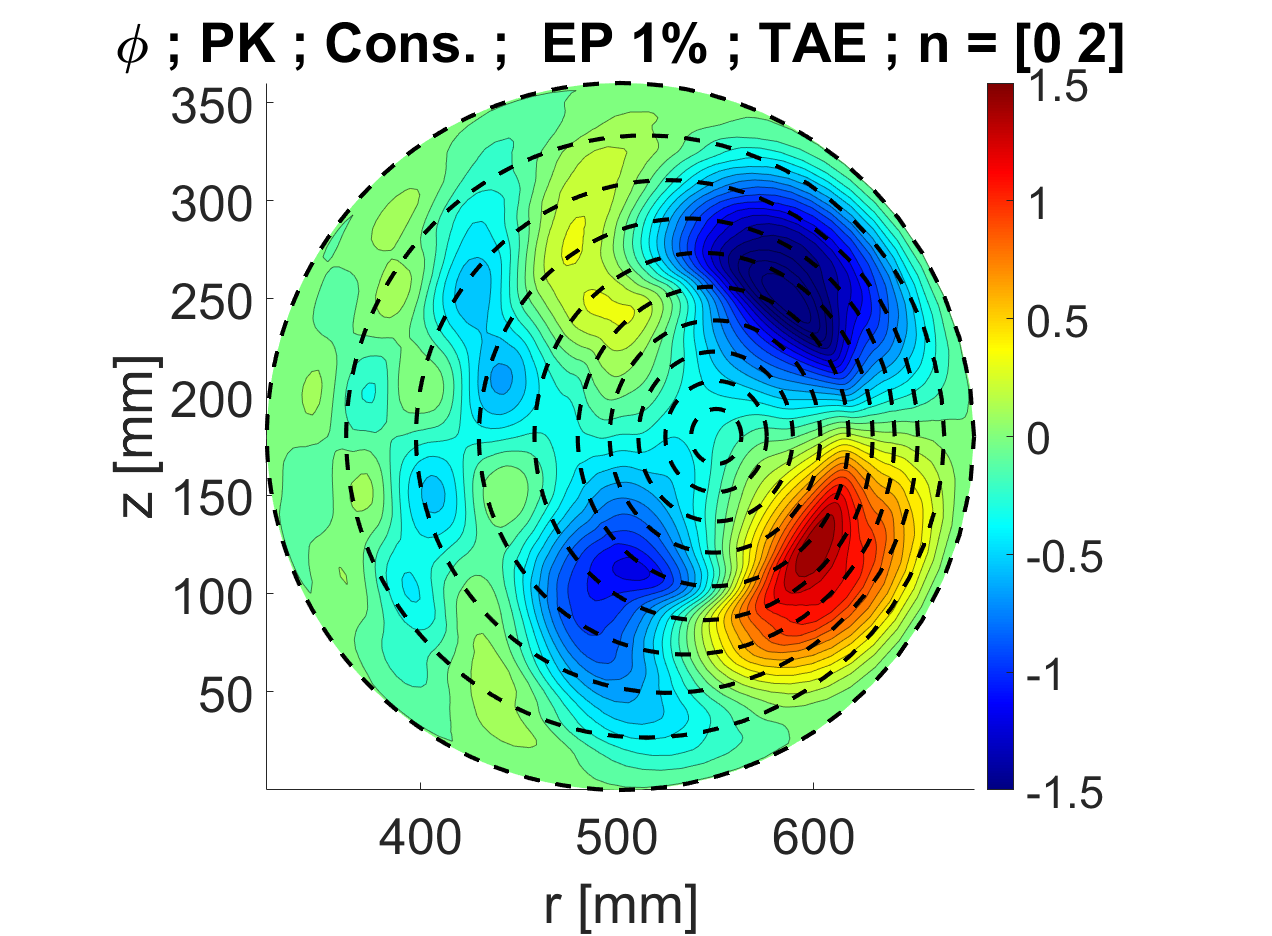}
\includegraphics[width=0.32\textwidth]{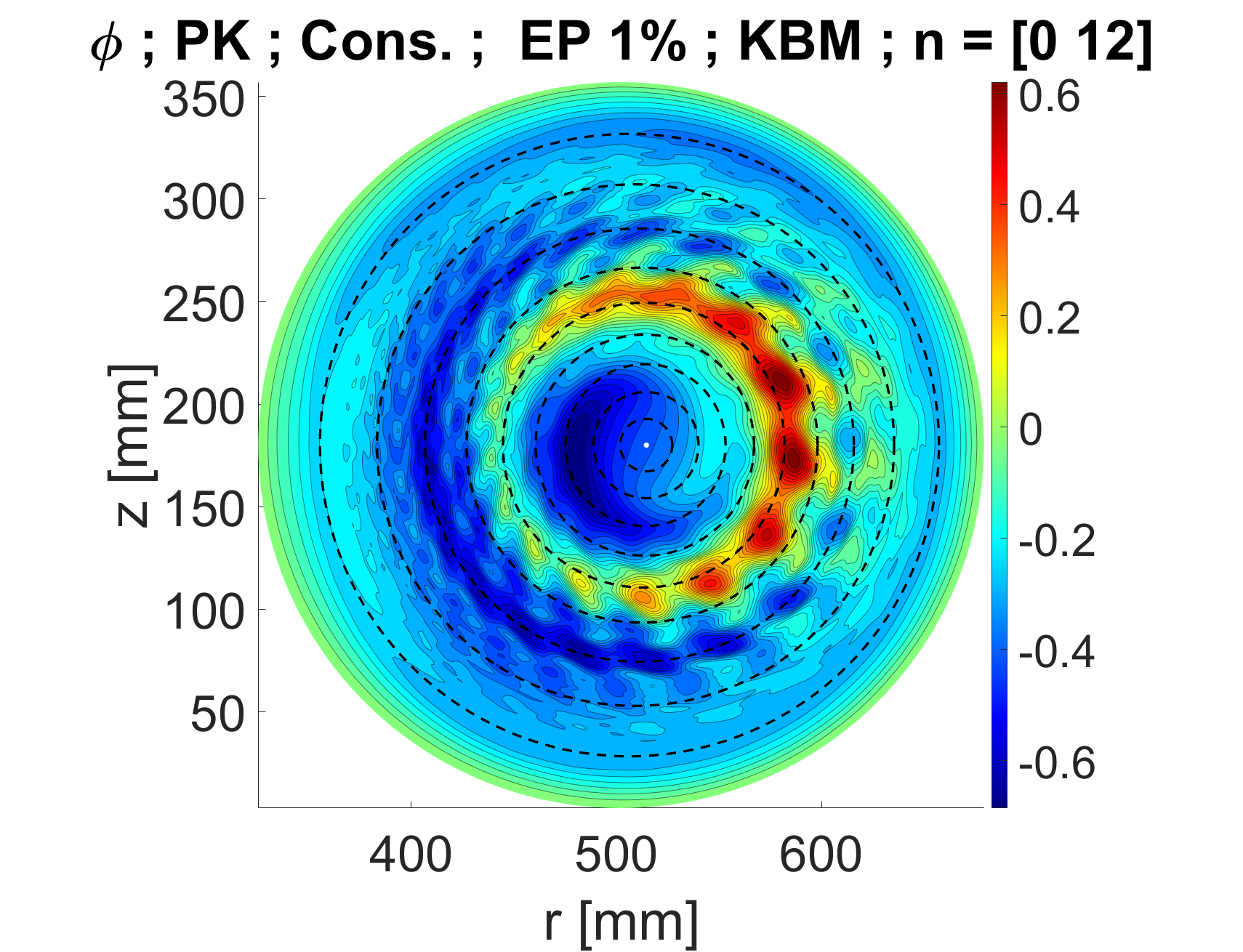}
\includegraphics[width=0.32\textwidth]{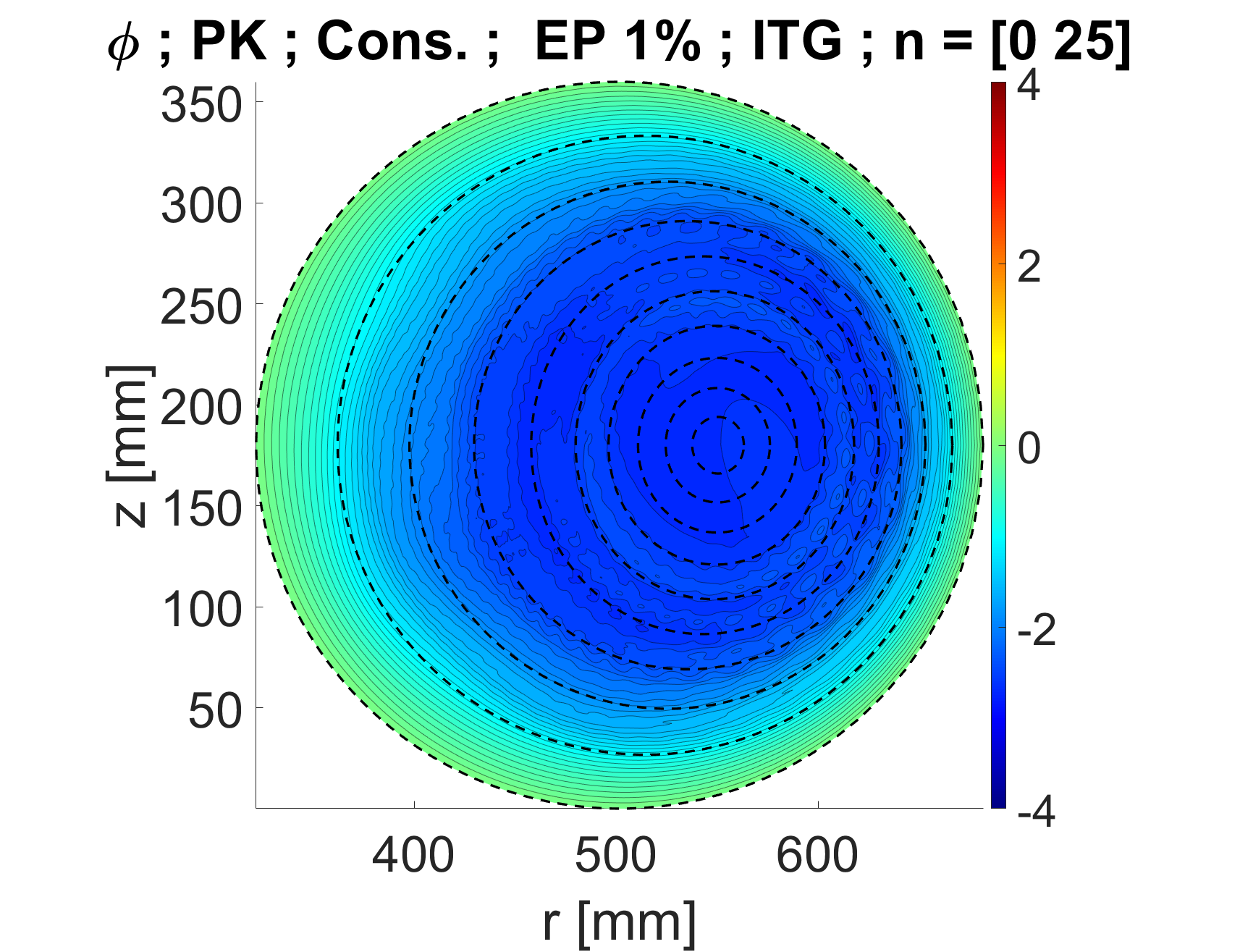}

\includegraphics[width=0.32\textwidth]{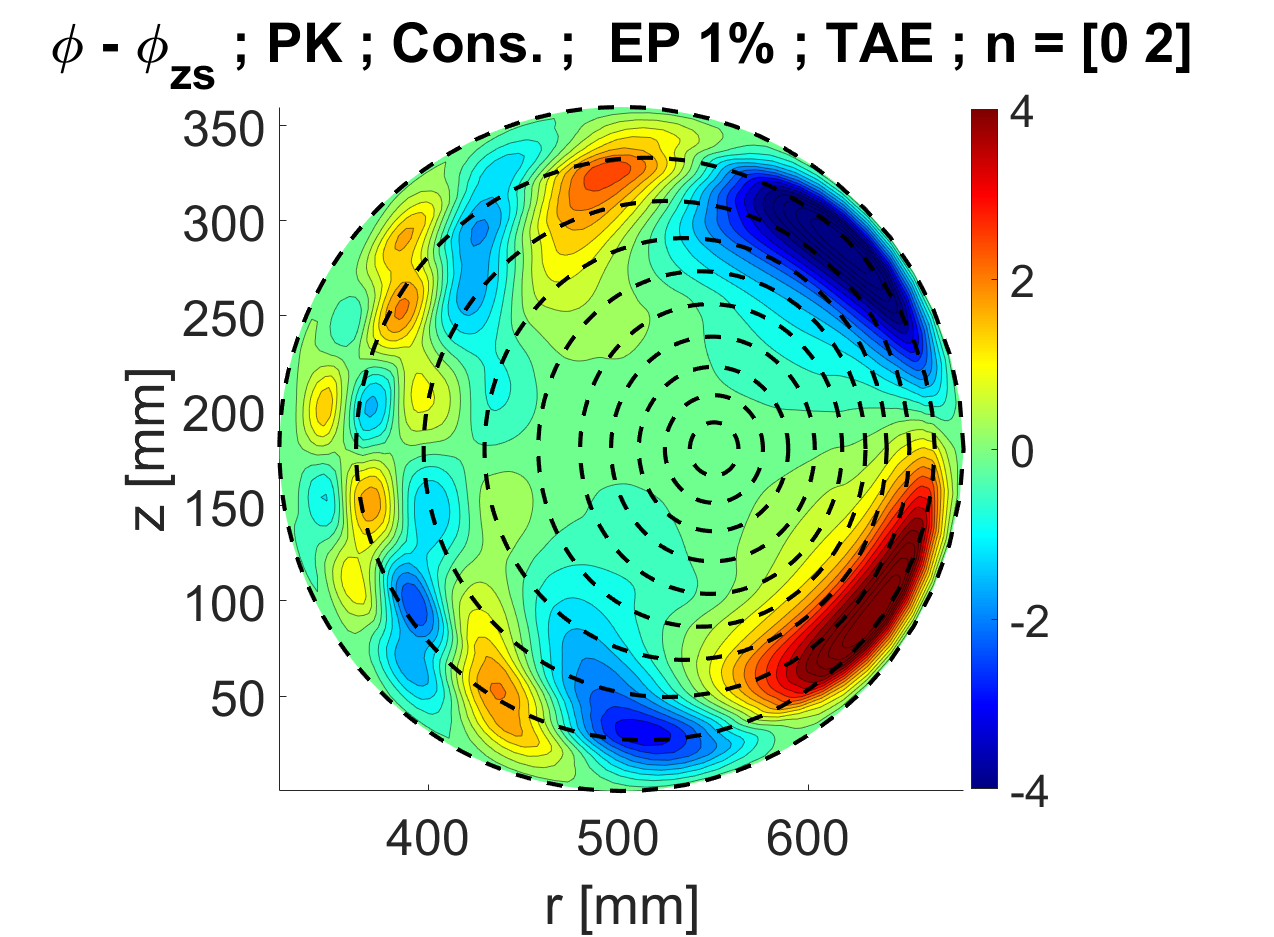}
\includegraphics[width=0.32\textwidth]{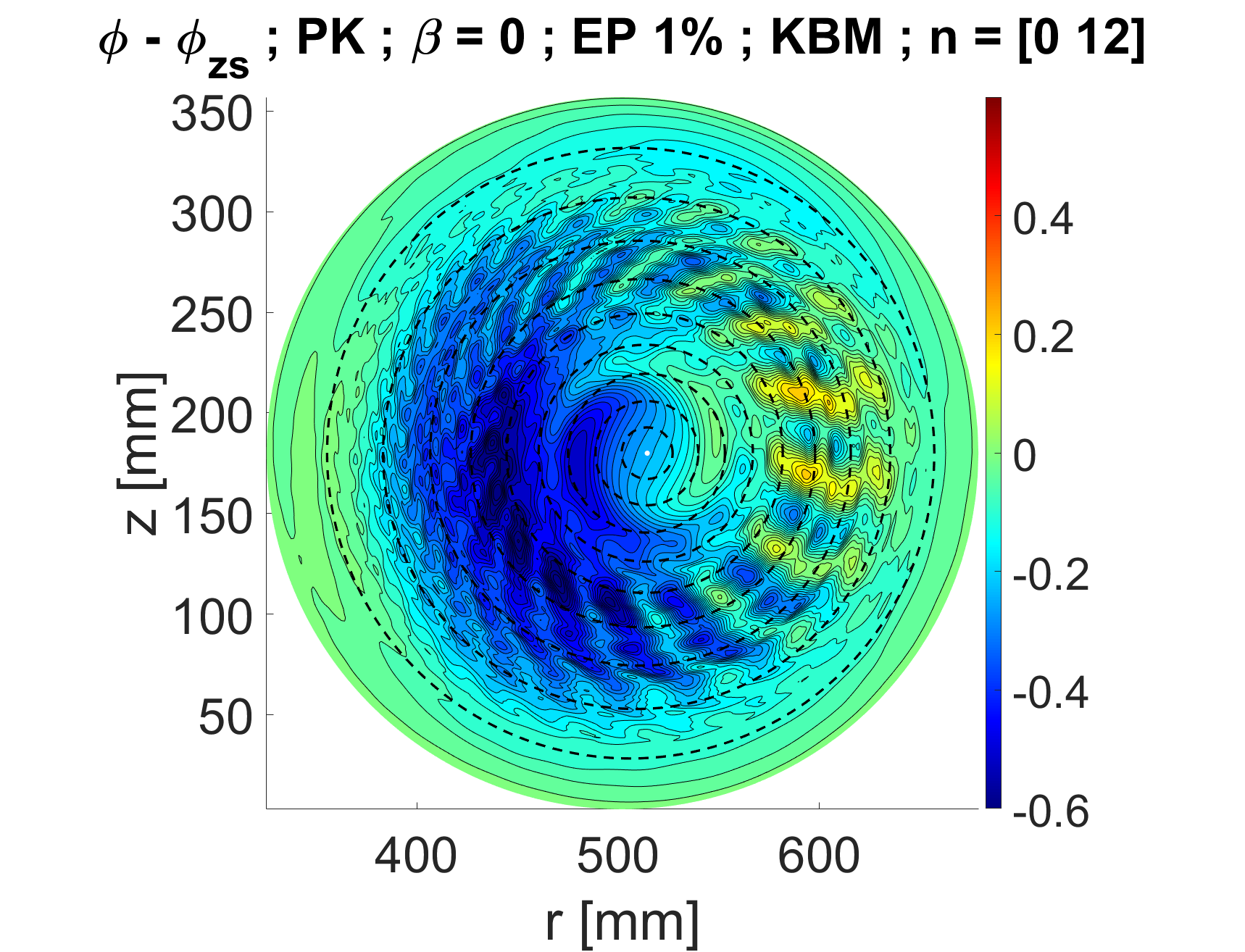}
\includegraphics[width=0.32\textwidth]{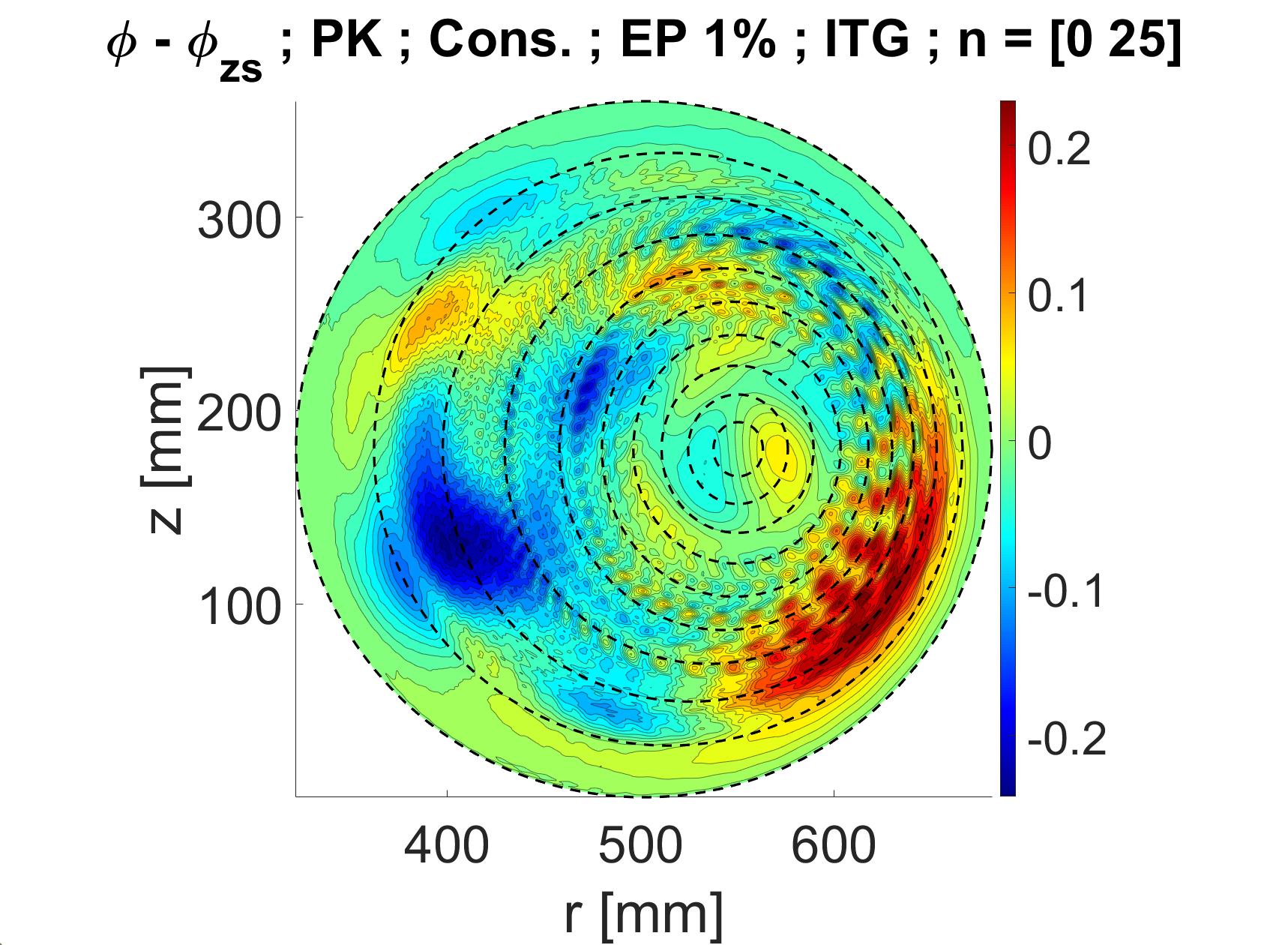}

\includegraphics[width=0.32\textwidth]{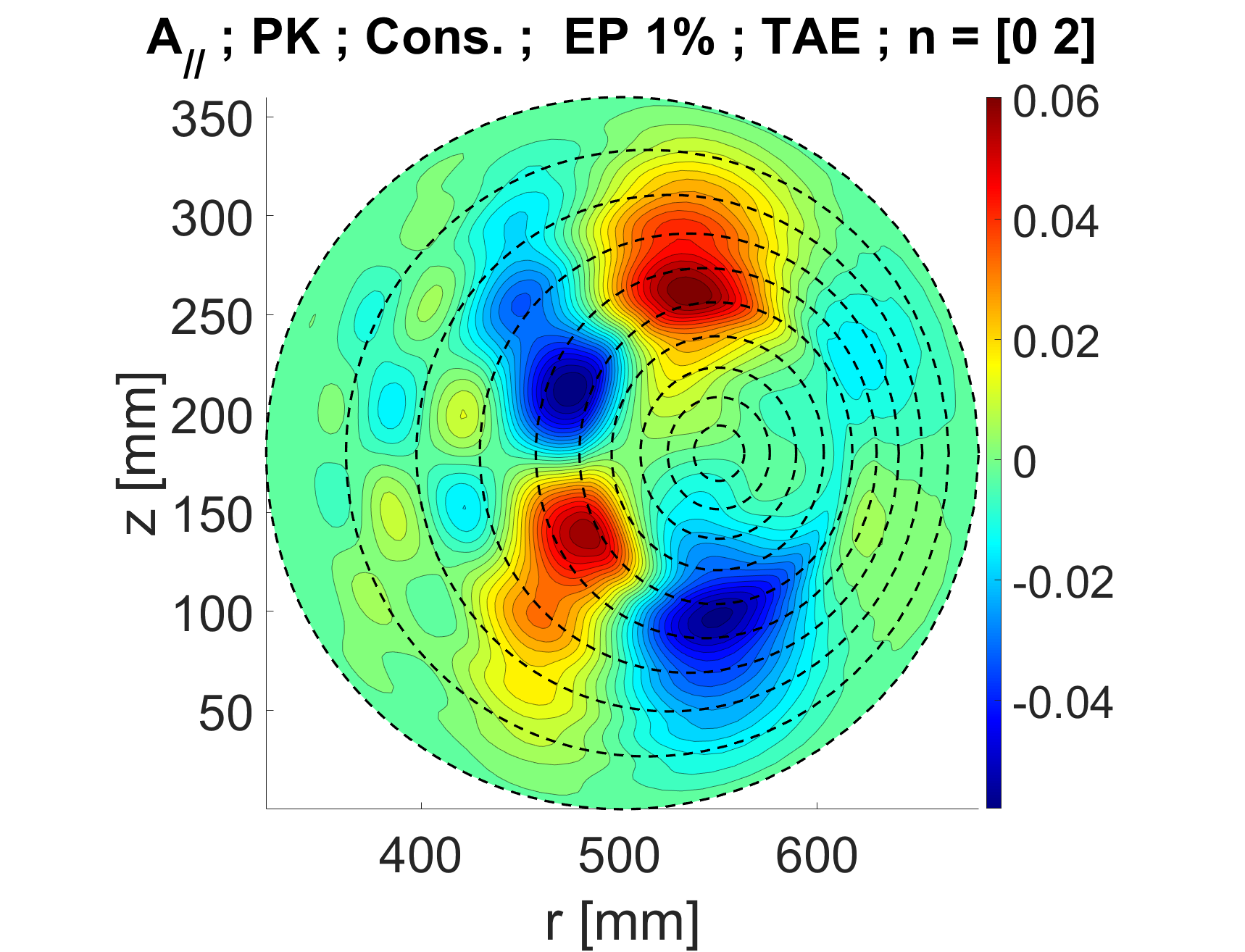}
\includegraphics[width=0.32\textwidth]{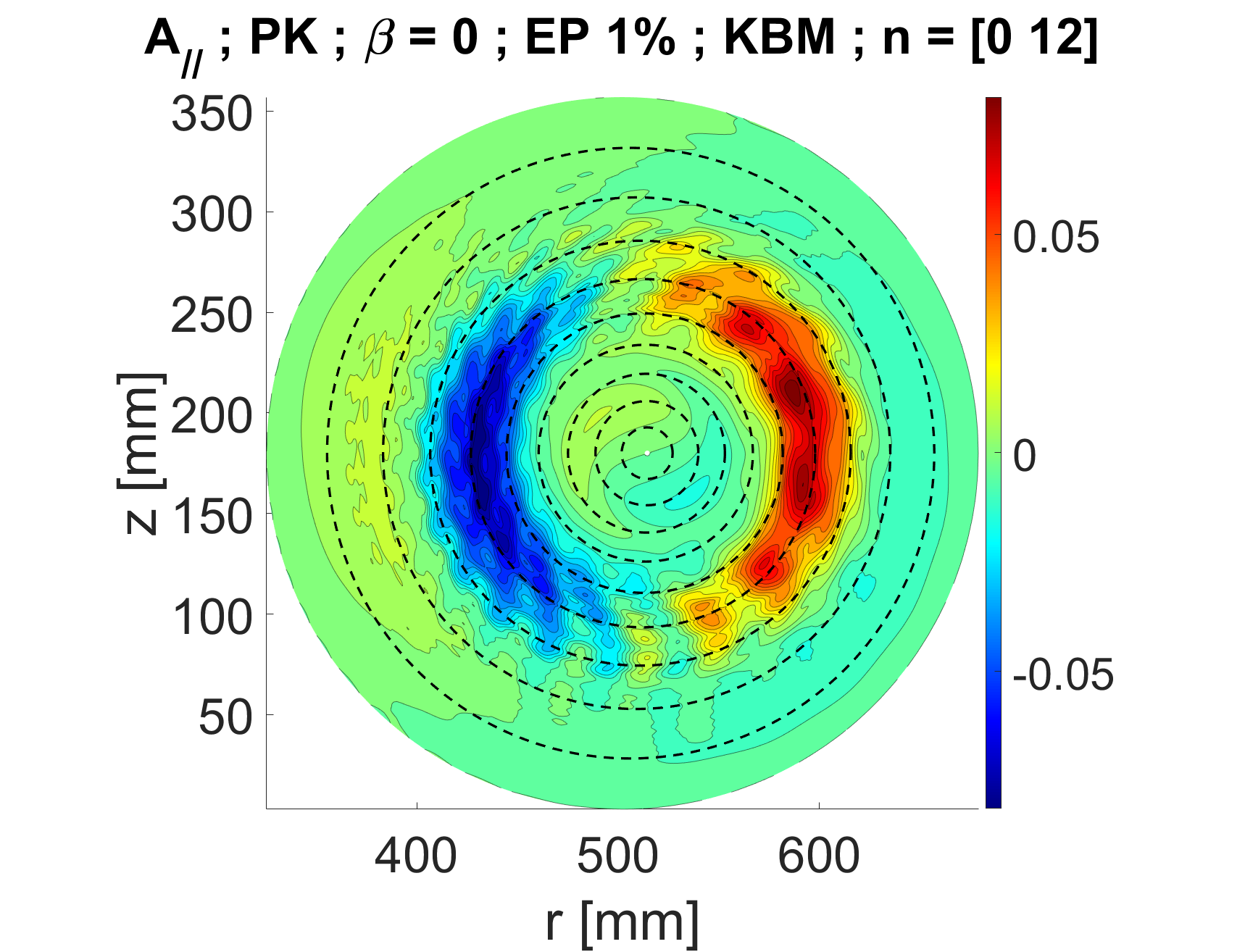}
\includegraphics[width=0.32\textwidth]{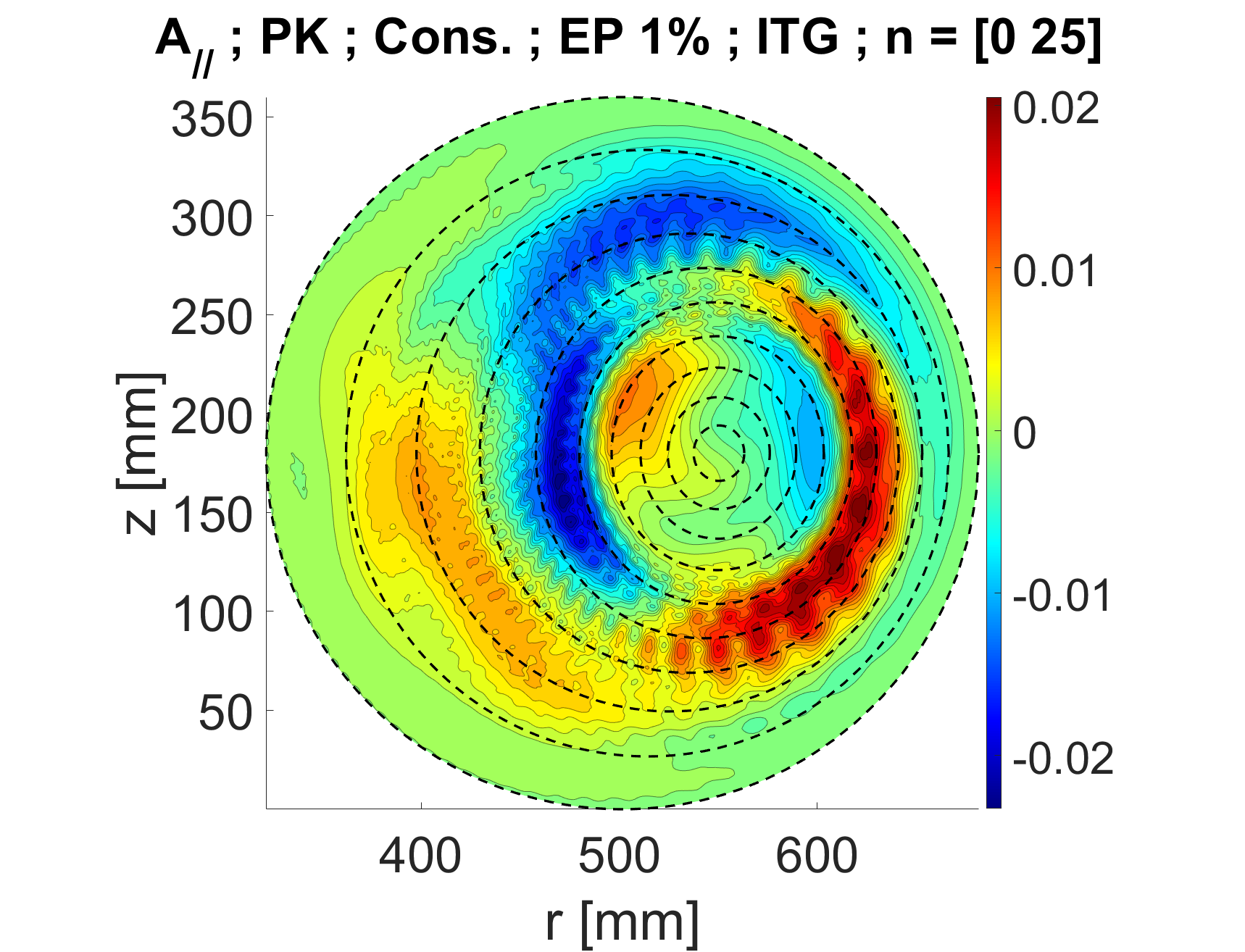}

\includegraphics[width=0.32\textwidth]{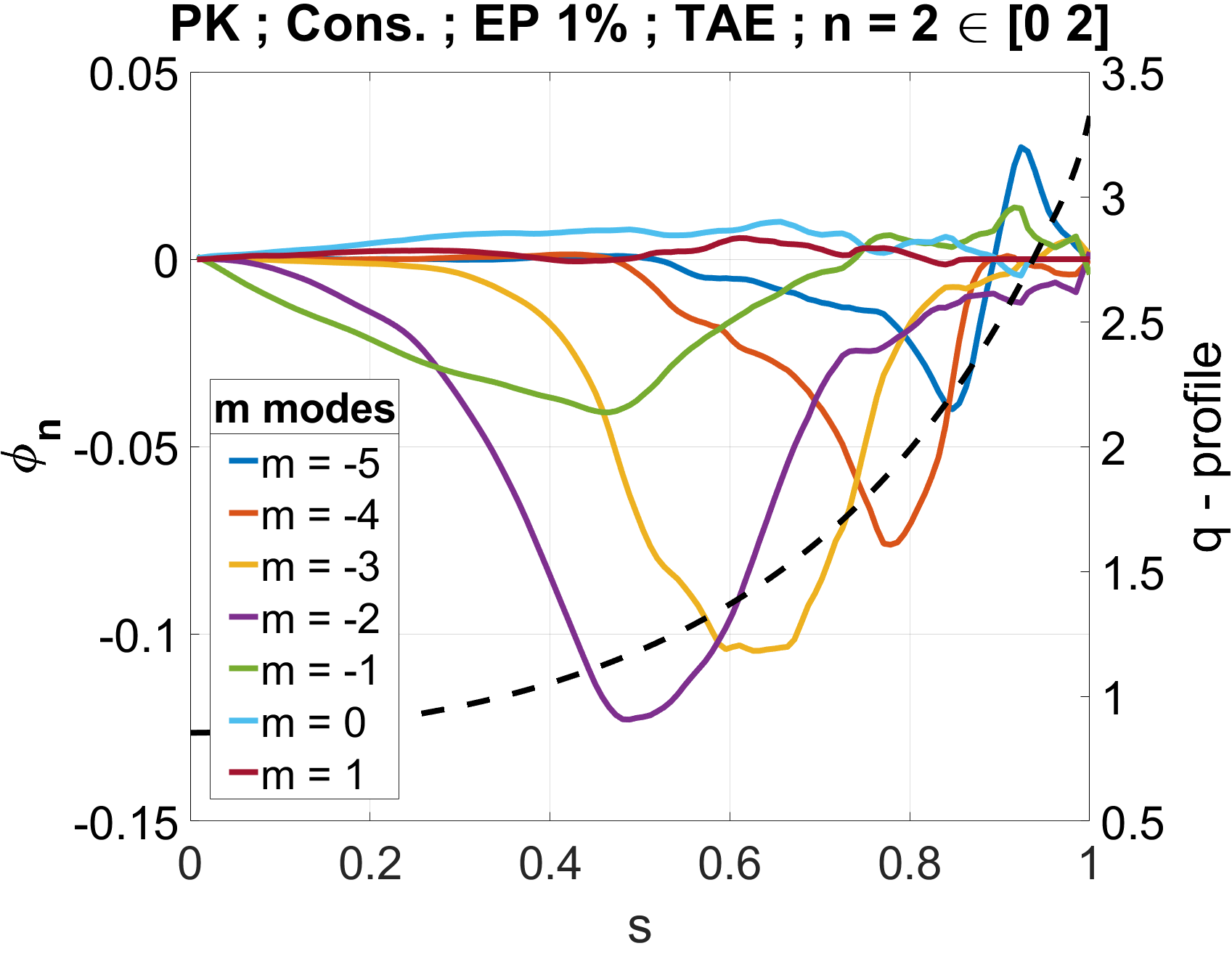}
\includegraphics[width=0.32\textwidth]{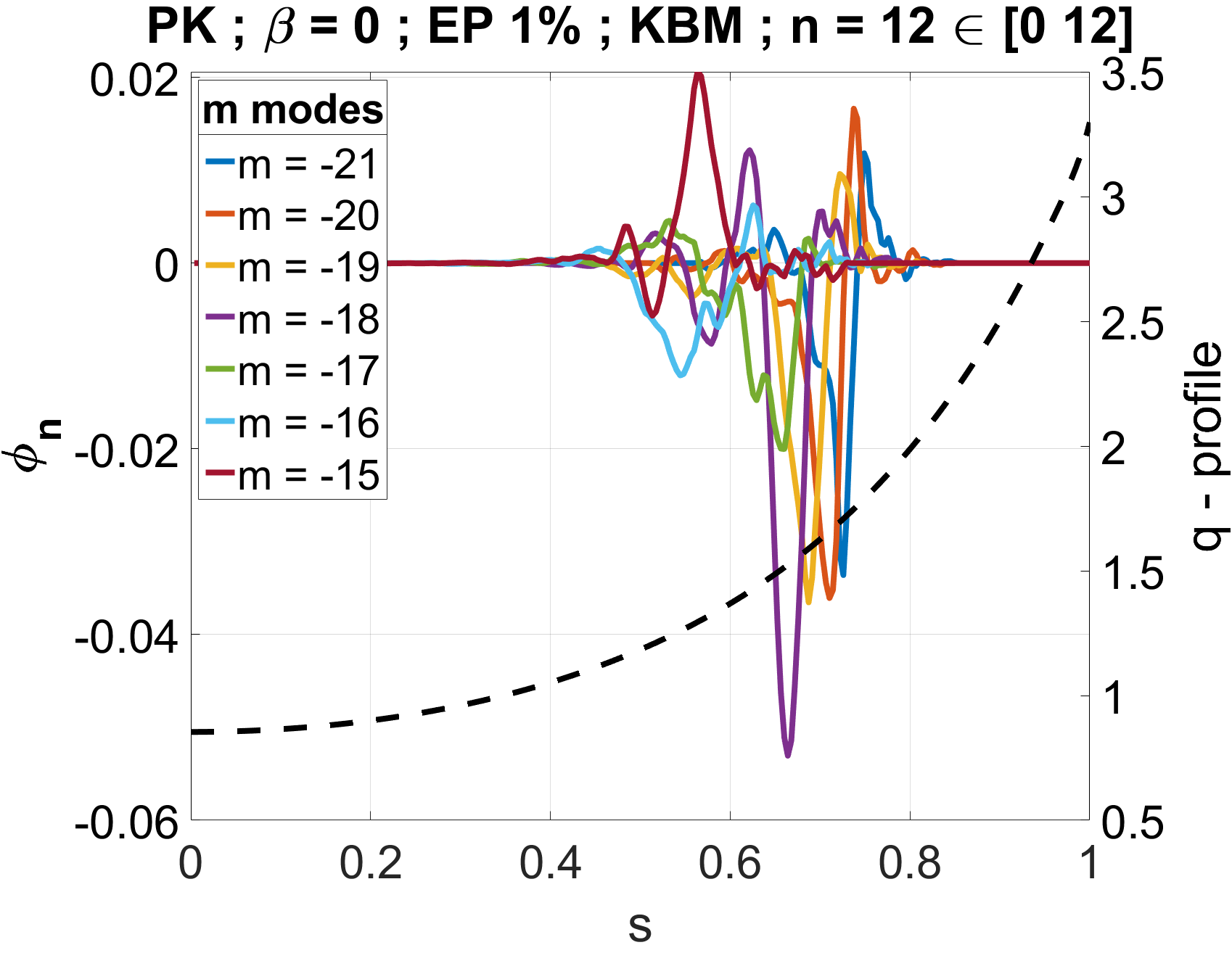}
\includegraphics[width=0.32\textwidth]{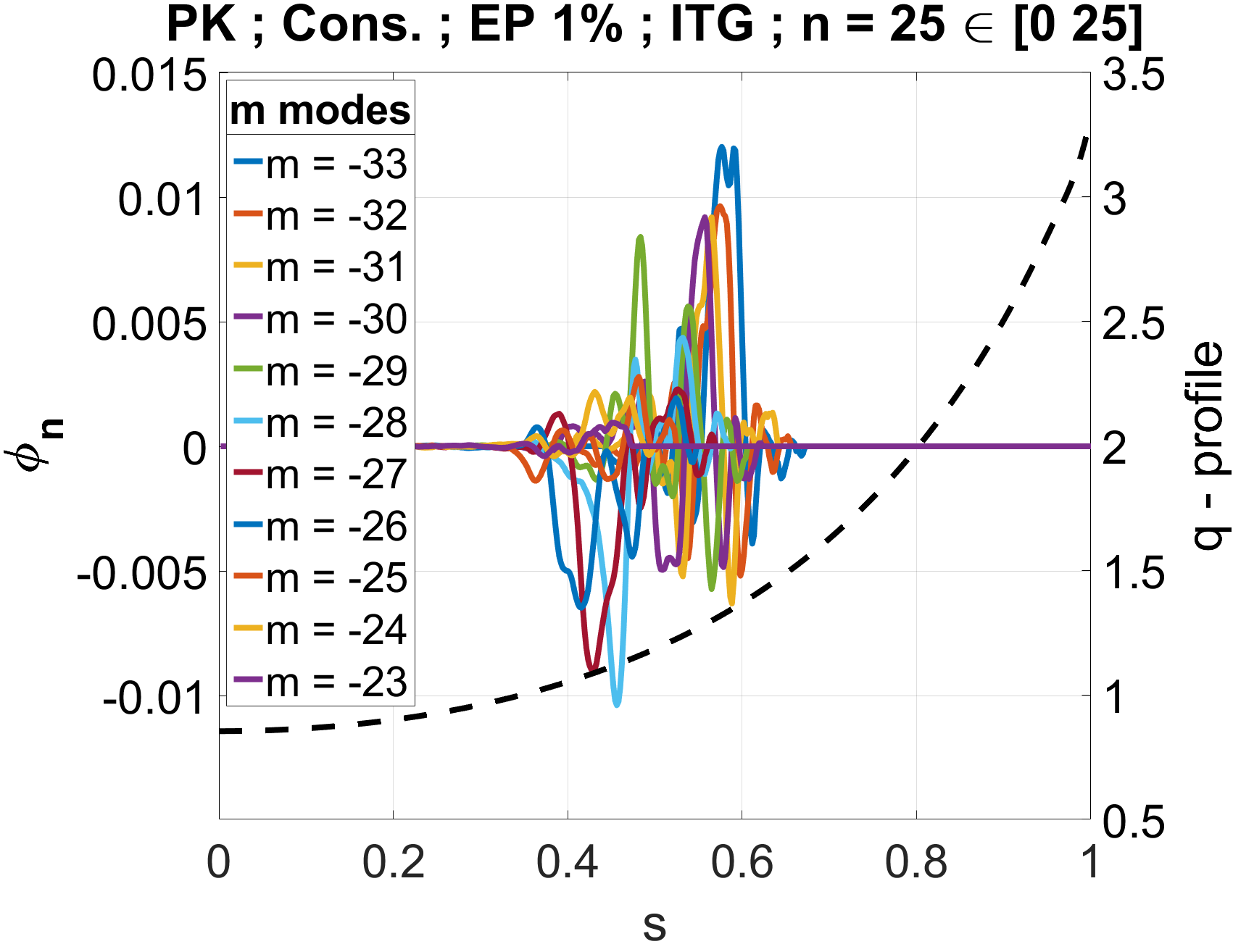}

\includegraphics[width=0.32\textwidth]{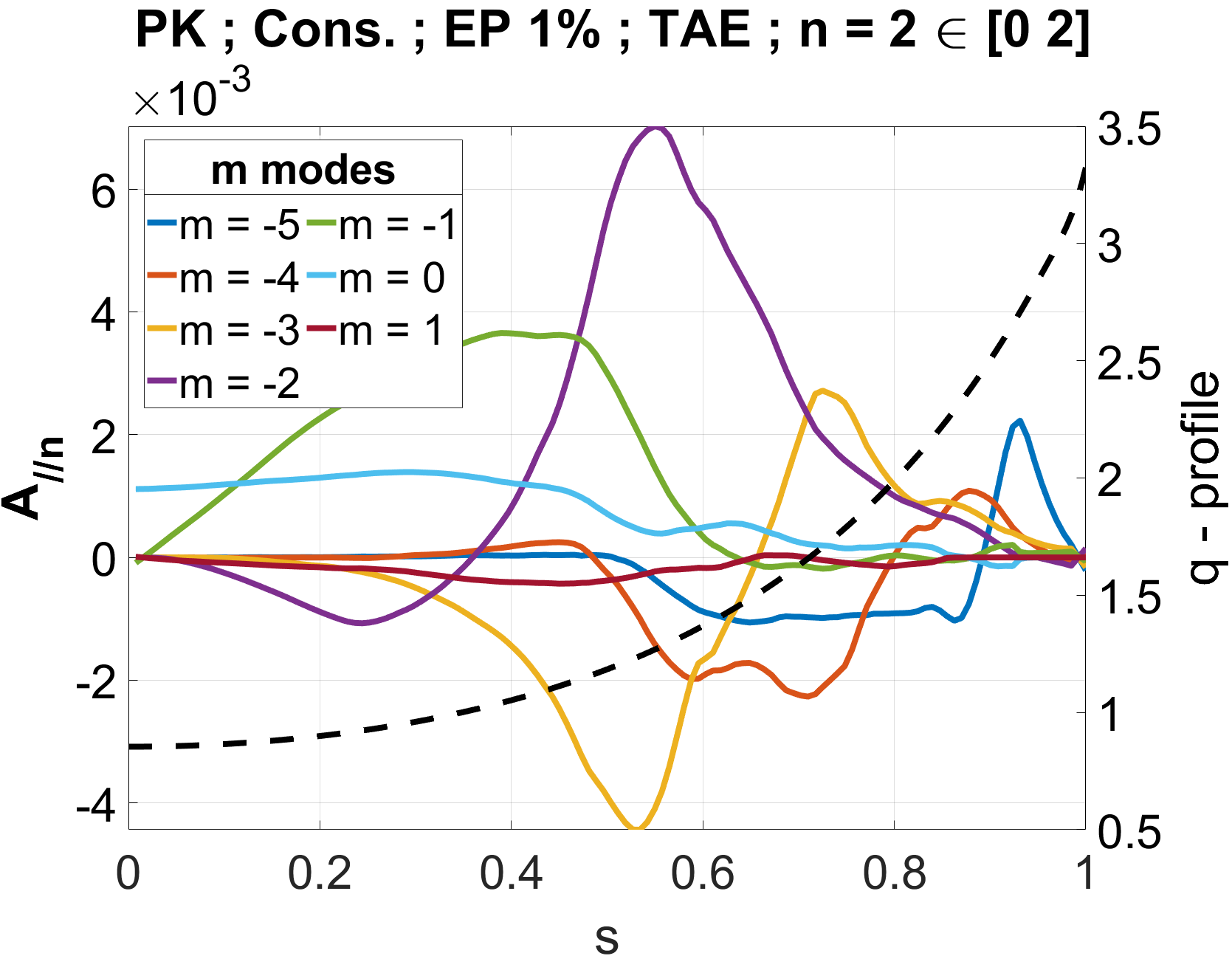}
\includegraphics[width=0.32\textwidth]{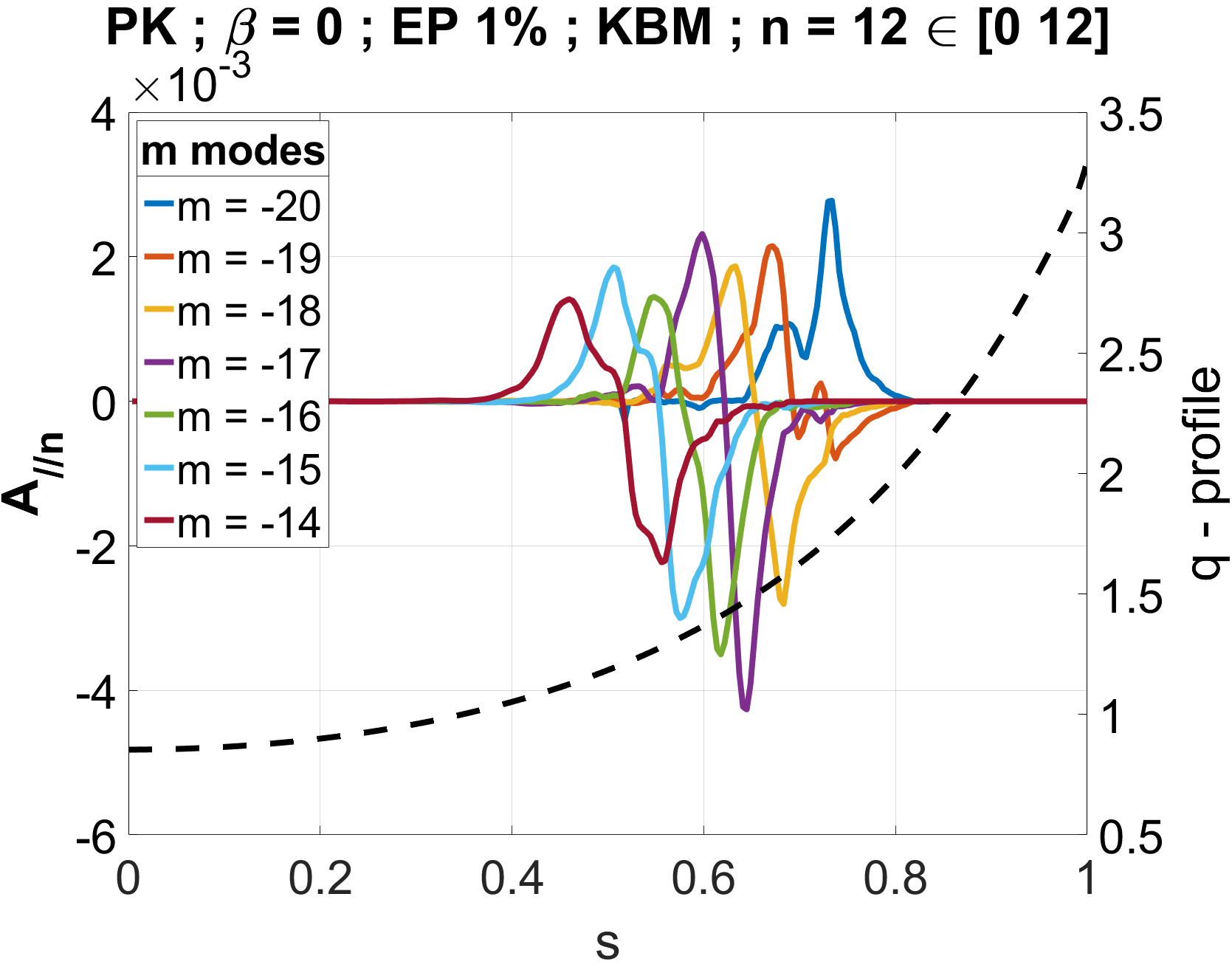}
\includegraphics[width=0.32\textwidth]{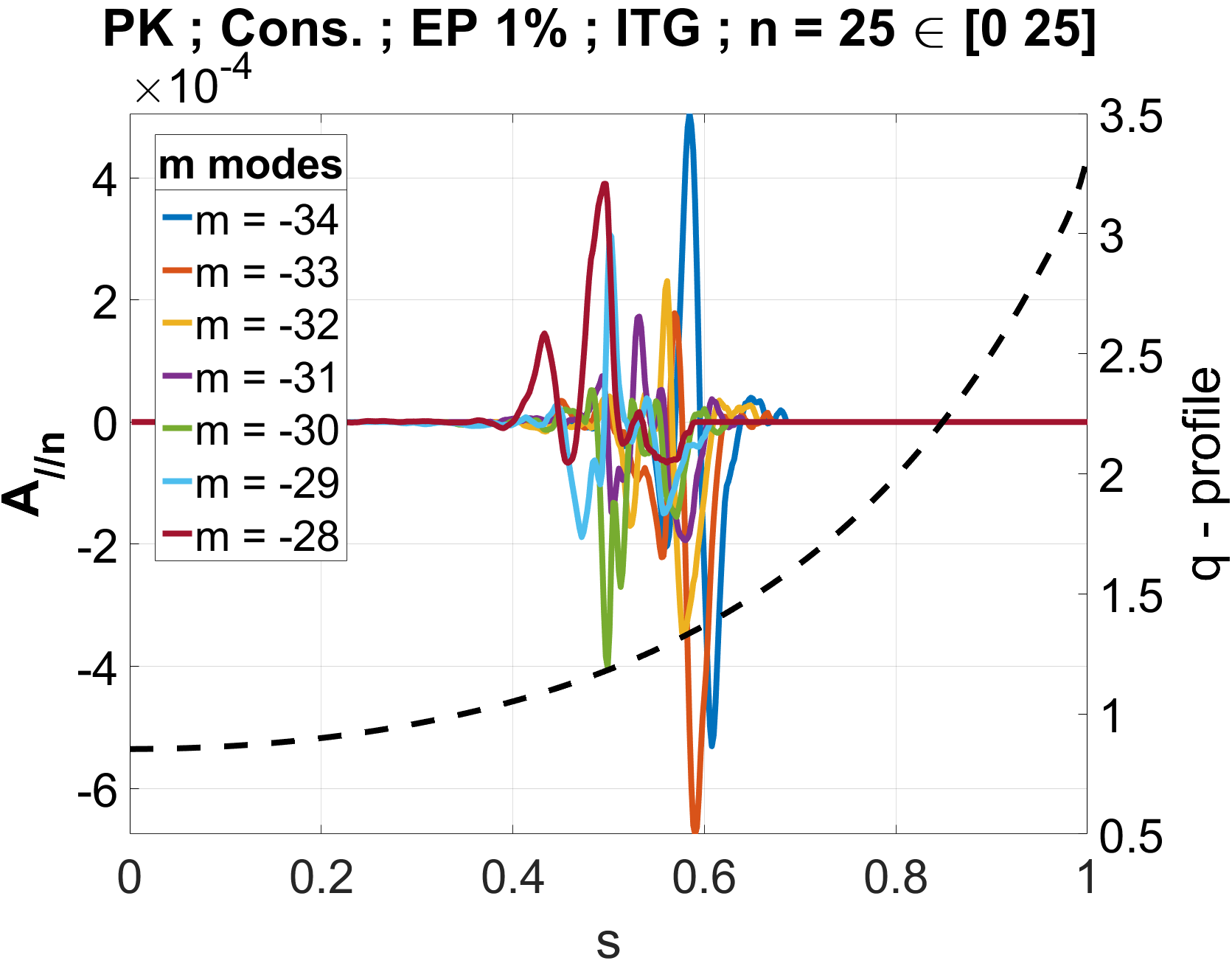}

\caption{\label{FIG:nonlinear_mode_structure} \it
The instantaneous characteristic nonlinear mode structure in $\phi$, $\phi-\phi_{ZS}$ and $A_{\parallel}$ in the middle of the early nonlinear stage. The left column presents a TAE, the middle a KBM and the right column an ITG. The dashed black circles are $s=[0;0.1;1]$.} 
\end{center}
\end{figure}

Figure \ref{FIG:nonlinear_timetrace} shows the time evolution of some measure of the total fluctuating electric field for a selected toroidal mode $n$, $ E_n(t) = \int (| \nabla \phi_n| + |\frac{\partial A_{\parallel_n}}{\partial t}|)J_{s,\theta^*} ds d\theta^* $. In this plot it is interesting to observe the effects of Shafranov shift on the self emerging ZS ($n=0$, mode) and the nonlinear saturation dynamics of an $n=2$ TAE and an $n=25$ ITG.

In PIC simulations, numerical noise tends to accumulate in the zonal response already during the linear phase, leading to a growth rate comparable to the main unstable mode. During the saturation stage the amplitude grows and nonlinear effects begin to dominate the system. The zonal response develops a physical drive and grows at a different rate than the mode, until the entire system reaches saturation.

In ITG cases, the zonal response tends to saturate with an amplitude that is an order of magnitude higher than the driving mode. Thus, we usually state that the physical drive begins once both modes have a comparable amplitude a bit before the saturation. In TAE cases in the other hand, the zonal response saturates with an amplitude comparable to that of the mode, making this criterion irrelevant. Nonetheless, we can study the nonlinear excitation of the zonal response comparing the growth rates, as seen in \ref{FIG:nonlinear_TAE_Sat_dynemics}.  

Interestingly, besides its effect on the linear growth rate, the Shafranov shift has little to no effect on the nonlinear saturation levels of both the unstable modes and associated ZS. The order of magnitude difference in saturation level between the ZS and the microinstability is in line with the paradigm that ZS store the majority of the free energy in the system.

Figure \ref{FIG:nonlinear_mode_structure} presents the nonlinear structure of the characteristic modes. Comparing the two top rows, representing contours of $\phi$ and $\phi - \phi_{ZS}$, respectively, we can see the relative magnitude of the developing ZF vs. the unstable mode. 

\subsection{Nonlinear TAE saturation dynamics}
\begin{figure}
\begin{center}
\includegraphics[width=0.32\textwidth]{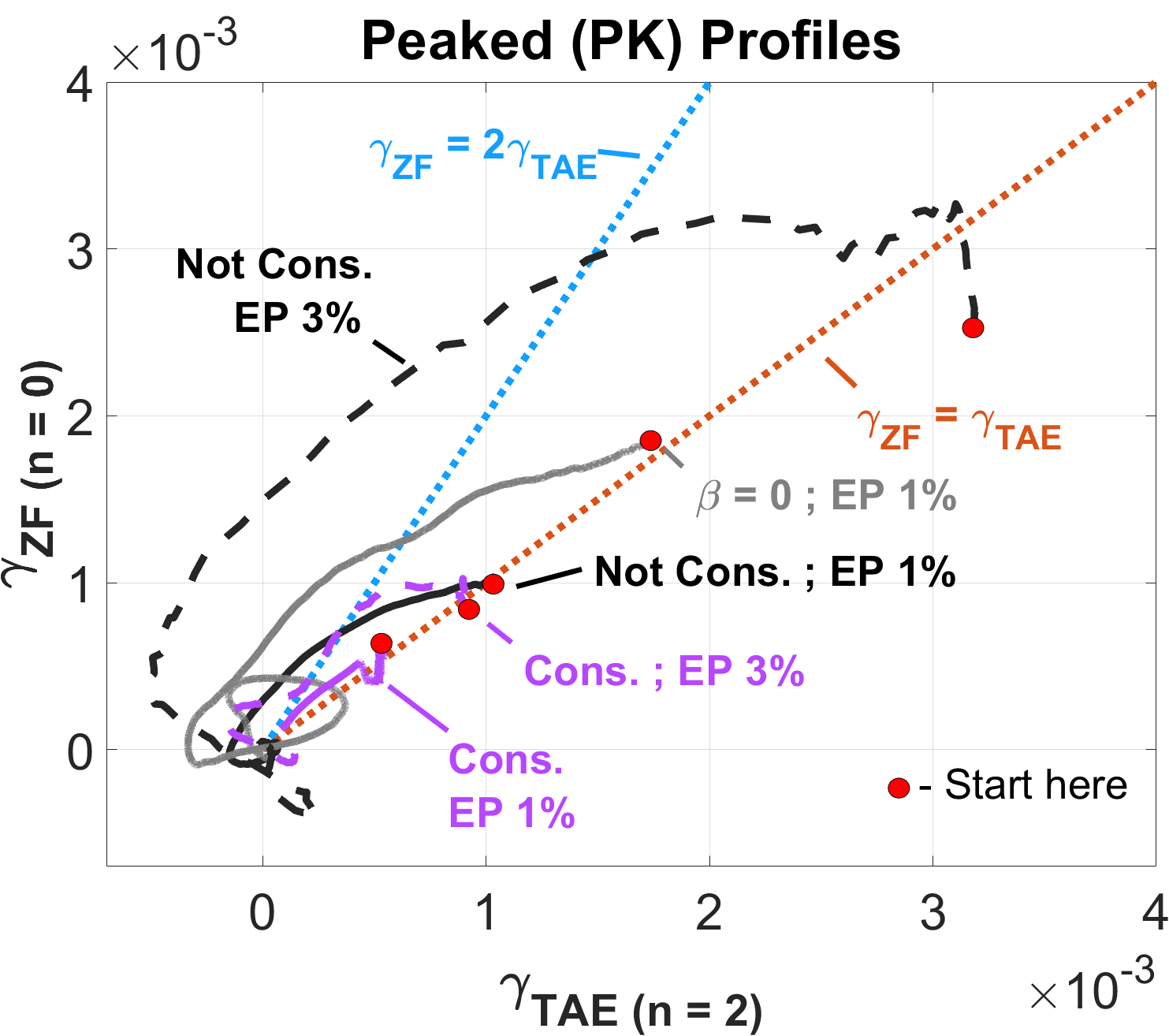}
\includegraphics[width=0.32\textwidth]{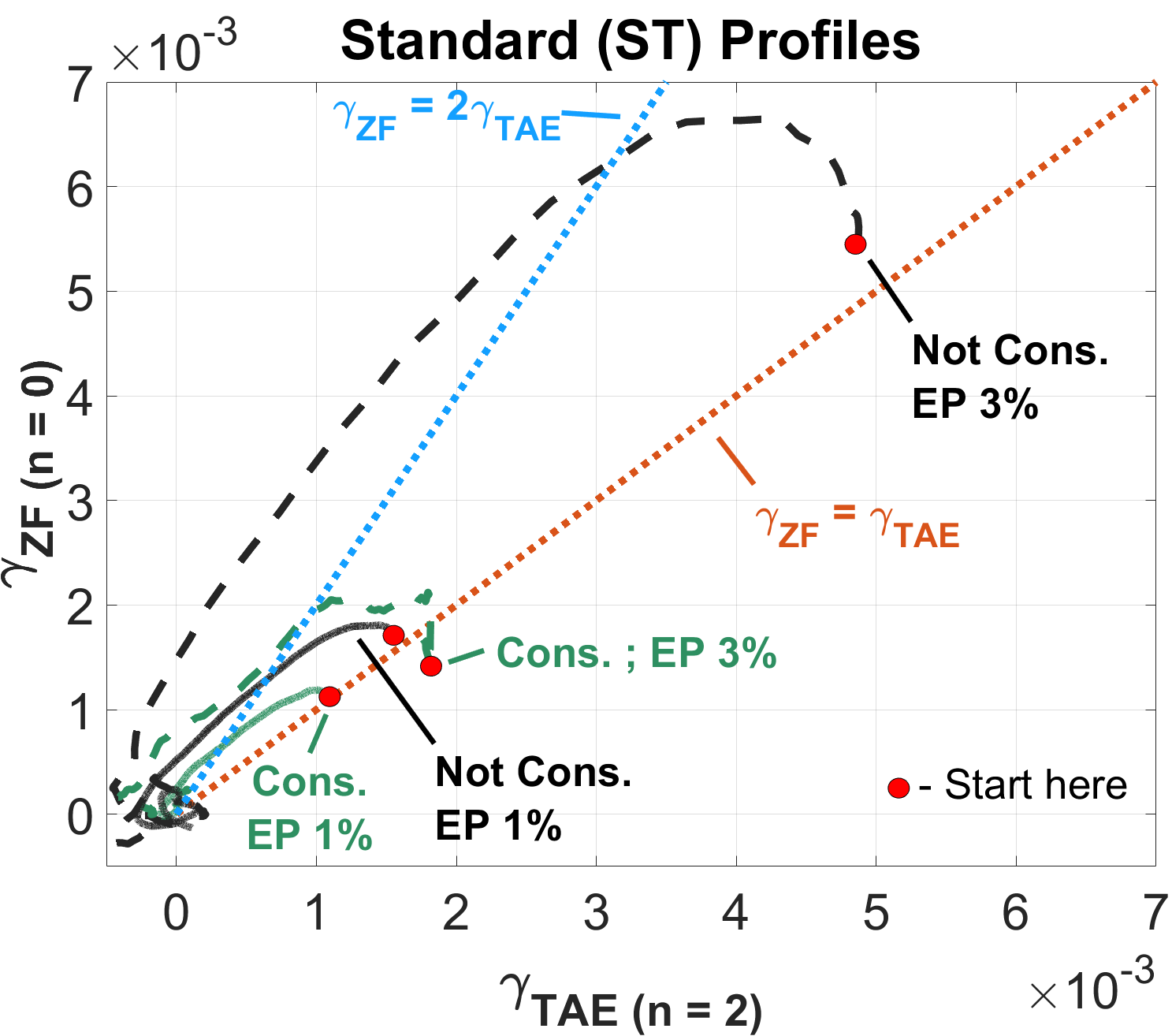}
\includegraphics[width=0.32\textwidth]{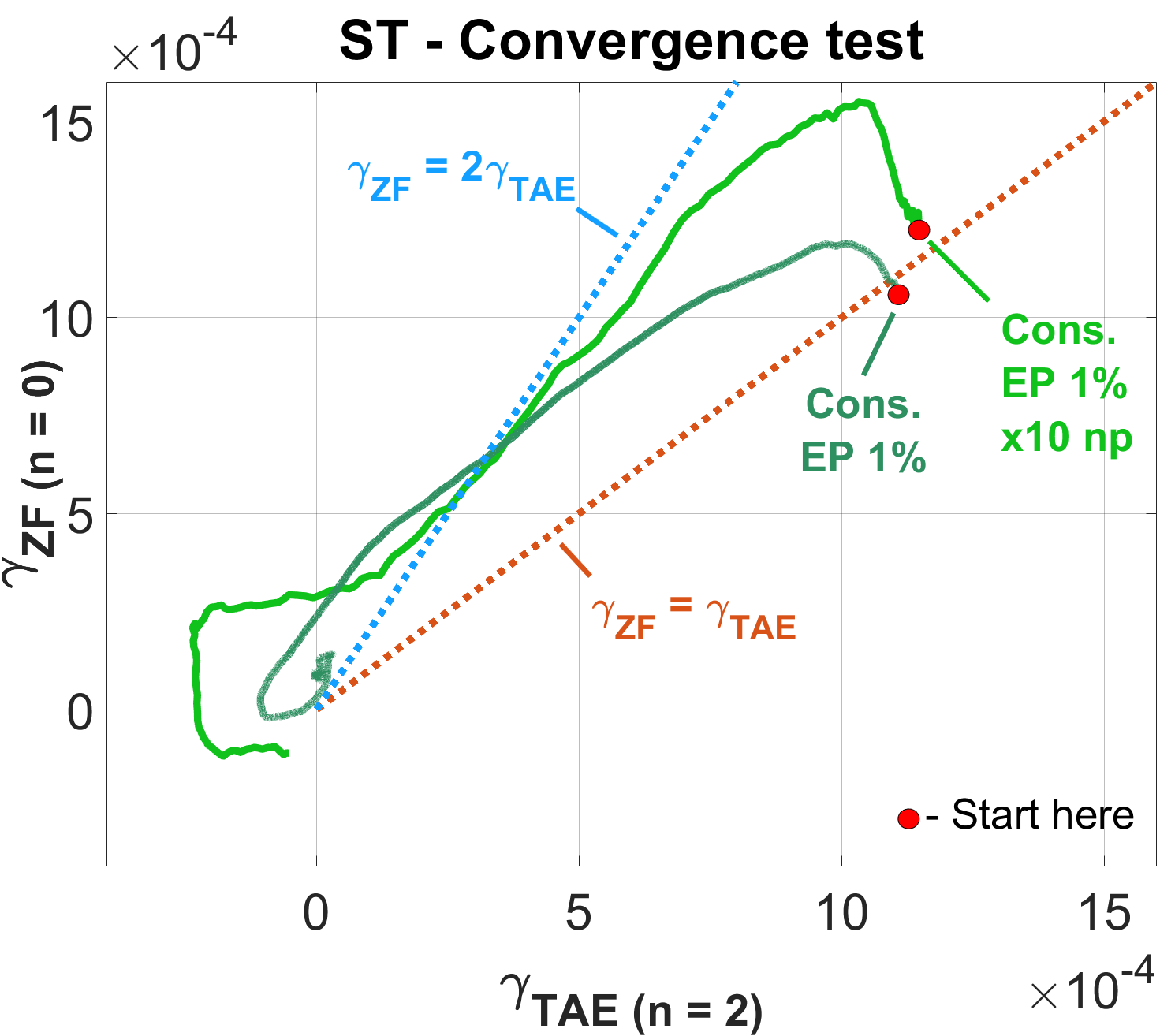}
\caption{\label{FIG:nonlinear_TAE_Sat_dynemics} \it
The instantaneous growth rate of the $n=0$ mode plotted versus that of the $n=2$ TAE. Each curve is a trajectory starting near the $\gamma_{ZF}=\gamma_{TAE}$ line in the linear phase, and converging to the origin at the end of the saturation. We add the $\gamma_{ZF}=\gamma_{TAE}$ and  $\gamma_{ZF}=2\gamma_{TAE}$ dashed lines to better distinguish the nonlinear mechanism driving the generation of ZS. 
Color coding: Gray - $\beta = 0$, Black - Not Consistent (Not Cons.), and Colored - Consistent (Cons.) MHD equilibria. Solid lines stand for $1\%$ EP, and dashed lines stand for $3\%$ EP.}
\end{center}
\end{figure}

In Figure \ref{FIG:nonlinear_TAE_Sat_dynemics} on the left we plot the $\gamma_{ZF}$ (the $n=0$ instantaneous growth rate) as a function of the instantaneous $n =2$ TAE growth rate, showing the system's trajectory during the nonlinear saturation phase. During the linear phase, the $n=0$ mode grows due to numerical noise with a growth rate very similar to that of the unstable mode. On this plot such behavior translates to a single point, which we indicate as the starting point. For all cases the starting point lies close to the $\gamma_{zs} = \gamma_{TAE}$ curve. 

Cases with a strong drive e.g. $3\%$ EPs, and artificially weak Shafranov shift e.g. cases with a not consistent or $\beta = 0$ MHD equilibria, have a nonlinear saturation dynamics dominated by the axisymmetric response . Where, first the TAE begins to saturate, while the axisymmetric response of the system grows for a while longer before saturating as well. Reducing the drive by lowering the EP fraction, or increasing the Shafranov shift results in a more restrained dynamical behavior exhibiting a "smoother" saturation. Meaning, the axisymmetric response  saturates more closely with the TAE mode, such that $\gamma_{zs}/\gamma_{TAE} \sim 1-2$ up to the final saturation phase.

In Figure \ref{FIG:nonlinear_TAE_Sat_dynemics} on the results of a convergence test in which we used ten time more numerical particles per species. We can see that both simulations exhibit a similar behavior, clearly showing a transition to a beat-driven instability with $\gamma_{zs} = 2\gamma_{TAE}$ in line with theory \cite{Qiu_PoP2016}. 

\begin{figure}
\begin{center}
\includegraphics[width=0.495\textwidth] {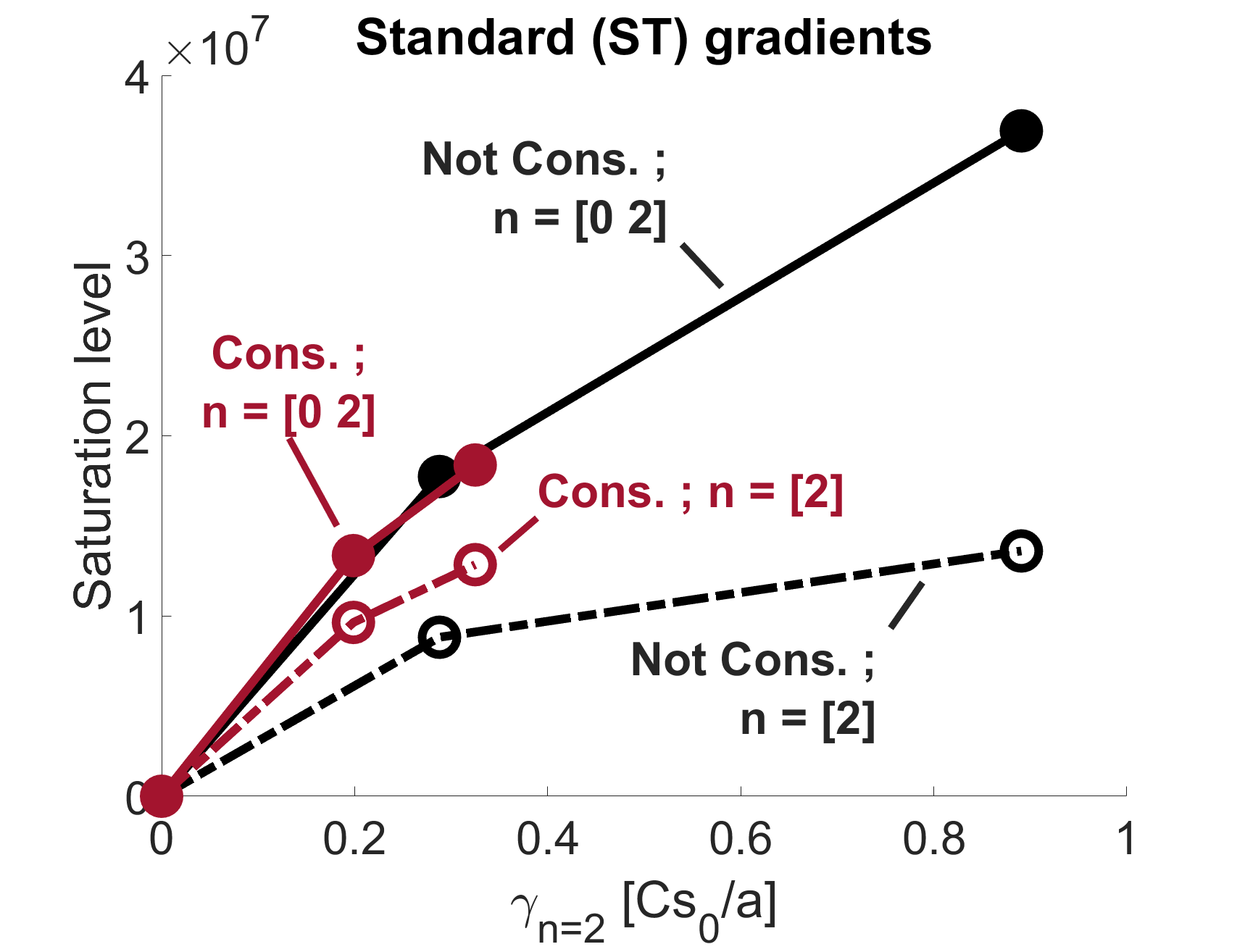}
\includegraphics[width=0.495\textwidth]{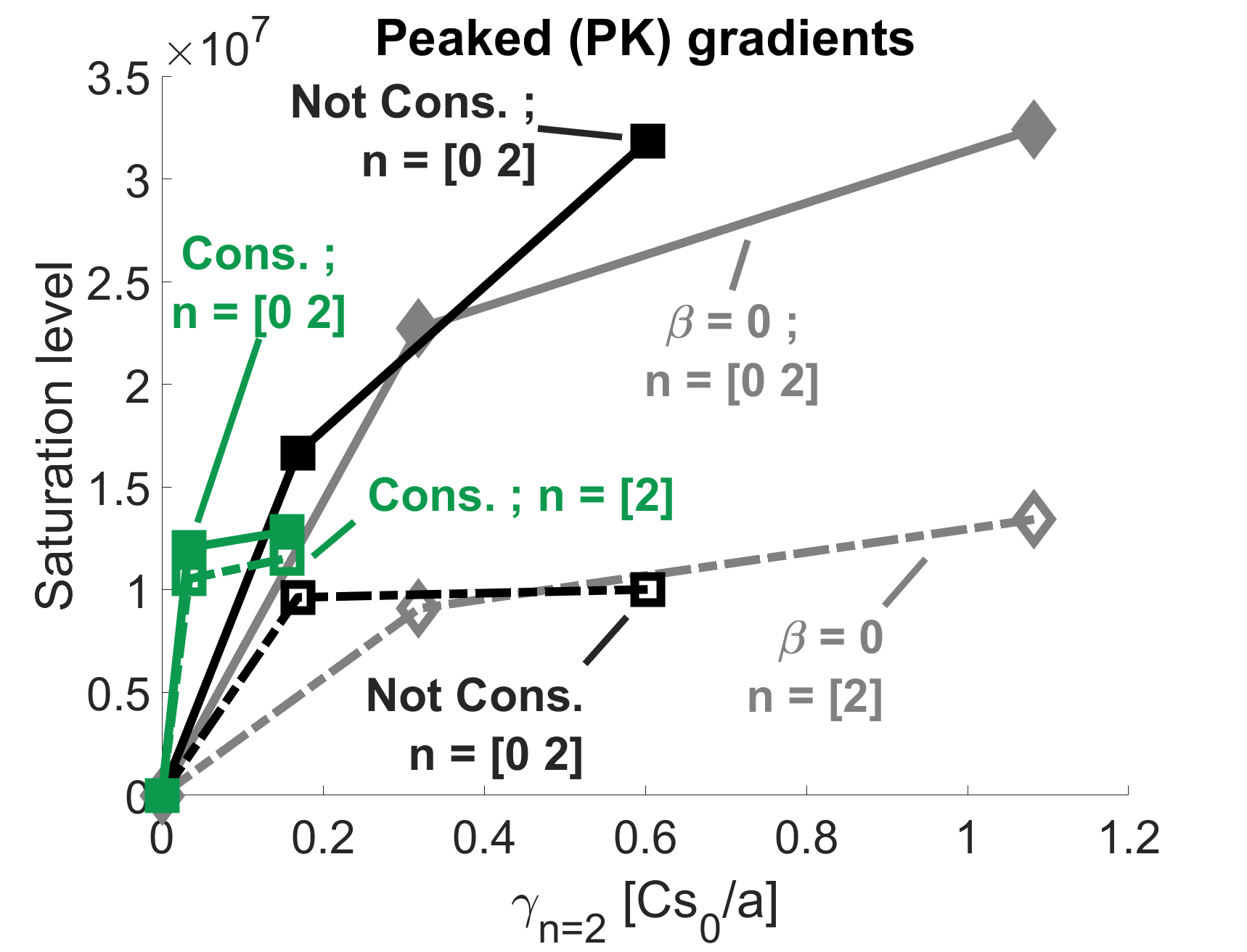}
\caption{\label{FIG:nonlinear_TAE_Sat_ratio} \it TAE saturation level vs. the linear growth rate as a function of EP fraction which increases from left to right: $0\%$ ; $1\%$ ; $3\%$. Color coding: Gray - $\beta = 0$, Black - Not Consistent (Not Cons.), and Colored - Consistent (Cons.) MHD equilibria.}
\end{center}
\end{figure}

Figure \ref{FIG:nonlinear_TAE_Sat_ratio} presents the saturation level of the $n = 2$ TAE as a function of the linear growth rate for three values of the EP fraction, $0\%$, $1\%$, $3\%$ (For the relation between growth rate and EP fraction, see Figure \ref{FIG:lin_TAE_GR_EP_scan}). The increase in the nonlinear saturation level with the growth rate (and therefore with EP fraction) levels off. For all presented cases, including the axisymmetric response  leads to a higher saturation level of the TAE. 

\begin{figure}
\begin{center}
\includegraphics[width=0.495\textwidth] {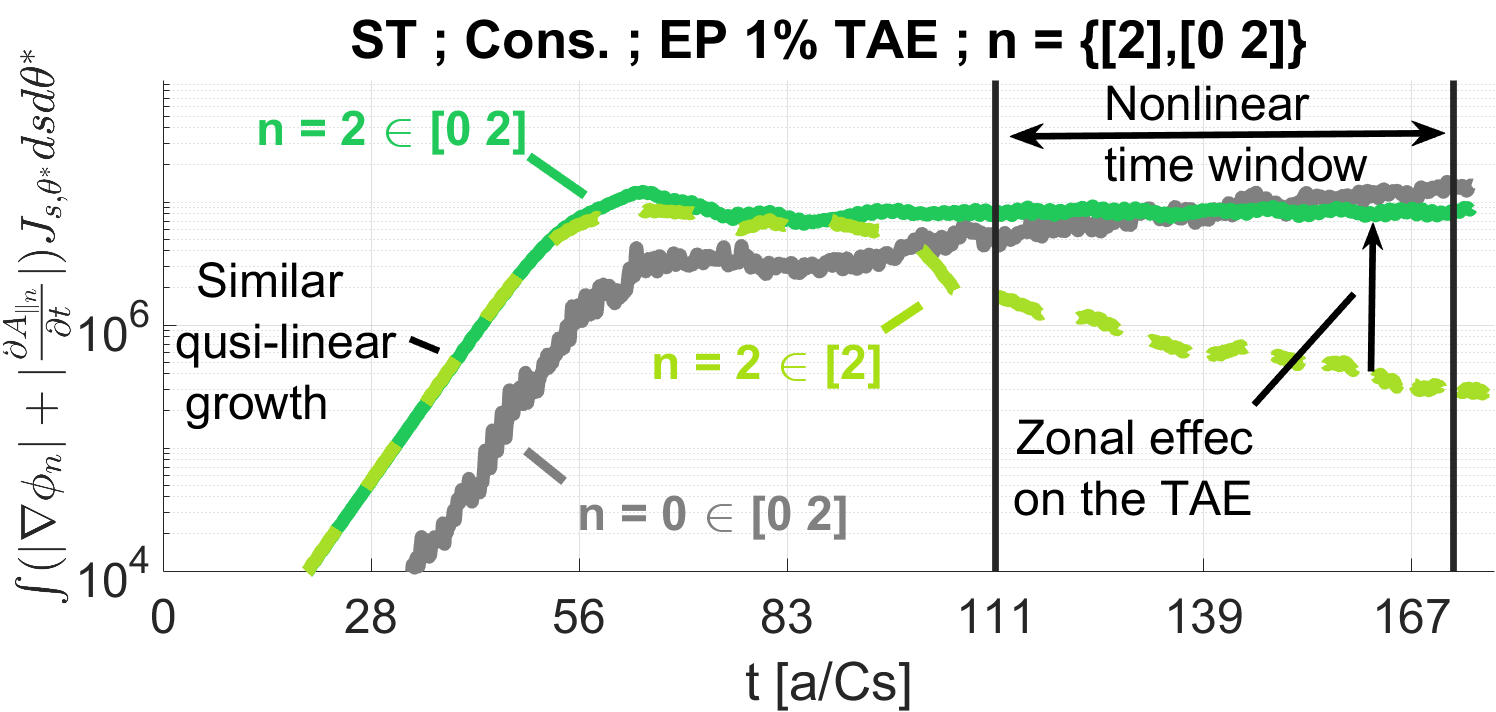}
\includegraphics[width=0.495\textwidth]{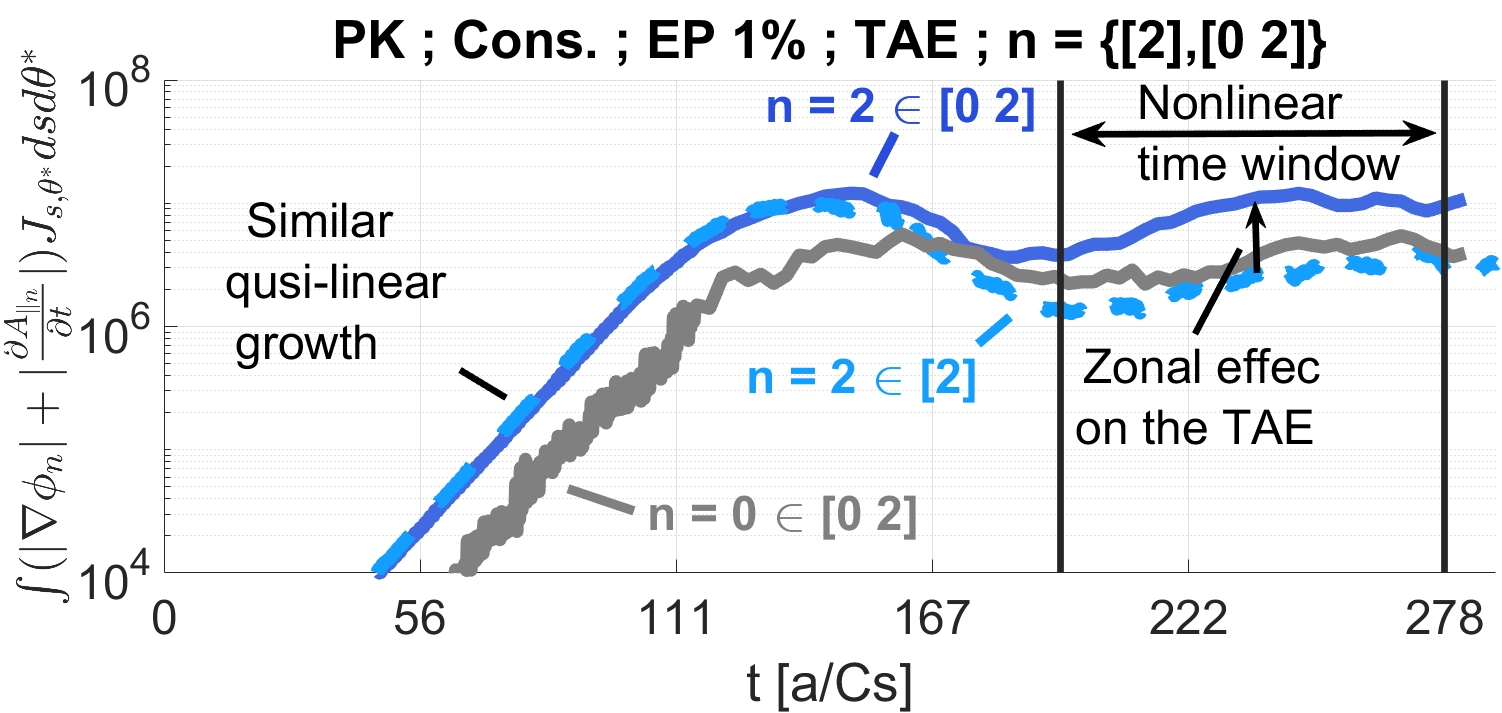}
\caption{\label{FIG:nonlinear_TAE_timewindow} \it For cases with consistent MHD equilibrium we show the early nonlinear phase on which we time average. We note the effect that excluding the zonal response ($n = 0$) has on the $n = 2$ TAE. We present cases with standard bulk gradients on the left and with peaked bulk gradients on the right. Where, 'ST' or 'PK' indicate standard or peaked bulk gradients respectively, and Cons. stands for self consistent MHD equilibria.}
\end{center}
\end{figure}

Figure \ref{FIG:nonlinear_TAE_timewindow} presents a time trace of the electric field amplitude which includes the nonlinear time window, a while after the saturation phase discussed above. Continuing the trend from before, including the zonal response, i.e. the $n = 0$ mode in the filter, has a destabilizing effect on the $n = 2$ TAE.

\subsection{Emerging Zonal Structures}

\begin{figure}
\begin{center}
\includegraphics[width=0.49\textwidth]{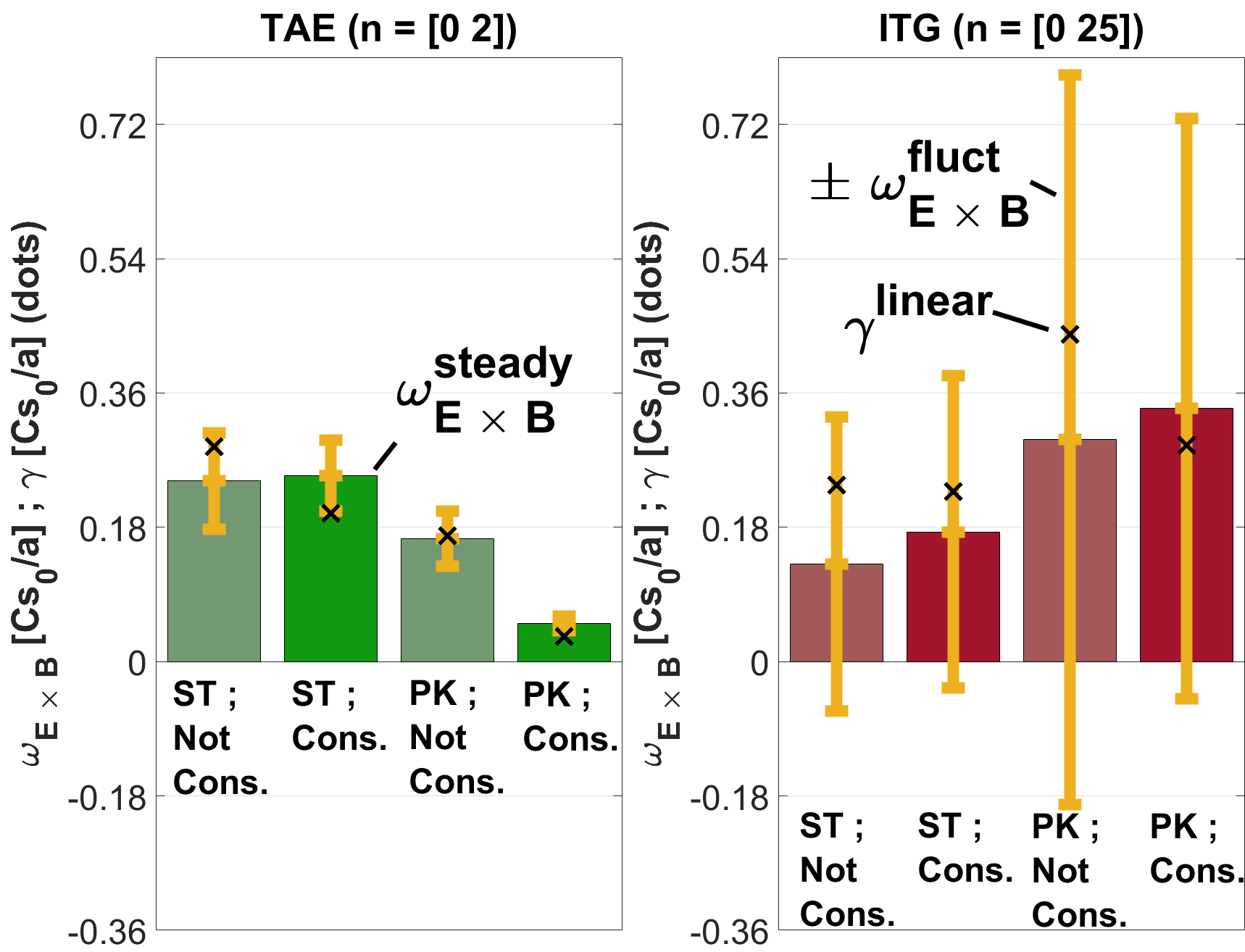}
\includegraphics[width=0.49\textwidth]{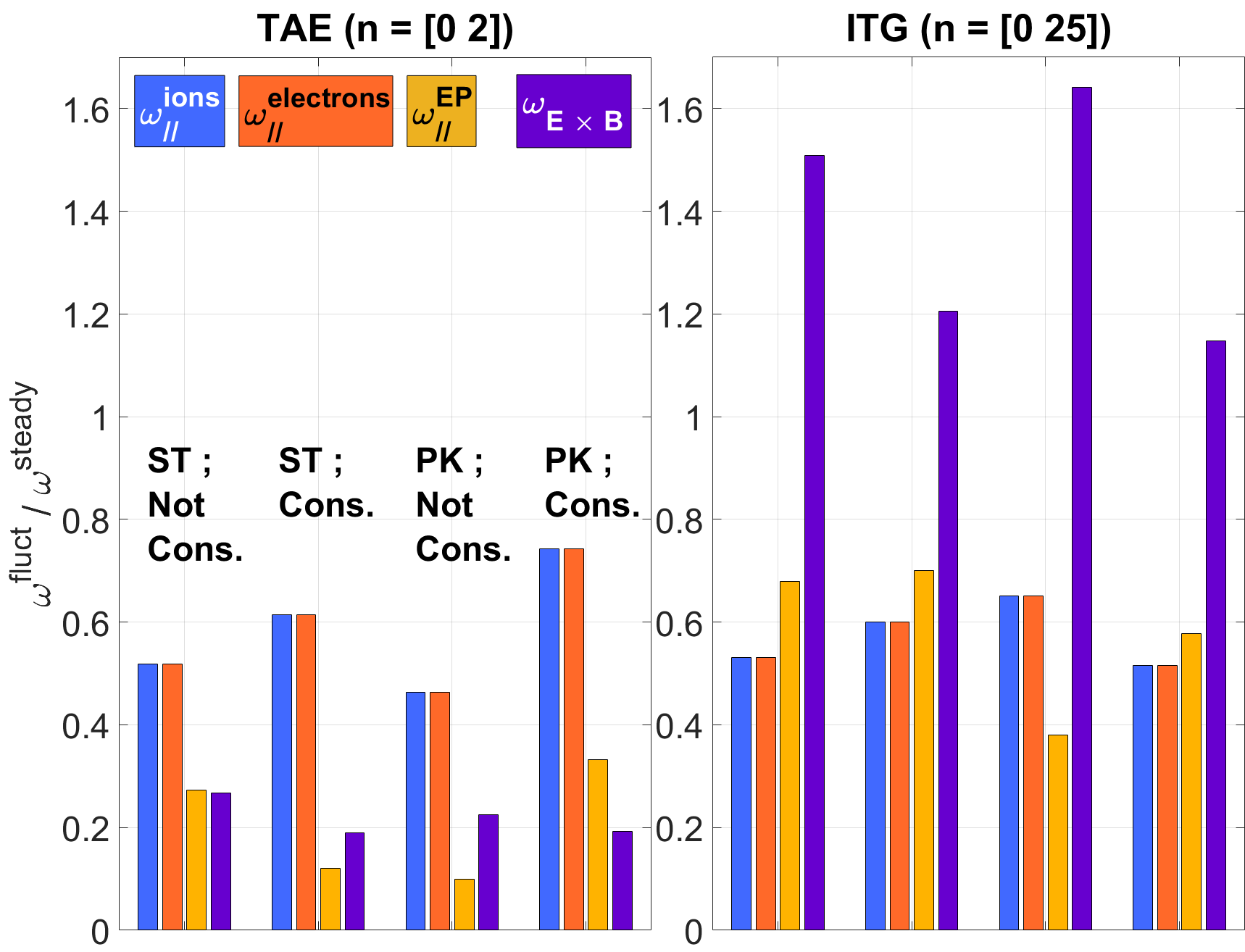}
\caption{\label{FIG:ExB_shear_steady_fluct} \it On the left:  $E \times B$ shearing rates, split between the steady $\omega_{E \times B}^{steady}$ (wide bars) and fluctuating $\omega_{E \times B}^{fluct}$ (error bars "like") parts and the linear growth rate of the unstable mode $\gamma^{linear}$ (black 'x's). 
On the right: The per species fluctuating to steady shearing rate ratio of $E \times B$ and parallel flows.} 
\end{center}
\end{figure}

In a nonlinear response to a growing instability, the $v_{E \times B}$ flows in the system self-organize into zonal structures with an $\omega_{E \times B}$ shearing rate comparable in magnitude to the growth rate of the unstable mode. To quantify this point, Figure \ref{FIG:ExB_shear_steady_fluct} represents the mode linear growth rate for comparison with the steady and fluctuating parts of the $\omega_{E \times B}$ shearing rate, defined as follows: 

\begin{equation}
\omega_{E \times B}(s,t) = \frac{s}{2 q \psi_{edge}}\frac{\partial}{\partial s}(\frac{1}{s}\frac{\partial \langle{ \phi}\rangle_{fsa}}{\partial s})
\end{equation}

where $\psi_{edge}$ is the poloidal magnetic flux on the last closed flux surface, and $\langle{ \phi}\rangle_{fsa}$ is the flux surface averaged electrostatic potential. Flux surface averaging is defined as

\begin{equation}
\langle{\cdot}\rangle_{fsa}(s,t) = \frac{\int (\cdot) J_{\phi,\theta^*} d\phi d\theta^*}{\int J_{\phi,\theta^*} d\phi d\theta^*}
\end{equation}

We further divide the total magnitude of the shearing rate \cite{Villard_PPCF2013}
\begin{equation}
    \omega_{E \times B}^{total} = \langle{\langle{|\omega_{E \times B}|}}\rangle_s\rangle_t.  
\end{equation}
 In to a steady state part 
 \begin{equation}
     \omega_{E \times B}^{steady} = \langle{|\langle{\omega_{E \times B}}\rangle_t|}\rangle_s,
 \end{equation}
 and a fluctuating part 
 \begin{equation}
     \omega_{E \times B}^{fluct} = \omega_{E \times B}^{total} - \omega_{E \times B}^{steady}.
 \end{equation}
 
Where $\langle . \rangle_s$ indicates an average over most of the radial domain $s \in [0.01,0.99]$, and the $\langle . \rangle_t$ represents time average over the nonlinear time window. Figure \ref{FIG:ExB_shear_steady_fluct} shows the steady part as solid bars, and fluctuating part as symmetric "error-bars". We recognize two opposite trends between the TAE and ITG instabilities. For the TAE on the left, we see a complex picture where for the ST profiles, the additional Shafranov shift from the EP population has little effect on both the steady and fluctuating parts of $\omega_{E \times B}$ but the mode is still stabilized (the linear growth rate decreases), while the stronger Shafranov shift in the PK cases reduces both the steady and fluctuating parts of $\omega_{E\times B}$ and stabilizes the mode.

The ITG cases, presented in the same tile to the right, show a monotonic increase of $\omega_{E \times B}^{steady}$ with Shafranov shift. The fluctuating part increases with bulk plasma profiles and decreases with Shafranov shift. Interestingly, the linear growth rate of the ITG cases is higher than the steady rate, except for the strong Shafranov shift found in the [PK, Cons.] case. The ratio between fluctuating and steady state shearing rates $\omega^{fluct}/\omega^{steady}$ in the ITG cases presented on the right in Figure \ref{FIG:ExB_shear_steady_fluct} is noticeably grater than 1, i.e. $[\omega^{fluct}/\omega^{steady}]_{E \times B}^{ITG} > 1$.

In the case of the TAE, to the left, bulk gradients contribute very little to the instability drive, but still increase the Shafranov shift. In this case $[\omega^{fluct}/\omega^{steady}]_{E \times B}^{TAE} \approx 0.2-0.3$ which is $6-8$ lower than the $[\omega^{fluct}/\omega^{steady}]_{E \times B}^{ITG}$. A question arises on the possible nonlinear cross-talk between TAE and ITG modes through the Zonal Structures they excite. Further studies will address this question by considering the simultaneous presence of these modes. 

Along side the fluctuating to steady ratio of the $E \times B$ flow, in Figure \ref{FIG:ExB_shear_steady_fluct} we present the same ratio for the shearing rate of the parallel flow $\omega_{\parallel}^{sp} = \partial v_\parallel/\partial s$. Here each species (sp) has its own contribution, and although the parallel shearing rate itself is orders of magnitude different between ions and electrons, we recover the same ratio for both. 

\begin{figure}
\begin{center}
\includegraphics[width=0.49\textwidth]{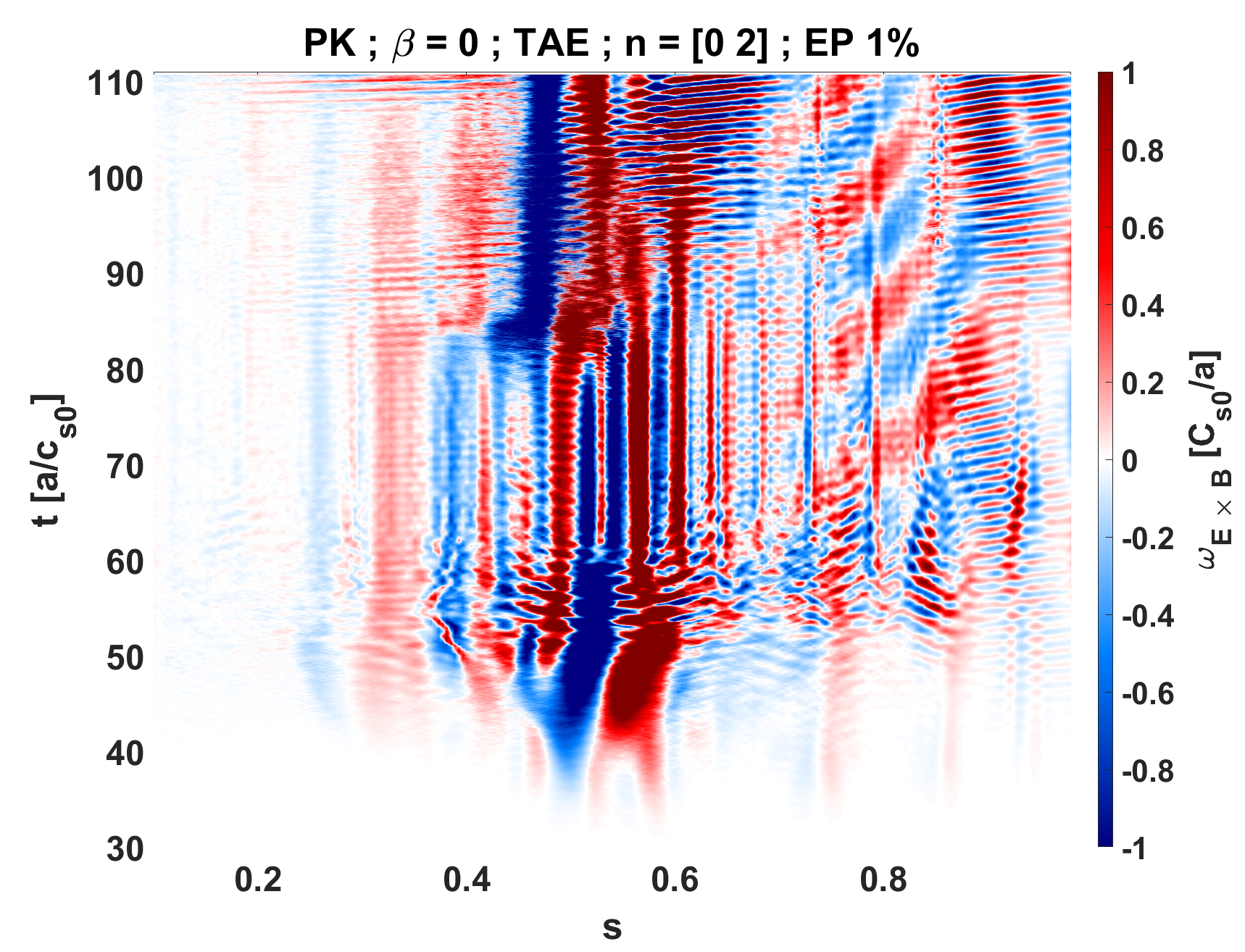}
\includegraphics[width=0.49\textwidth]{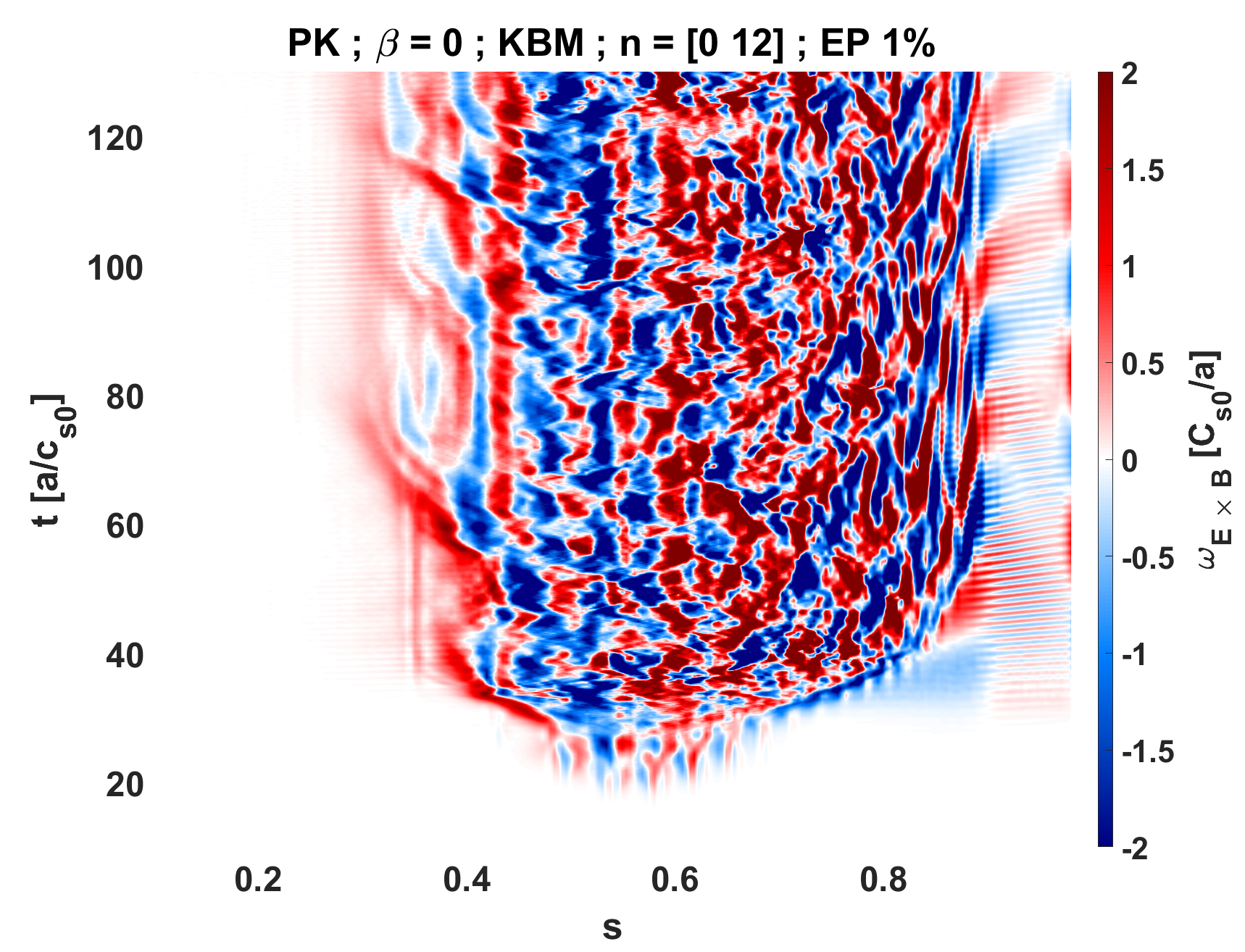}
\includegraphics[width=0.49\textwidth]{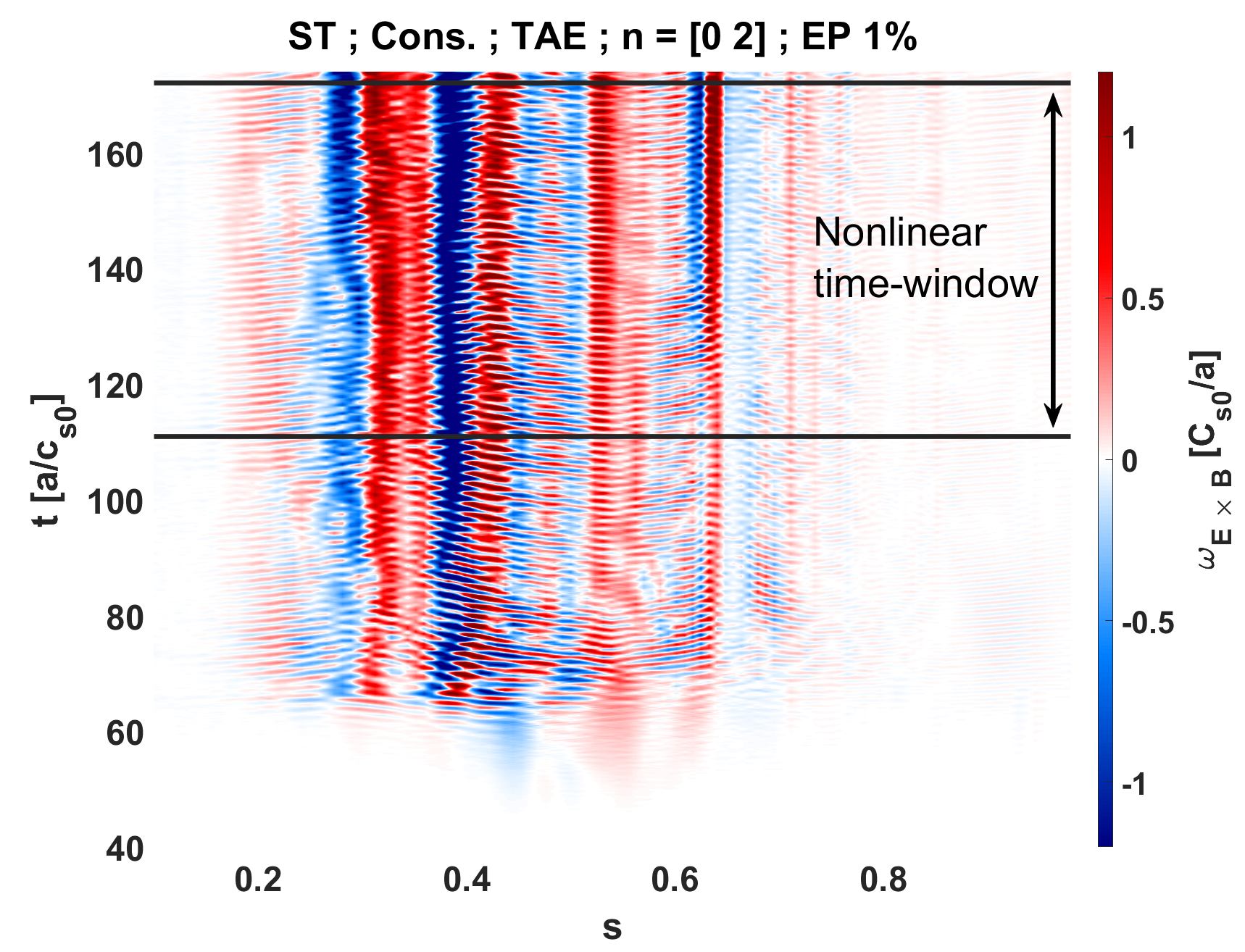}
\includegraphics[width=0.49\textwidth]{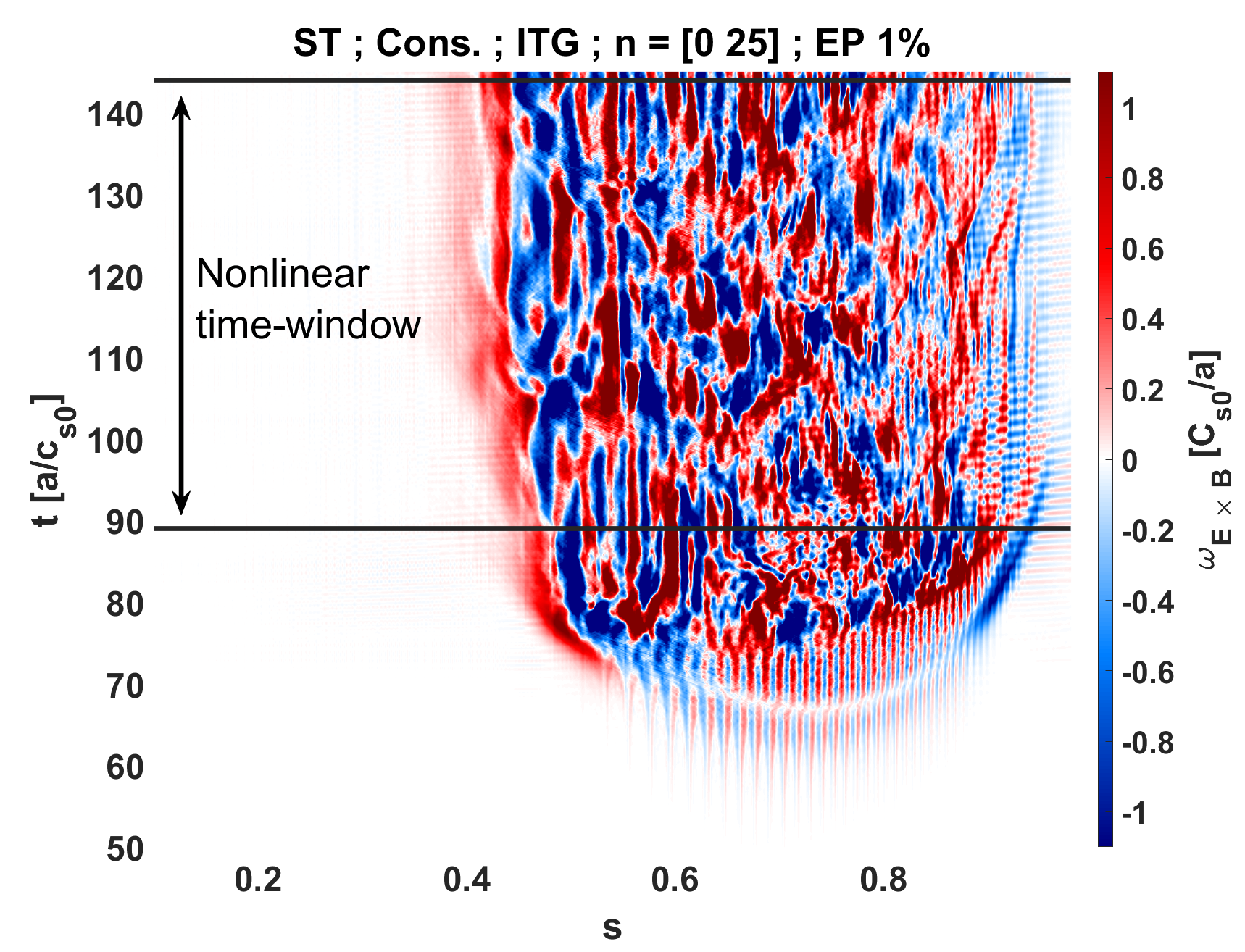}
\includegraphics[width=0.49\textwidth]{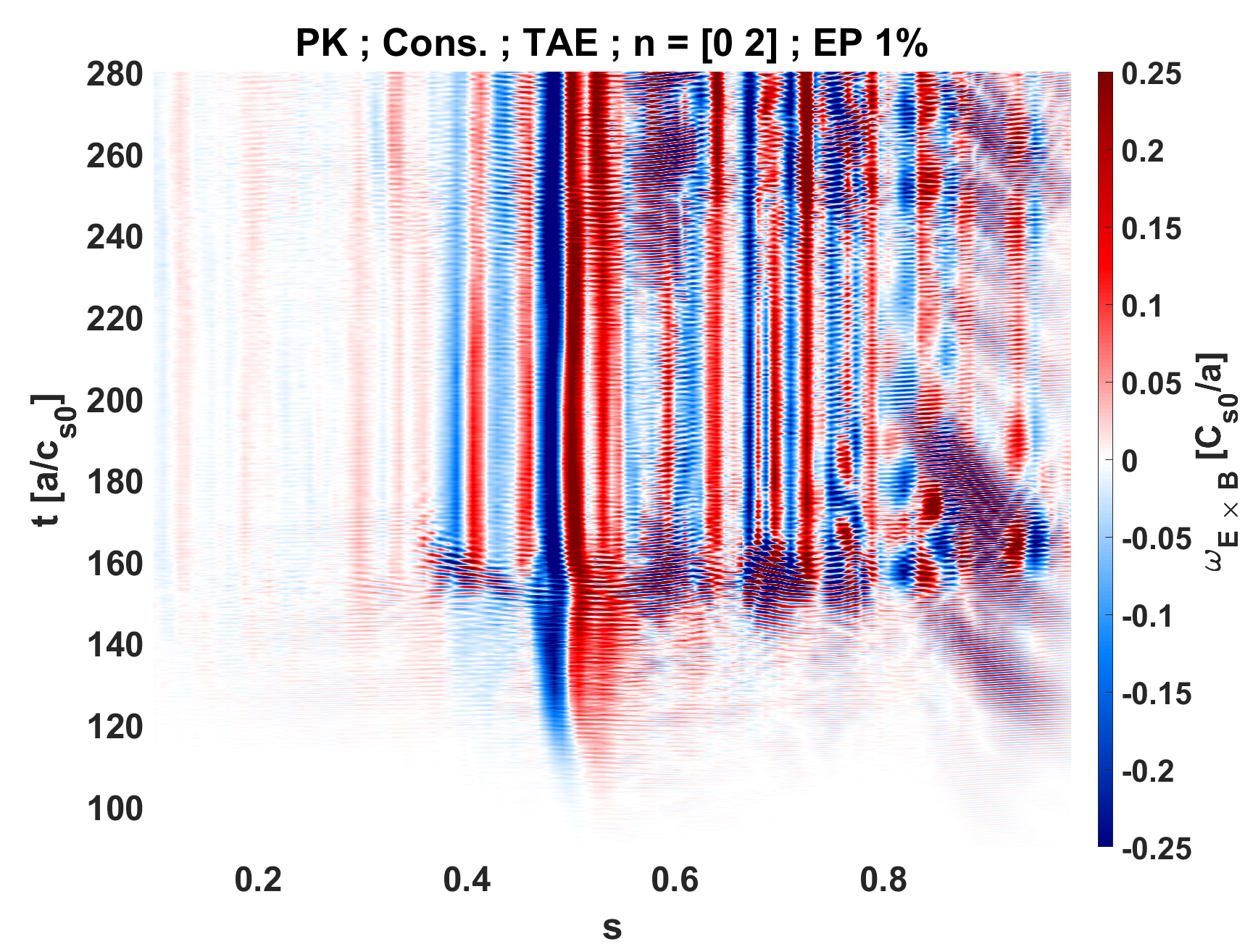}
\includegraphics[width=0.49\textwidth]{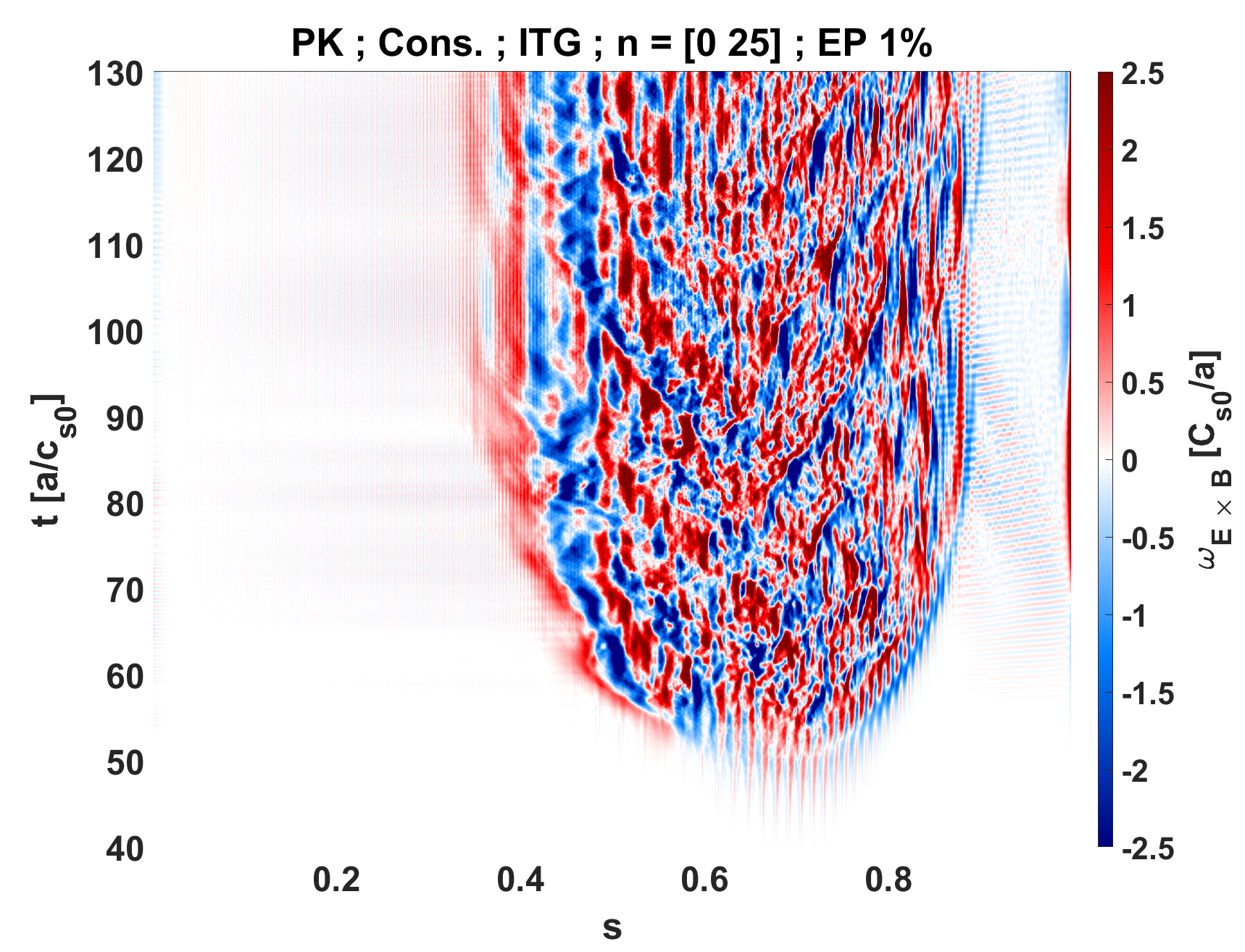}
\caption{\label{FIG:nonlinear_Omega_ExB_2D} \it
Nonlinear $\omega_{E \times B}$ shearing rate of the self emerging ZS arising in 3 different equilibria with $1\%$ EPs. The results are organized as follows: left coulomb shows the TAE mode and the right coulomb shows the ballooning mode of interest, i.e. a KBM in the top row and toroidal ITG in the middle and bottom row. Each row is dedicated to a different MHD equilibrium in an increasing order of Shafranov shift, with $\beta =0$ on the top row, and self-consistent MHD equilibria with standard (middle row) and peaked (bottom row) bulk profiles.}
\end{center}
\end{figure}

Figure \ref{FIG:nonlinear_Omega_ExB_2D} shows $\omega_{E \times B}$ contours vs. radius and time for several representative cases. First we observe a qualitative difference in the nature of the self-organized shearing pattern between TAE, KBM and ITG cases. The TAE $n=[0,2]$ cases appear radially coherent with long-lived radially situated bands, whereas the shearing patterns of the KBM $n=[0,12]$ and ITG $[0,25]$ cases appear chaotic and turbulent.

\begin{figure}
\begin{center}
\includegraphics[width=0.32\textwidth]{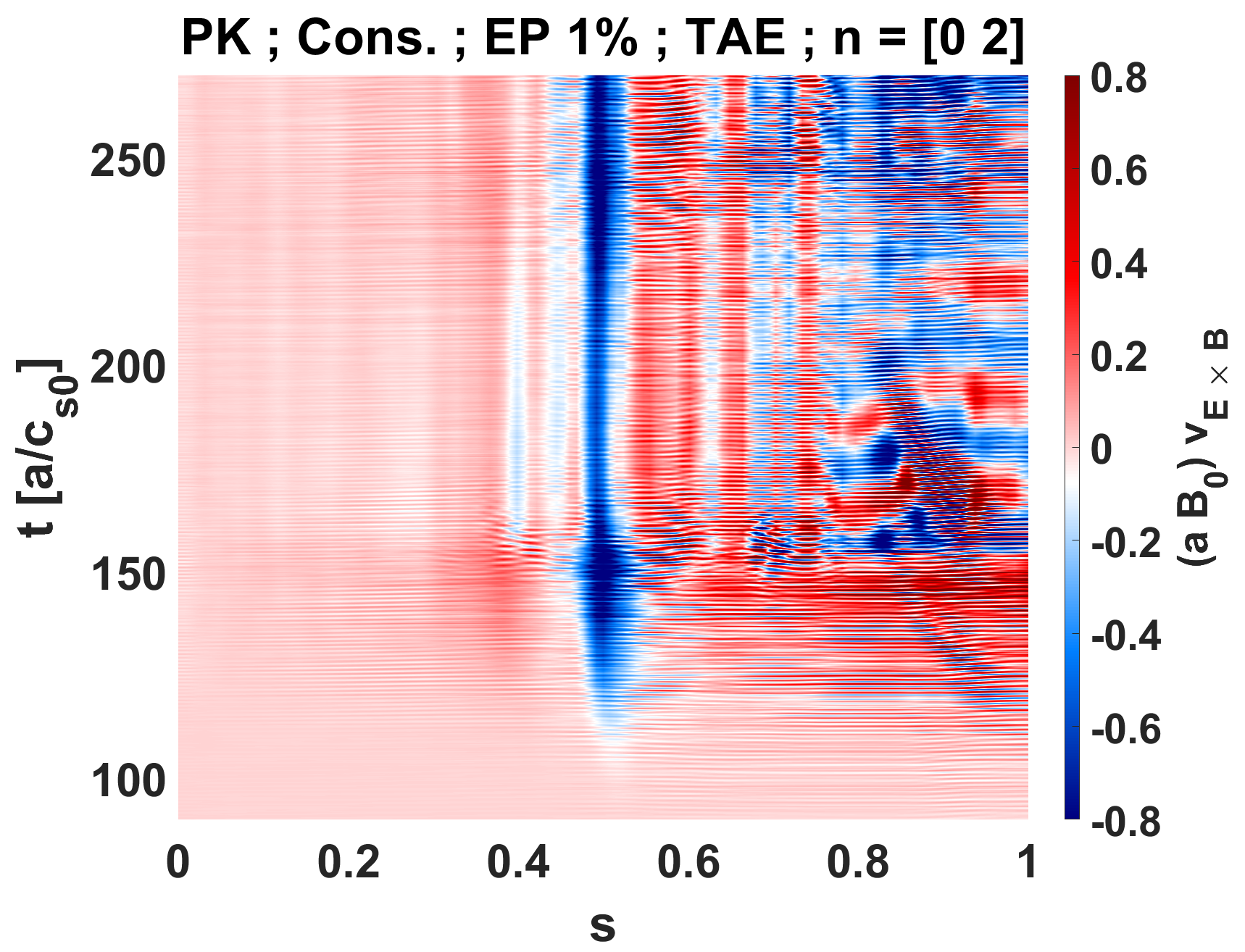}
\includegraphics[width=0.32\textwidth]{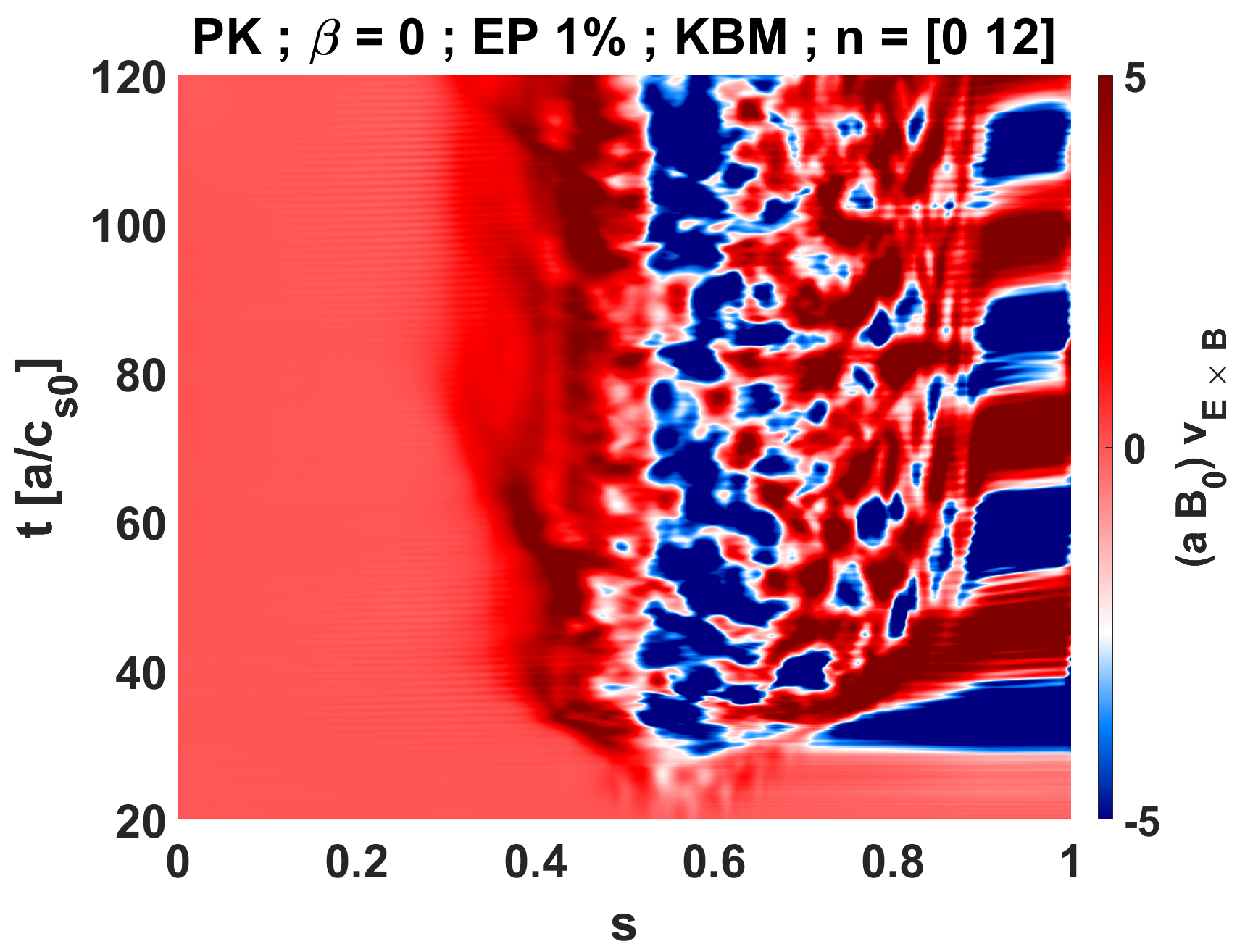}
\includegraphics[width=0.32\textwidth]{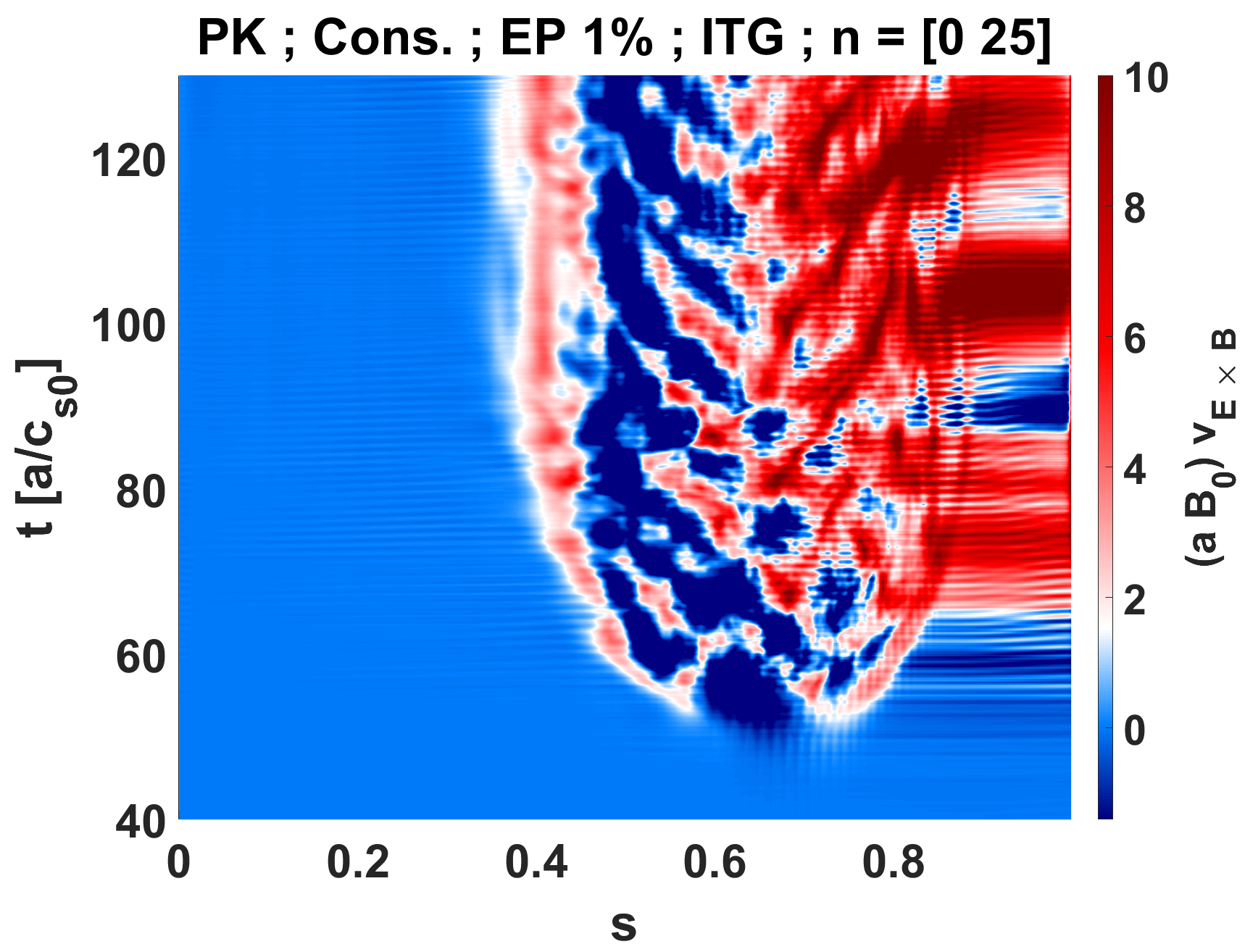}
\caption{\label{FIG:nonlinear_v_ExB_2D} \it
The $E \times B$ ($v_{E \times B}$) velocity for the TAE, KBM and ITG modes in the self consistent cases with peaked bulk gradients and $1\%$ EPs. Note that in the KBM and ITG cases, the $v_{E \times B}$ is asymmetrical.}
\end{center}
\end{figure}

In both Figures \ref{FIG:nonlinear_Omega_ExB_2D} and \ref{FIG:nonlinear_v_ExB_2D} we identify, on top of the main shearing bands and streamers, low frequency oscillations (most visually prominent in the $\beta = 0$ cases near the edge) that match the GAM dispersion relation \cite{Angelino_PoP2008}. And, high frequency oscillations (clearly apparent in the TAE cases) are prominent across the entire width of the plasma. The observed frequency matches the $n = 0$ GAE, a.k.a Axisymmetric Alfv\'en Eigenmode (AAE) dispersion relation \cite{Villard_NF1997}. The bottom plots in Figure \ref{FIG:Linear_disperssion} include the frequency of the AAE and figure \ref{FIG:TAE_conttinuum} places it in relation to the Alfv\'en continuum.

\begin{figure}
\begin{center}

\includegraphics[width=0.32\textwidth]{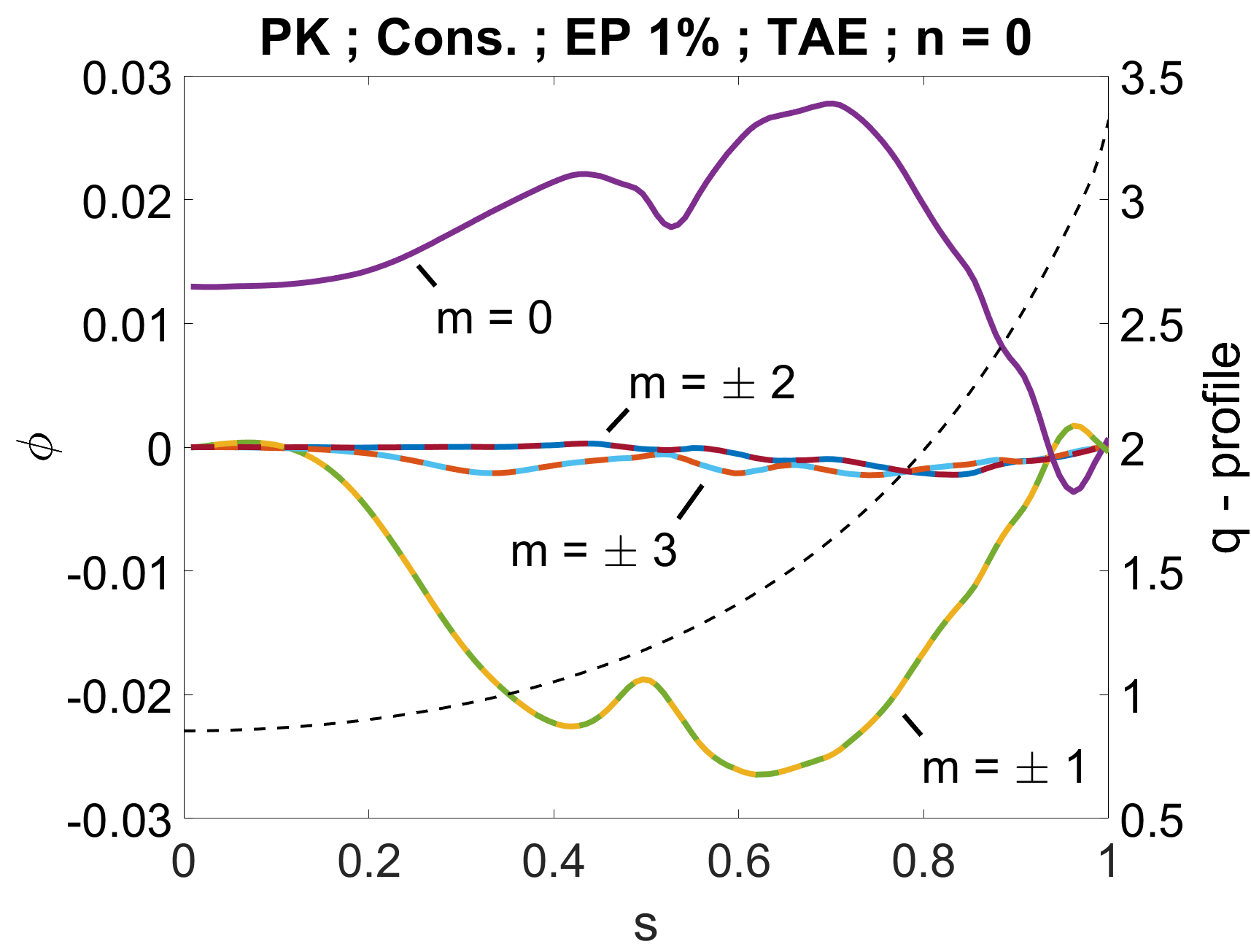}
\includegraphics[width=0.32\textwidth]{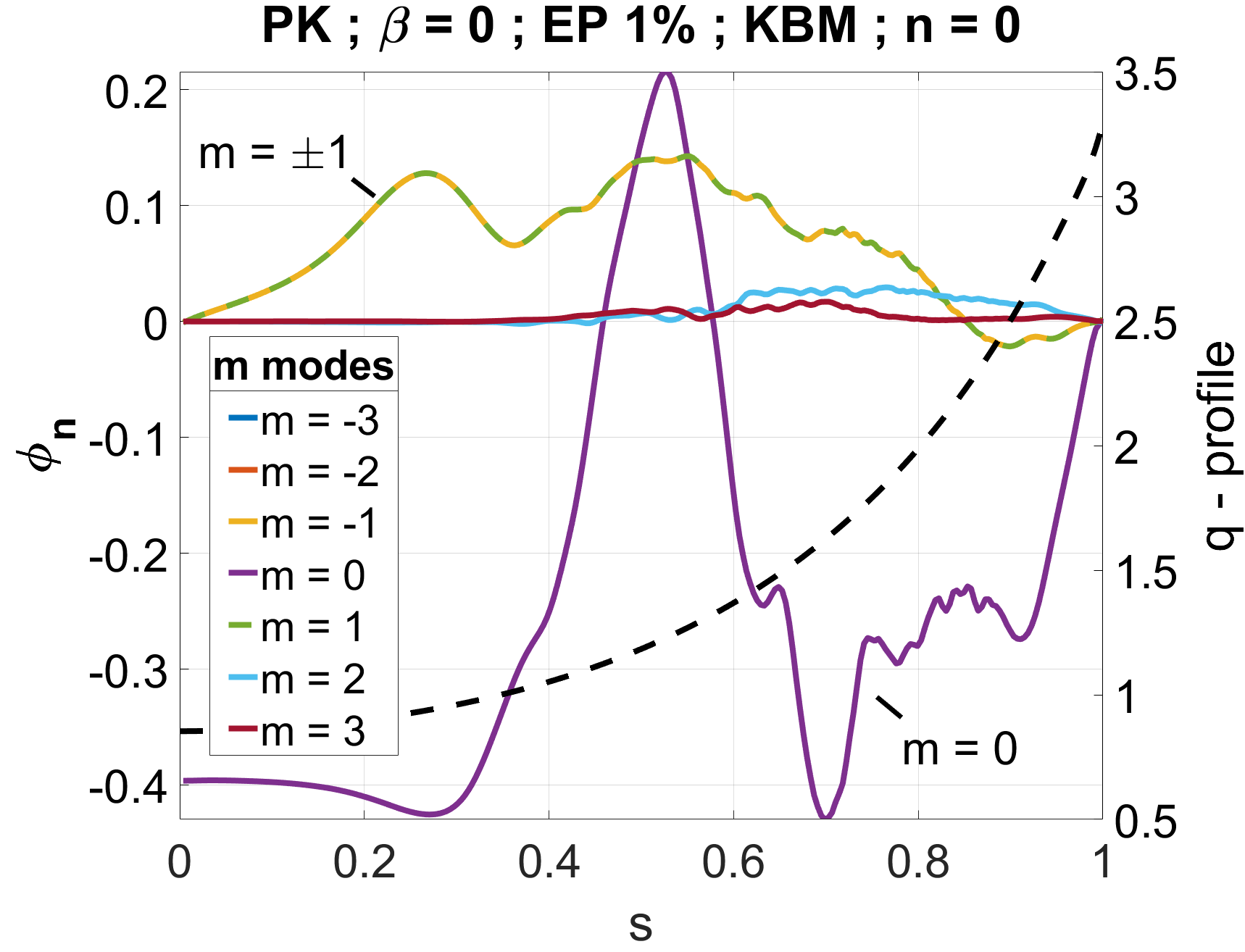}
\includegraphics[width=0.32\textwidth]{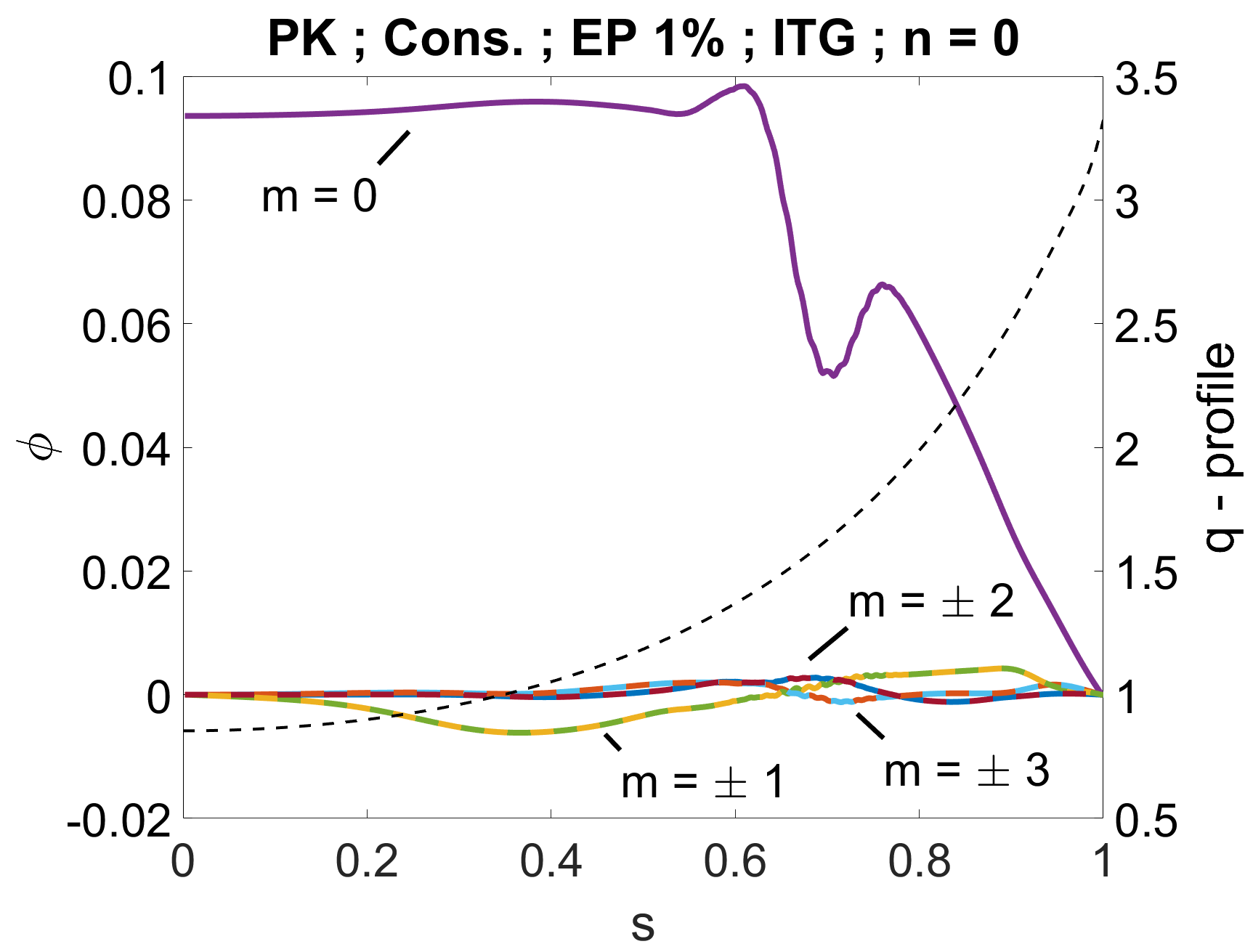}

\includegraphics[width=0.32\textwidth]{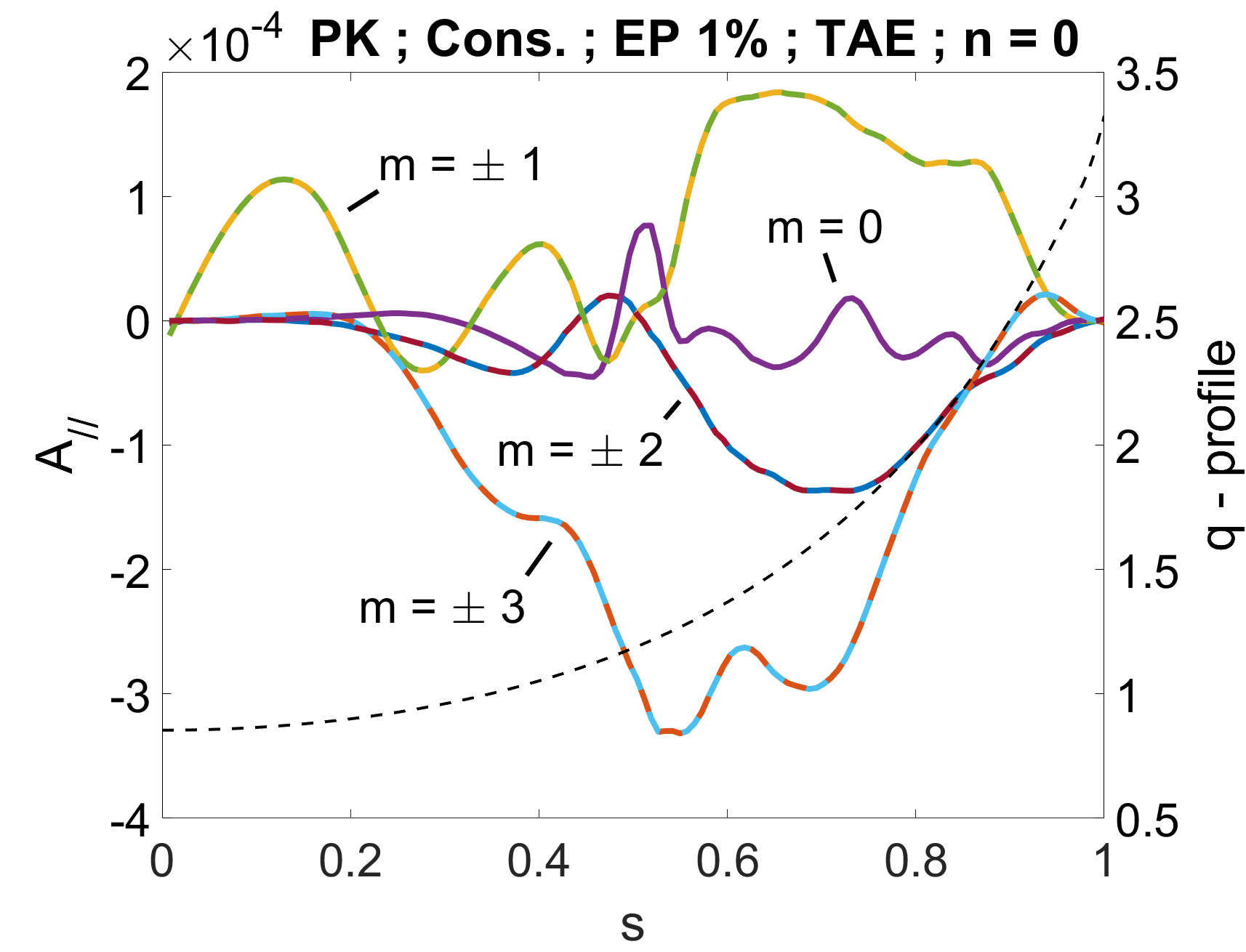}
\includegraphics[width=0.32\textwidth]{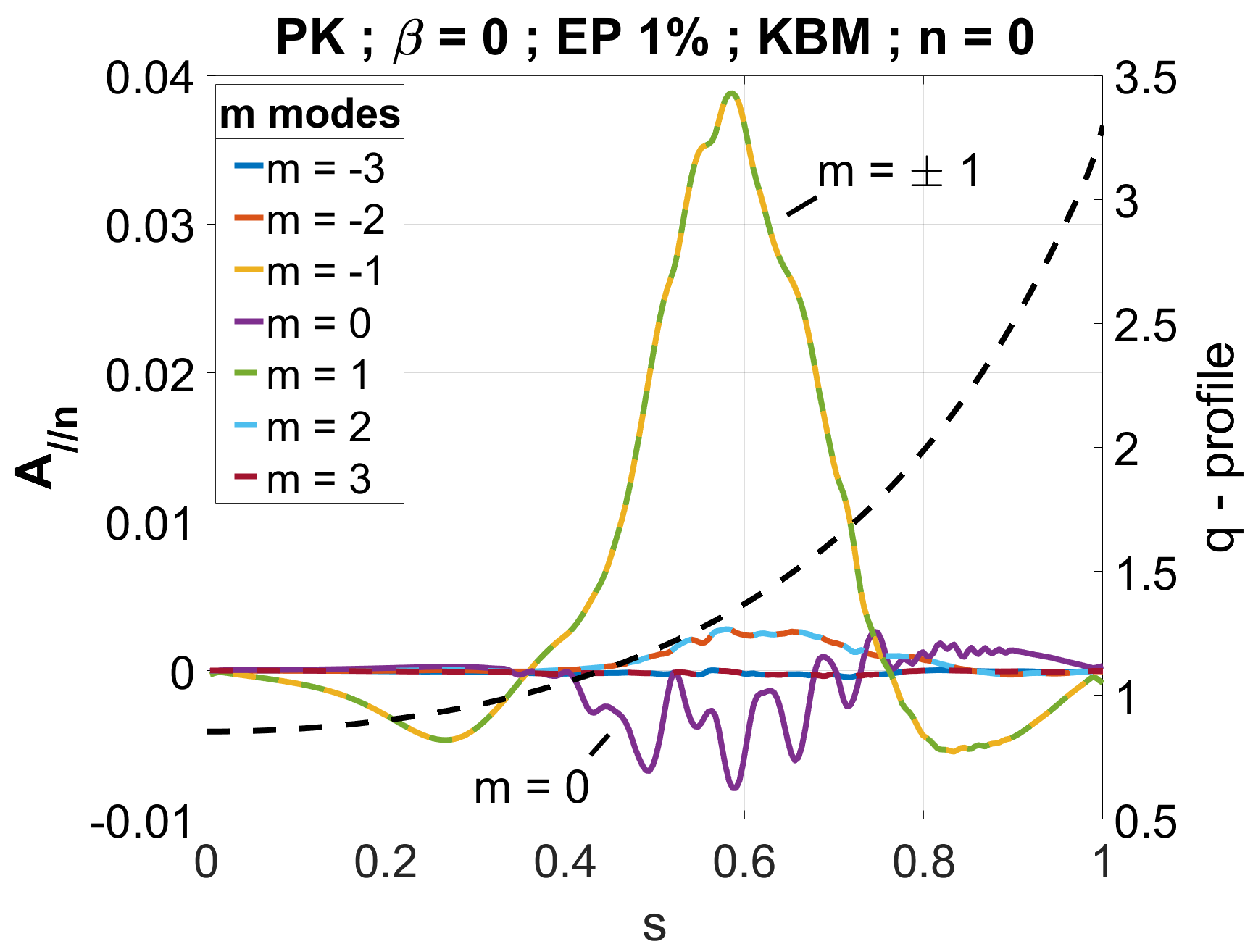}
\includegraphics[width=0.32\textwidth]{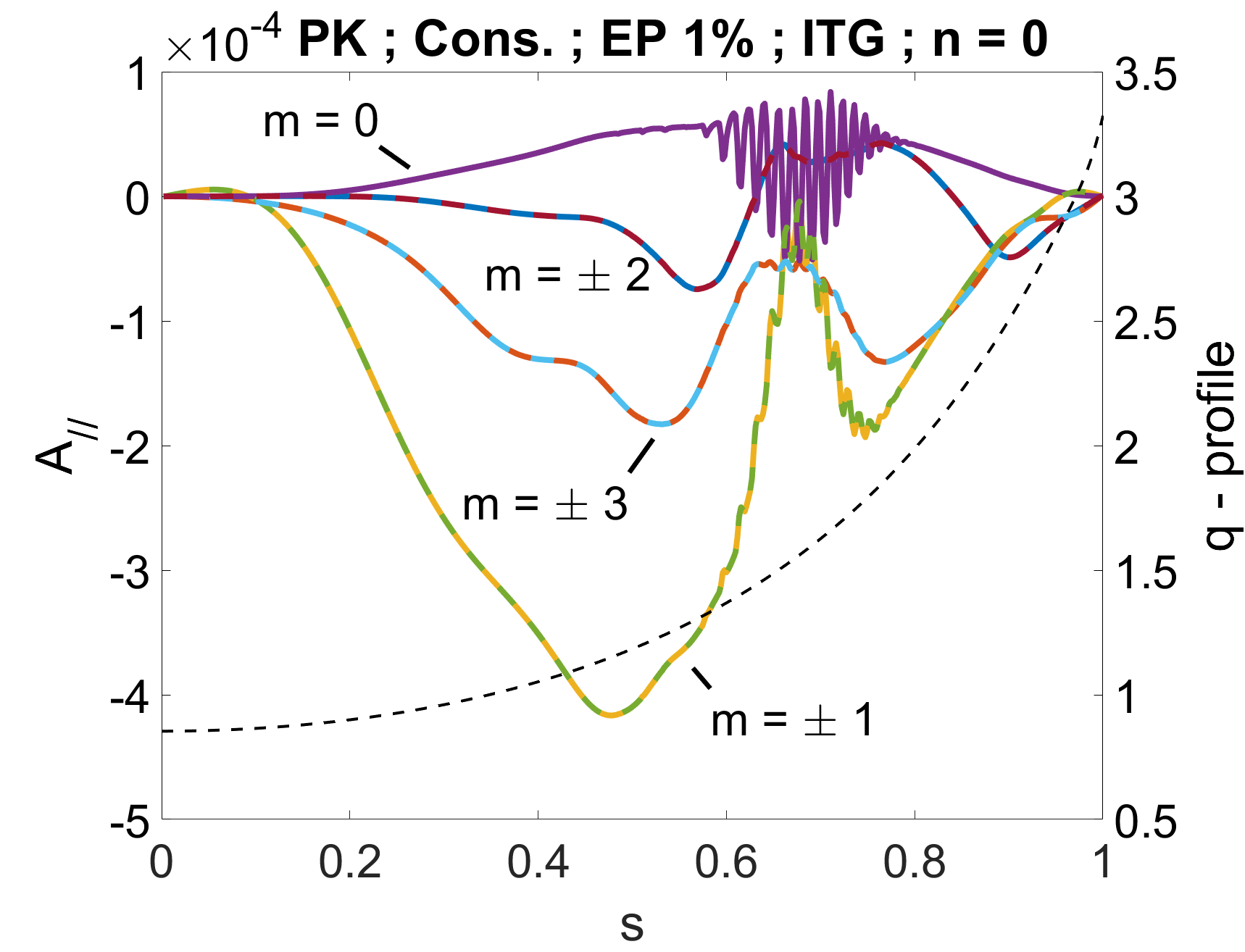}

\caption{\label{FIG:nonlinear_mode_n_0} \it
Characteristic nonlinear mode structure of the axisymmetric $n = 0$ mode in $\phi_0$ and $ A_{\parallel,0}$ for the $[ \text{PK, Cons., EP $1\%$}], \  n = 0 \in n = [0,2]$ TAE and $n = [0,25]$ ITG, and the $[ \text{PK, $\beta = 0$, EP $1\%$}], \ n = 0 \in n = [0,12]$ KBM. The dashed black circles are $s=[0;0.1;1]$.}
\end{center}
\end{figure}

As we notice the AAE presence in the time-traces of all the instabilities (TAE,KBM and ITG), we present in Figure \ref{FIG:nonlinear_mode_n_0} the instantaneous poloidal mode radial structure of the $n = 0$ mode. Alongside a dominant $m = 0$ harmonic, we find the $m = \pm 1$ and even $m = \pm 2$ \& $\pm3$ modes, consistent with a typical AAE structure. We can observe a nonlinear coupling between the $m = 0$ harmonic and higher harmonics, e.g. $m = \pm 1$, in TAE and ITG plots of $\phi$ and $A_{\parallel}$ respectively.  

\begin{figure}
\begin{center}
\includegraphics[width=0.325\textwidth]{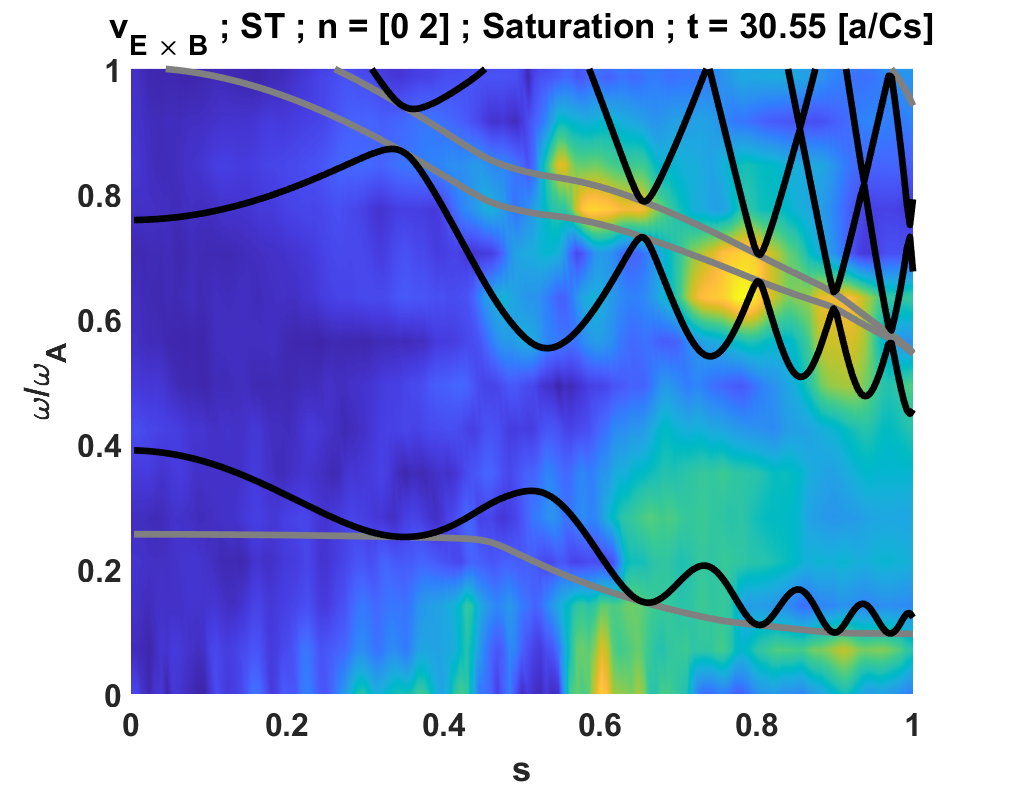}
\includegraphics[width=0.325\textwidth]{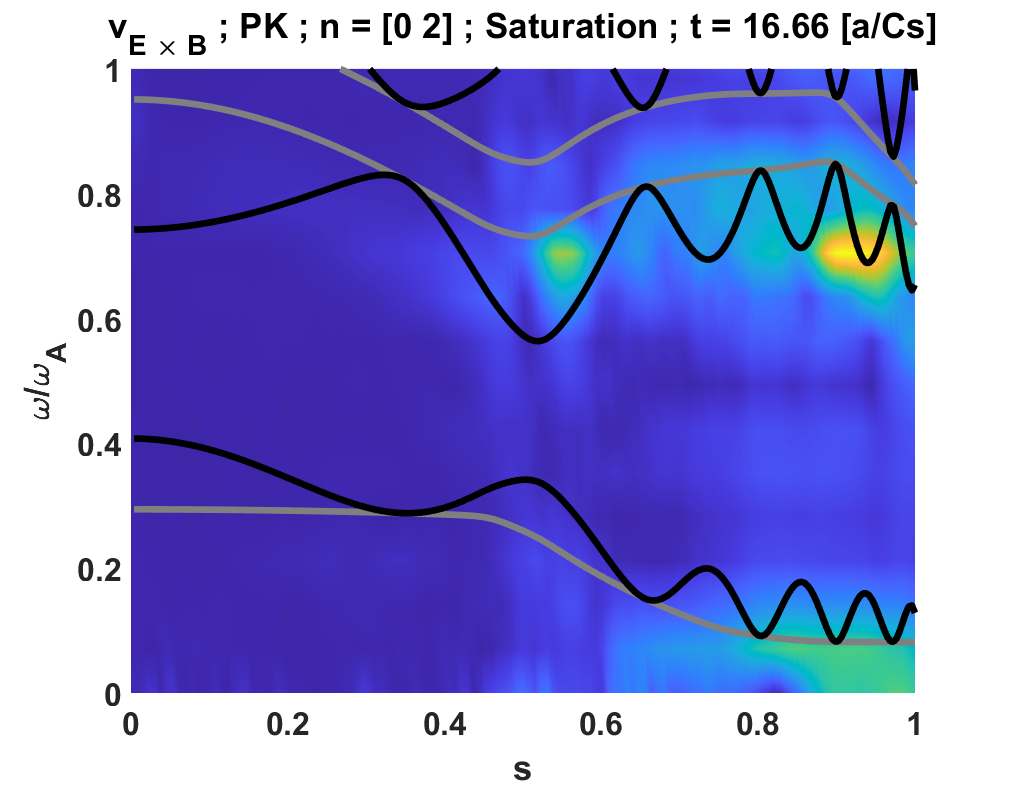}
\includegraphics[width=0.325\textwidth]{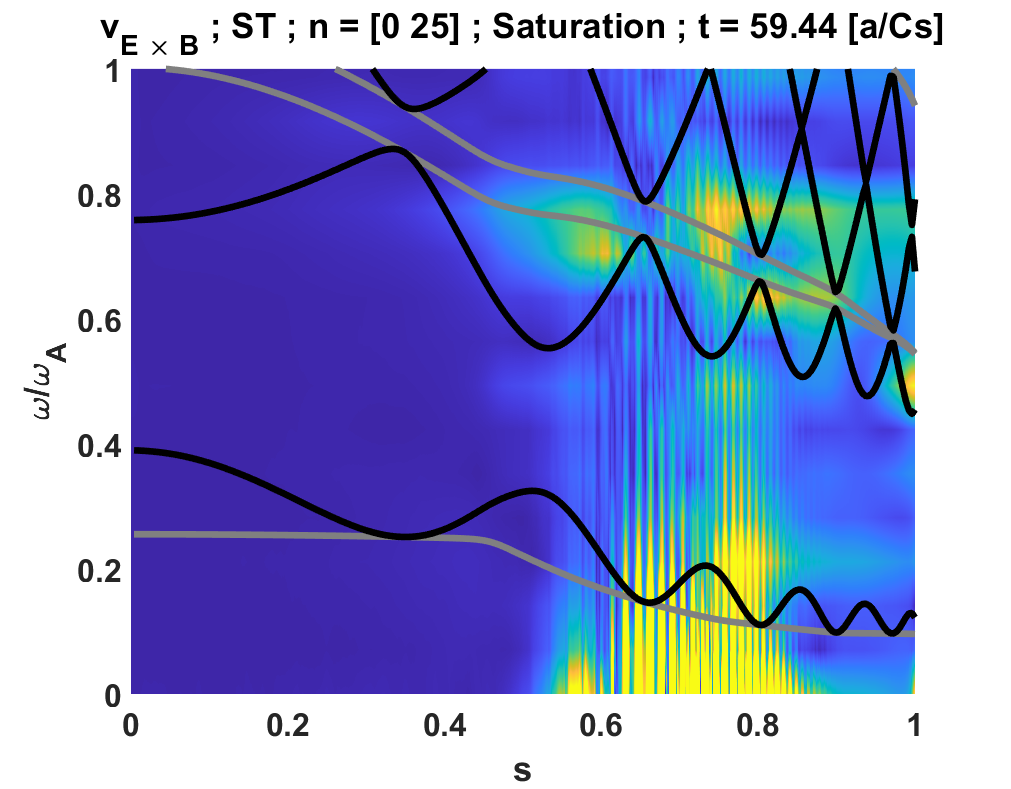}

\includegraphics[width=0.325\textwidth]{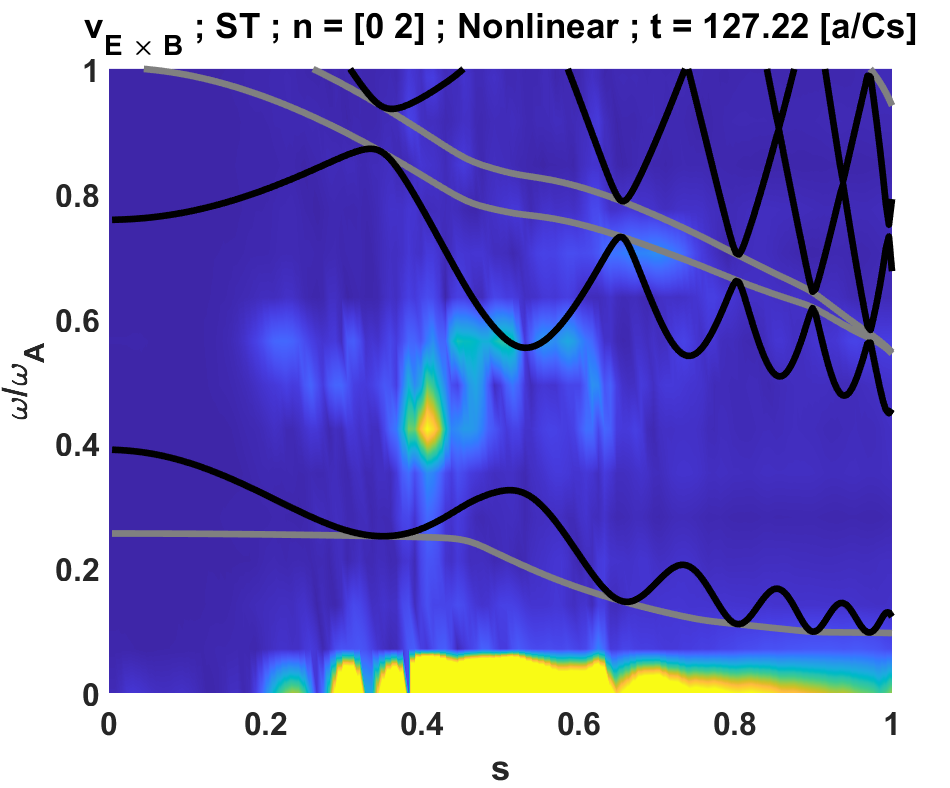}
\includegraphics[width=0.325\textwidth]{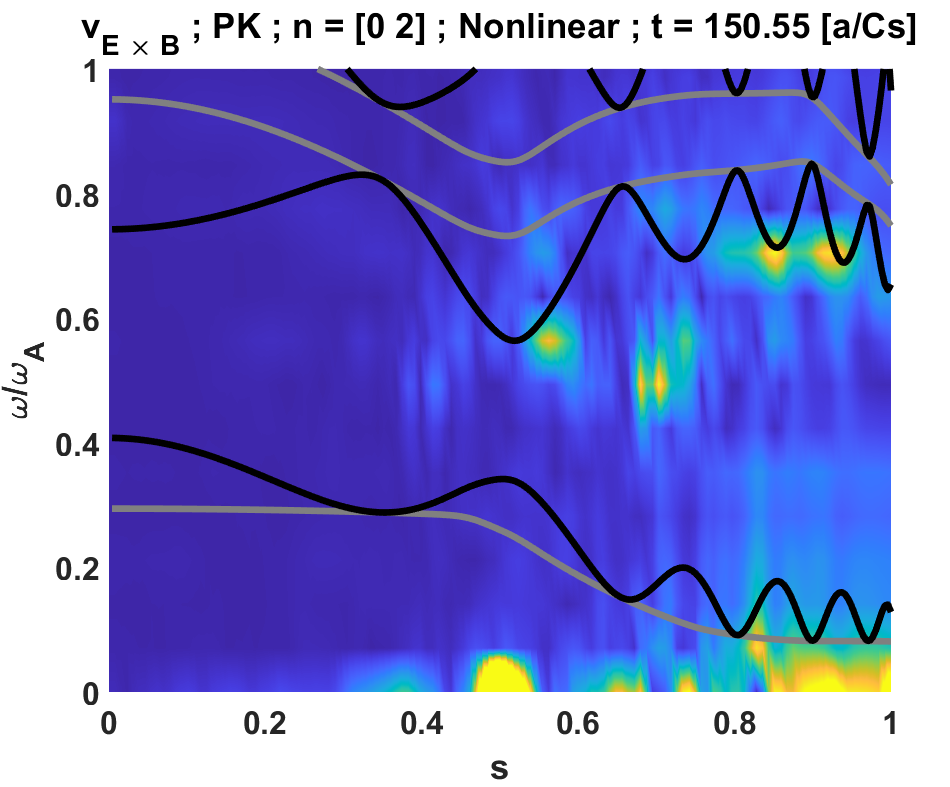}
\includegraphics[width=0.33\textwidth]{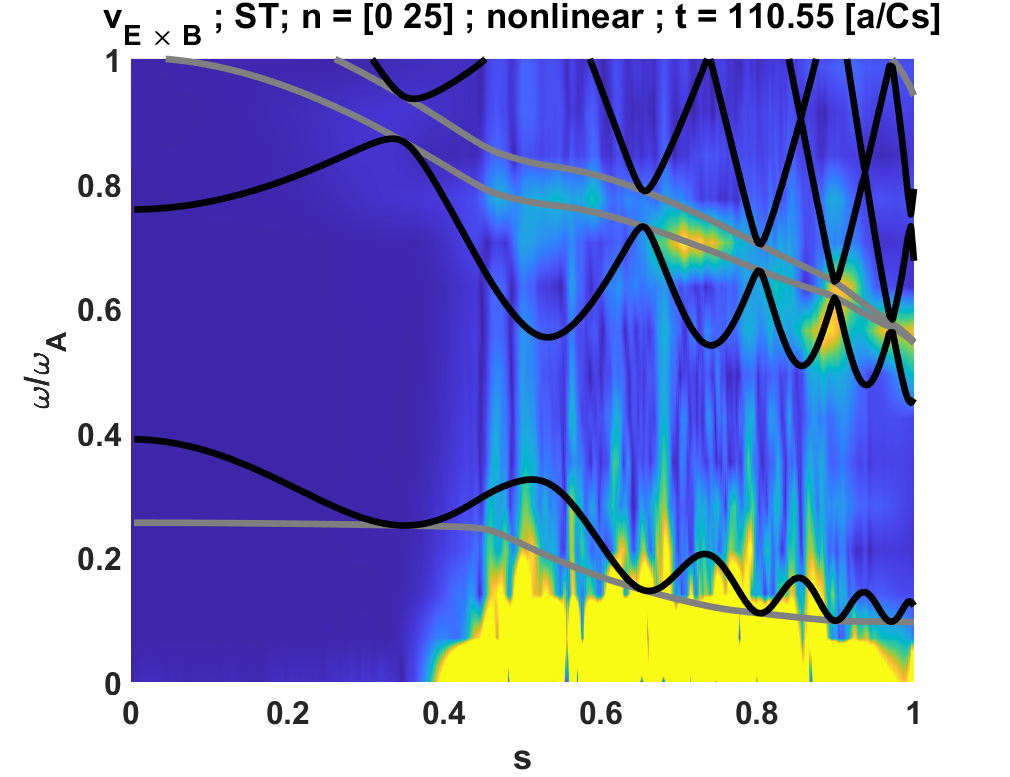}

\caption{\label{FIG:nonlinear_ExB_fft2D_timelaps} \it Frequency spectrum of $v_{E \times B}$ flows normalized to Alfv\'enic frequency $\omega_A$ with an overlay of $n = 0$ Alfv\'en continuum in gray and an $n = 2$ Alfv\'en continuum in black. The "Quasi-linear" and "Nonlinear" label indicates the system's stage at that timestamp.
Top row: quasi-linear stage. Bottom row: nonlinear stage. Left column: TAE with ST bulk profiles. Middle column: TAE with PK bulk profiles. Right column: ITG with ST bulk profiles.}
\end{center}
\end{figure}

We examine the frequency spectrum of the $E \times B$ flow, and in Figure \ref{FIG:nonlinear_ExB_fft2D_timelaps} we present a collection of snapshots capturing some of the nonlinear dynamics of the system. Both the TAE and the ITG show strong AAE activity during the saturation phase (Top row). For both the standard and peaked profiles, the AAE in $v_{E \times B}$ is excited at the edge. The minimum of the $n = 0$ Alfv\'en continuum, Figure \ref{FIG:TAE_conttinuum}, for the cases with standard profiles is located at the edge, and from that point inwards, the AAE chirps-up along the continuum branch. This is true for both the TAE (top left tile) and the ITG (top right tile). The $n = 0$ Alfv\'en continuum for the cases with peaked bulk gradients, has two very similar local minima at the edge and the global minimum around $s=0.55$. Nonetheless, the AAE gets excited at the edge before propagating inwards while chirping up along the continuum branch. As seen. e.g. in the middle top tile where both minima light up simultaneously, at about $s = 0.95$ the mode deviates from its spectrum as it moves deeper into the core with a constant frequency, perhaps due to long range interactions with the global minimum. 

On the bottom row of Figure \ref{FIG:nonlinear_ExB_fft2D_timelaps} we see the system at a much later nonlinear stage. Here the system establishes a TAE for the $n=2$ mode, which for the ST cases remains close to the linear frequency. However for the PK cases the inner minimum in AAE continuum results in a beating between the AAE and the TAE which chirps up to the top of the continuum. During the saturation the ITG loses its linear mode structure and spreads (radially) while continuously exciting AAE activity as seen in the right column.

\begin{figure}
\begin{center}
\includegraphics[width=\textwidth]{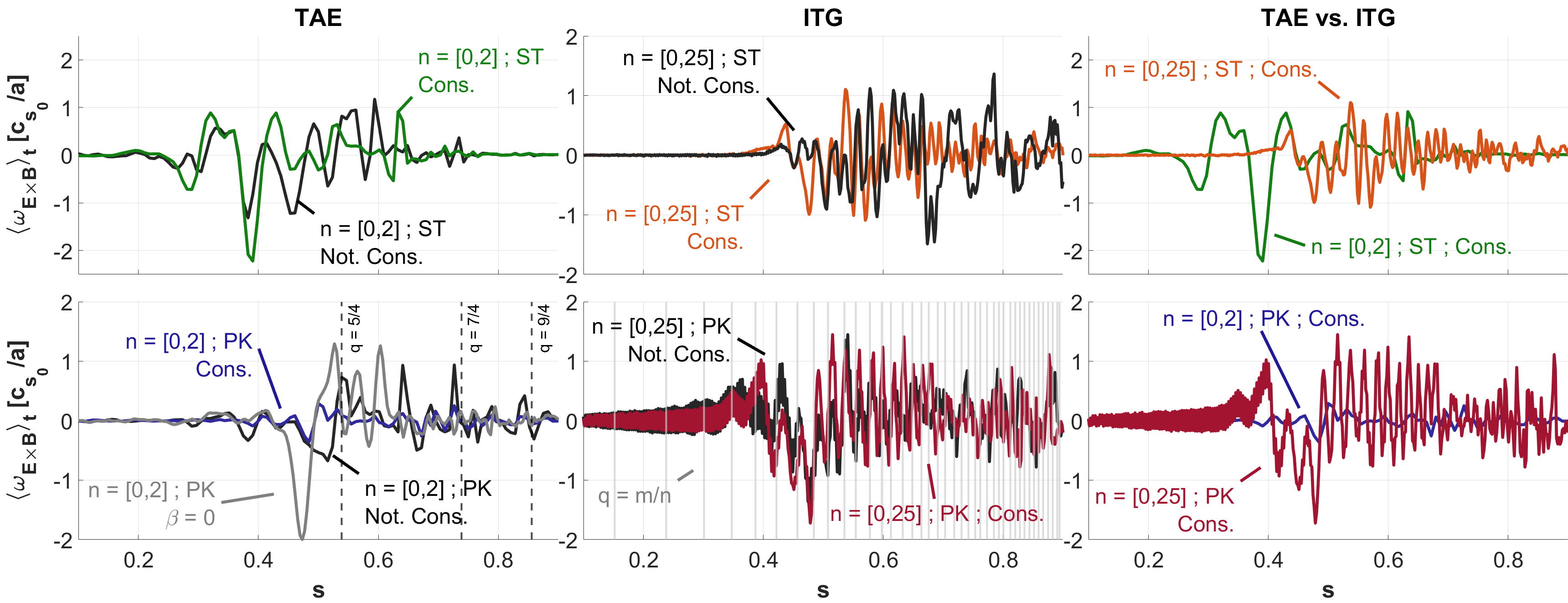}
\caption{\label{FIG:Omega_ExB} \it
The radial profiles of the time averaged $E \times B$ shearing rate - $\langle \omega_{E \times B} \rangle_t$. 
Color coding: Gray - $\beta = 0$, Black - Not Consistent (Not Cons.), and Colored - Consistent (Cons.) MHD equilibria.}
\end{center}
\end{figure}

\begin{figure}
\begin{center}
\includegraphics[width=\textwidth]{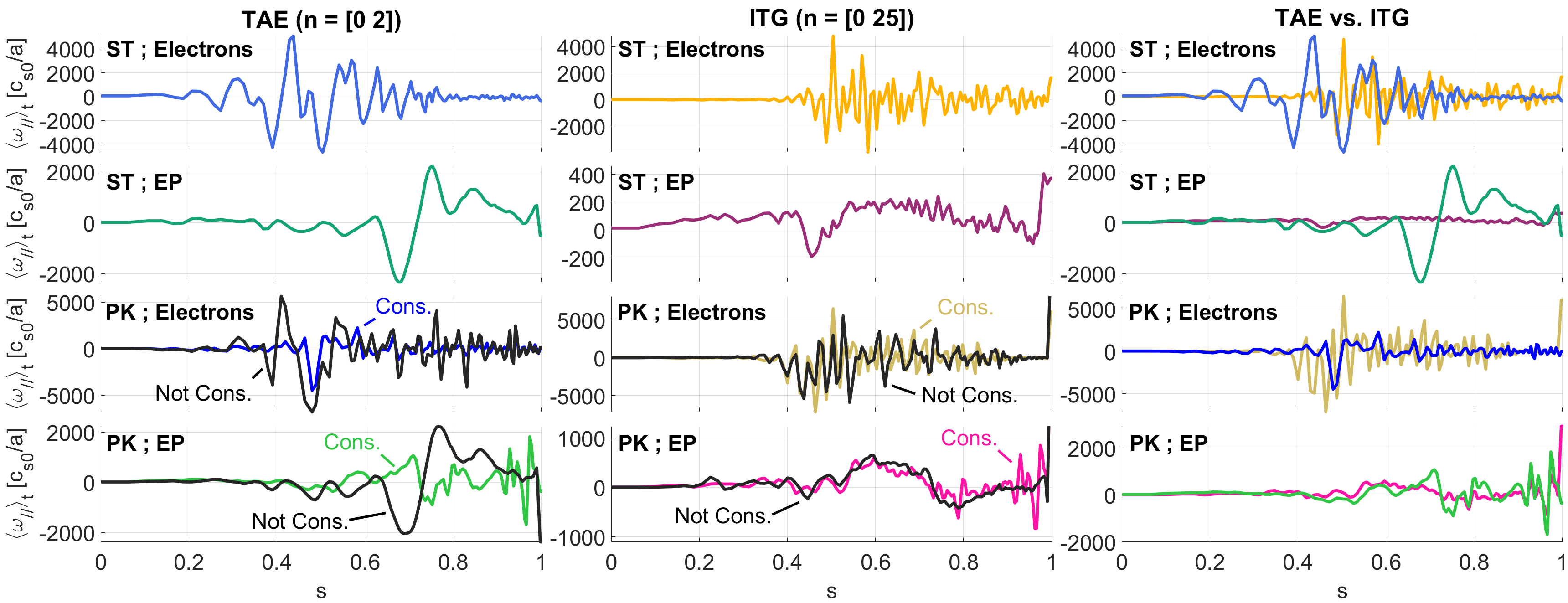}
\caption{\label{FIG:Omega_para} \it
The radial profiles of the time averaged parallel shearing rate - $\langle \omega_{\parallel} \rangle_t$ - for the electron and EP channels. For the PK cases we add the not consistent results in black.}
\end{center}
\end{figure}

To expand our understanding of $E \times B$ and parallel flows we plot in Figures \ref{FIG:Omega_ExB} for $\omega_{E \times B}$ and \ref{FIG:Omega_para} for $\omega_{\parallel}$, the radial profile of the steady state shearing rates $\omega_{k}^{steady}(s) = \langle{\omega_{k}}\rangle_t\ |\ k=\{E \times B,\parallel\}$. The $n=2$ TAE and $n=25$ ITG self-organize into different ZS patterns that only partially align with the mode rational surfaces $q = (m + 1/2)/n$ for the TAE or $q = m/n$ for the ITG. The radial patterns are consistent with the lower $n$ TAE modes showing larger (in amplitude and width) perturbations compared to the higher $n$ ITG. Thus both modes can have comparable steady state sharing rates like Figure \ref{FIG:ExB_shear_steady_fluct} shows for the ST cases, while having a different radial profile as illustrated in Figures \ref{FIG:Omega_ExB} and \ref{FIG:Omega_para} in the right column. The direct effects of Shafranov shift are hard to interpret and do not show as a clear trend in the radial patterns, unlike in Figure \ref{FIG:ExB_shear_steady_fluct}. There is also no clear correlation between the radial pattern of the EP parallel velocity $\langle \omega_{\parallel} \rangle^{EP}_t(s)$ and electron parallel velocity $\langle \omega_{\parallel} \rangle^{elec}_t(s)$.    

\subsection{Modified $q$ profile}

\begin{figure}
\begin{center}
\includegraphics[width=\textwidth]{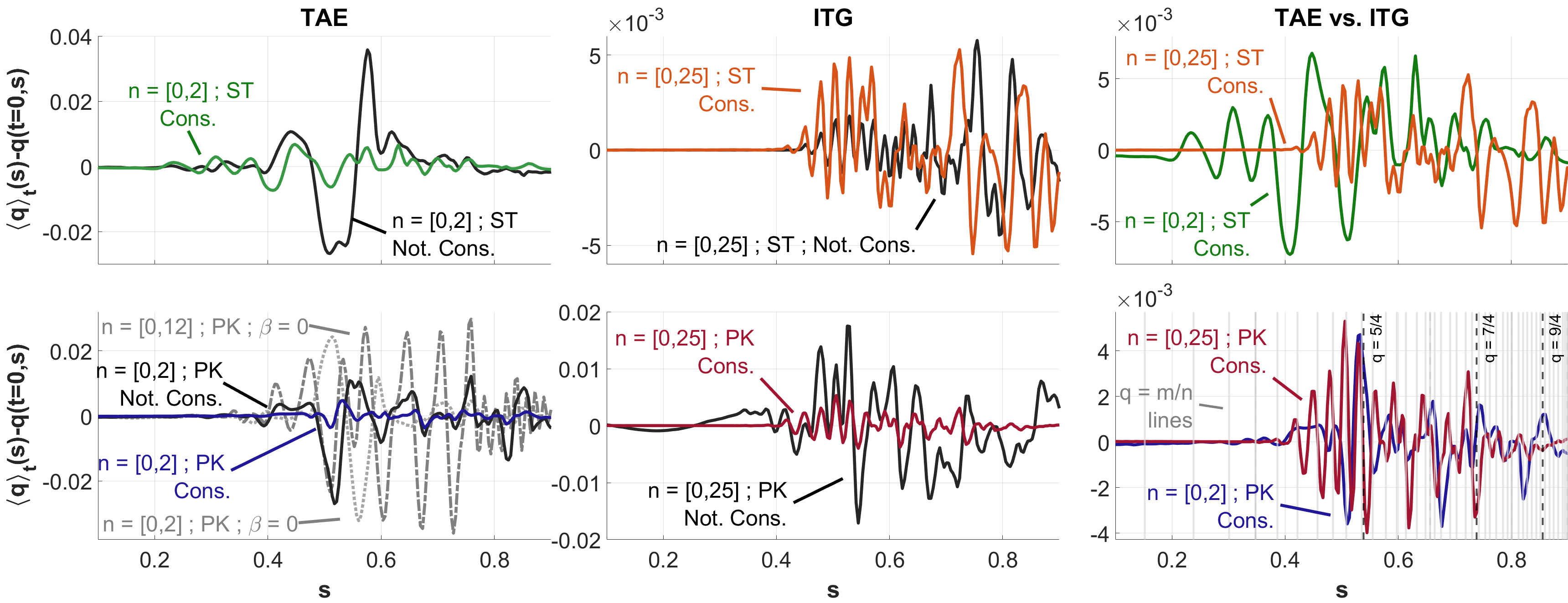}
\includegraphics[width=0.50\textwidth]{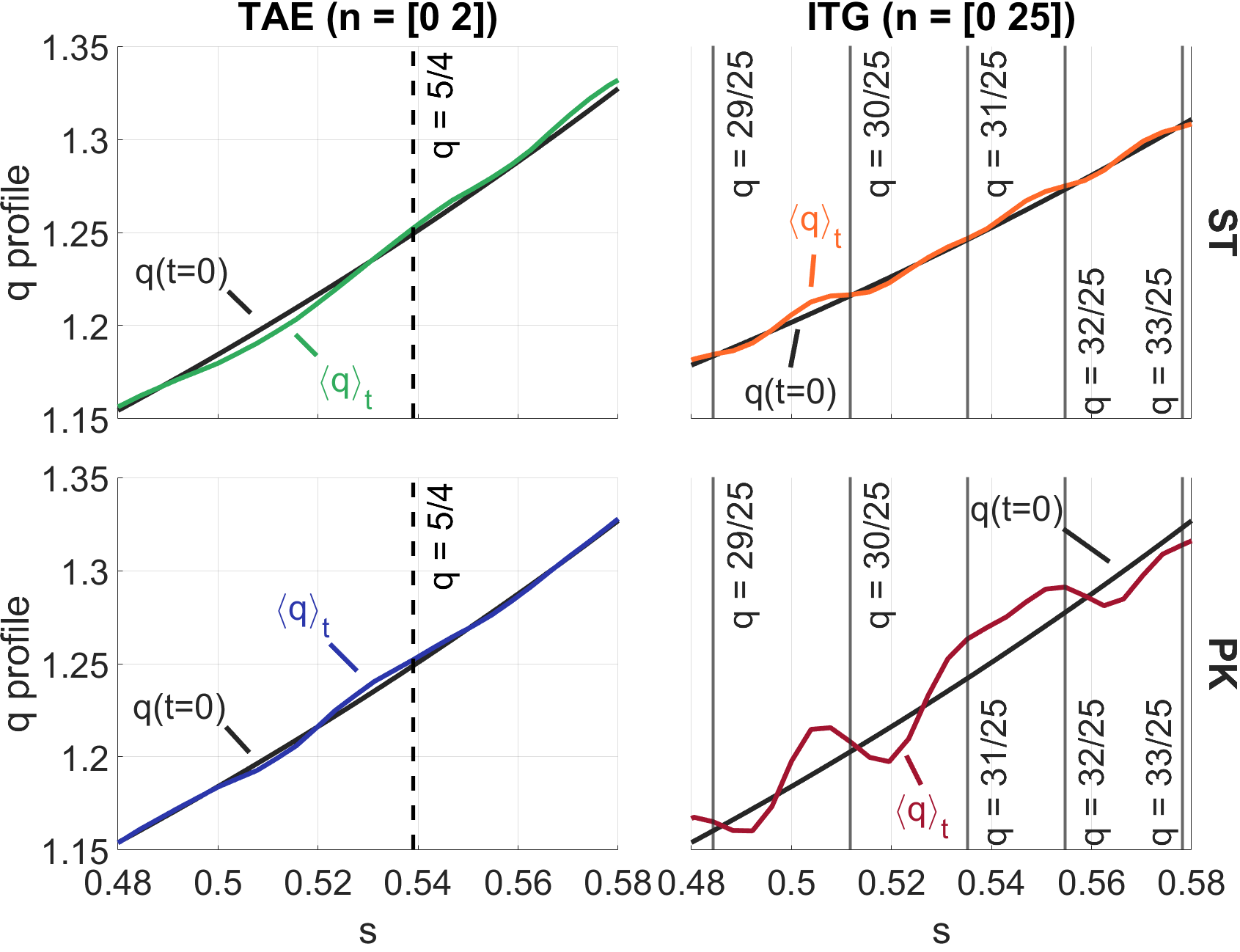}
\includegraphics[width=0.49\textwidth]{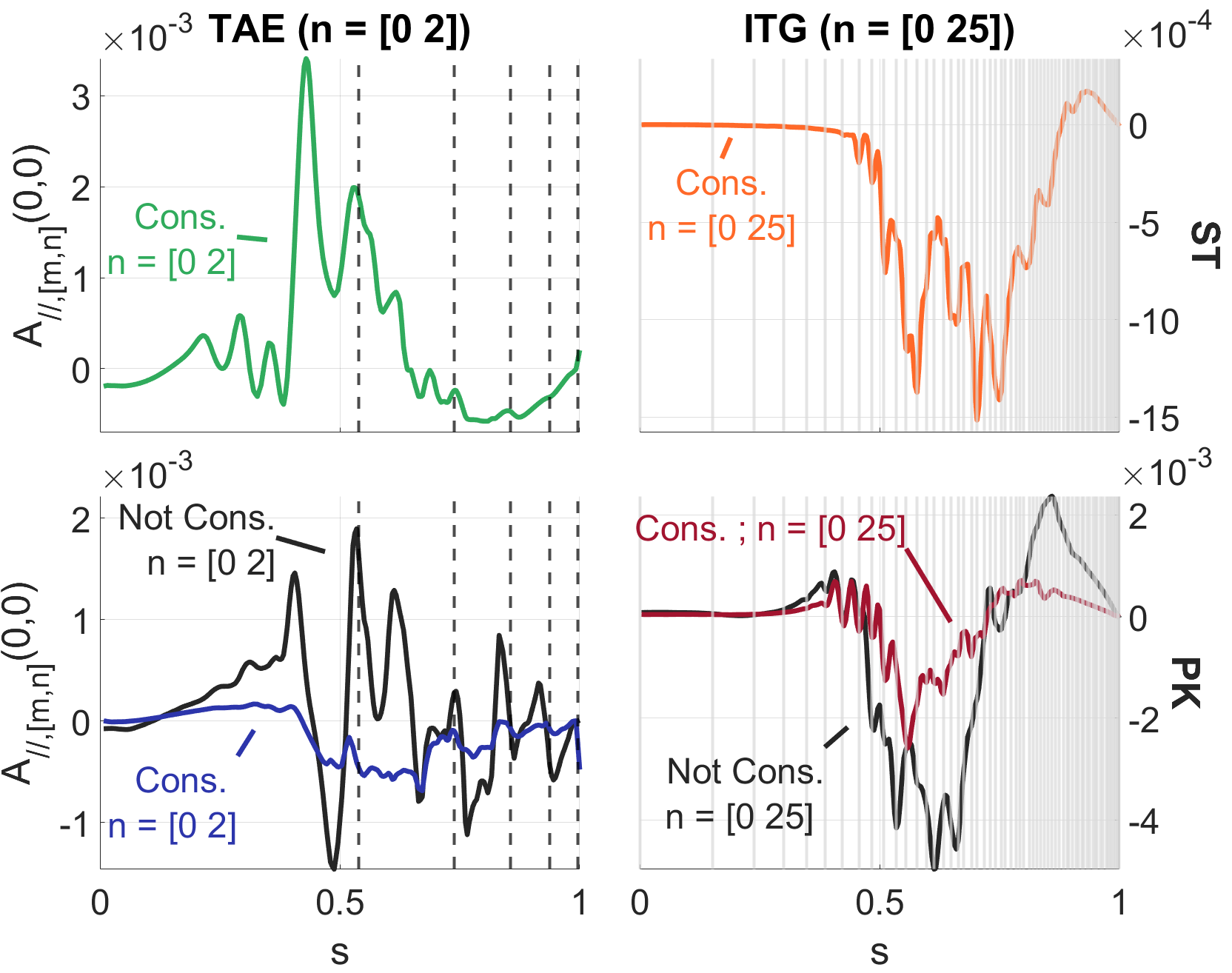}
\caption{\label{FIG:modified q-profile} \it
Top plots: The difference between the initial and the nonlinear time average "final" $q$-profiles. 
Color coding: Gray - $\beta = 0$, Black - Not Consistent (Not Cons.) and Colored - Consistent (Cons.) MHD equilibria. In addition, the mode rational surfaces where $q = m/n$ are plotted in gray, and the TAE resonant surfaces, where $nq=m+1/2$ , are marked in black dashed lines.
Bottom plots: Zoomed view on the initial and corrugated $q$ profiles.}
\end{center}
\end{figure}

The electromagnetic zonal currents can modify the magnetic fields and as a result the safety factor \cite{Volčokas_PRE2025}. This effect was explored by Di Giannatale \cite{Di_Giannatale_PPCF2025} in a reverse shear scenario, in global, gyrokinetic, electromagnetic simulations using ORB5, and showed flattening of the $q$ profile, for $q_{min}$ near a mode rational value $(q=2)$ in a turbulent plasma. Here we further explore this effect on a monotonically increasing $q$-profile, in the presence of either $n=[0,2]$ TAE or $n=[0,25]$ ITG. Figure \ref{FIG:modified q-profile} shows the modification of the $q$ profile $\langle{q\rangle}_t(s)-q(t=0,s)$, and the zonal component of $\langle A_\parallel \rangle_t $ with $\langle \cdot \rangle_t$ the time average taken over the quasi-steady nonlinear phase. In our cases, the $q$ profile begins to corrugate and flatten, in some correlation to the mode rational surfaces. This corrugation increases with bulk gradients, and is more prominent for the $n=25$ ITG than the $n = 2$ TAE. As before, the Shafranov shift has a stabilizing effect on the TAE and KBM (not shown here), and a mixed effect on the ITG, where it is exciting for the ST and stabilizing for the PK cases, such that both self-consistent ITG cases have a similar amplitude.

\subsection{Turbulent transport and relaxation of kinetic profiles}


\begin{figure}
\begin{center}
\includegraphics[width=\textwidth]{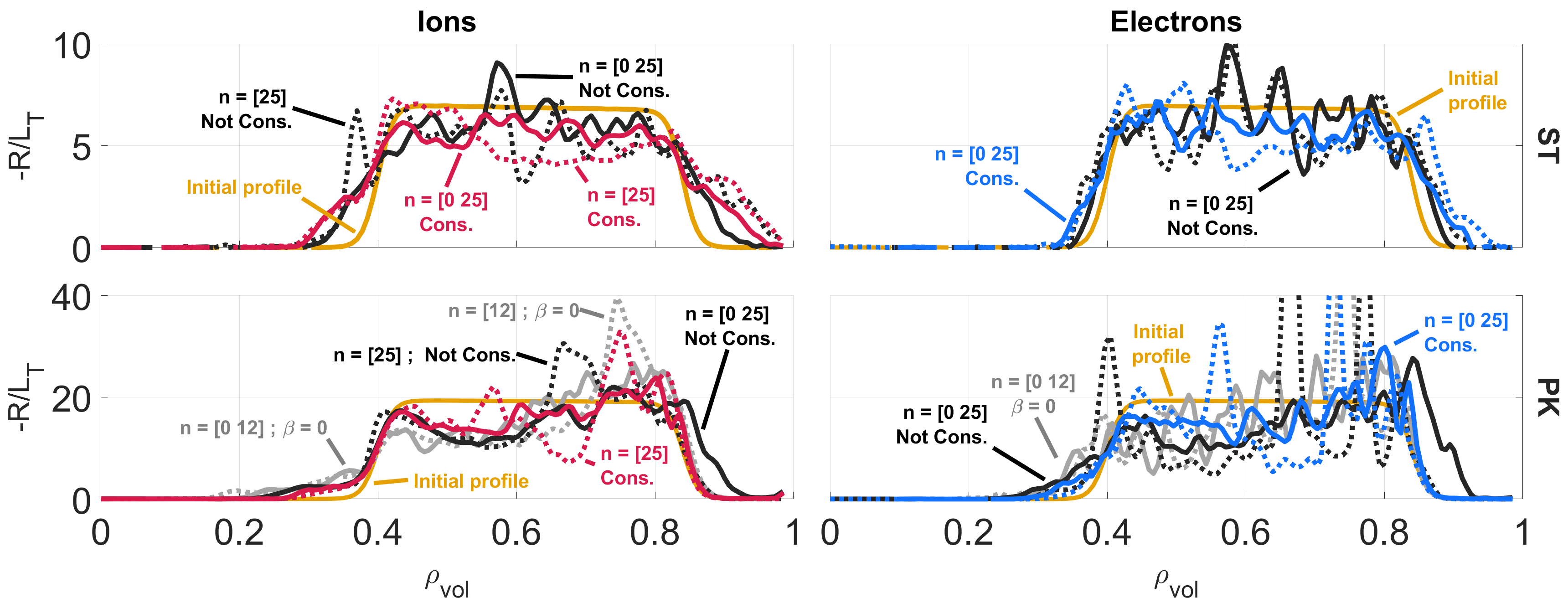}
\caption{\label{FIG:RLT_nonlin} \it 
The flux surface averaged logarithmic temperature gradient $R/{L_T}_{i,e}$ evolution after the initial saturation phase, separated between the ion (left) and electron channels (right), the standard (top) and peaked (bottom) bulk profiles, and with (solid line) or without (dashed line) zonal flows.}
\end{center}
\end{figure}

Figure \ref{FIG:RLT_nonlin} shows the nonlinear evolution of the normalized logarithmic temperature gradient for the ITG cases, $\{R/\langle{ {L_T} \rangle}_{i,e}\ |\ n=[0,25]\}$. Including the $n = 0$ response has a stabilizing effect on turbulence, as evident in the reduced magnitude of the corrugation. The very strong corrugation in the electron temperature gradient for a plasma with peaked gradients (PK) emphasizes the importance of including the $n = 0$ mode for obtaining the correct final self-organized state of the plasma. It might also be connected to the electron channel frequency spectrum not broadening during the nonlinear saturation stage unlike the ion channel (Fig. \ref{FIG:nonlinear_ITG_pflux_fft}). The Shafranov shift has a stabilizing effect as well. However, the large variation in profile relaxation under the combined effects of the two makes it harder to point out a clear trend.   

\begin{figure}
\begin{center}
\includegraphics[width=0.49\textwidth]{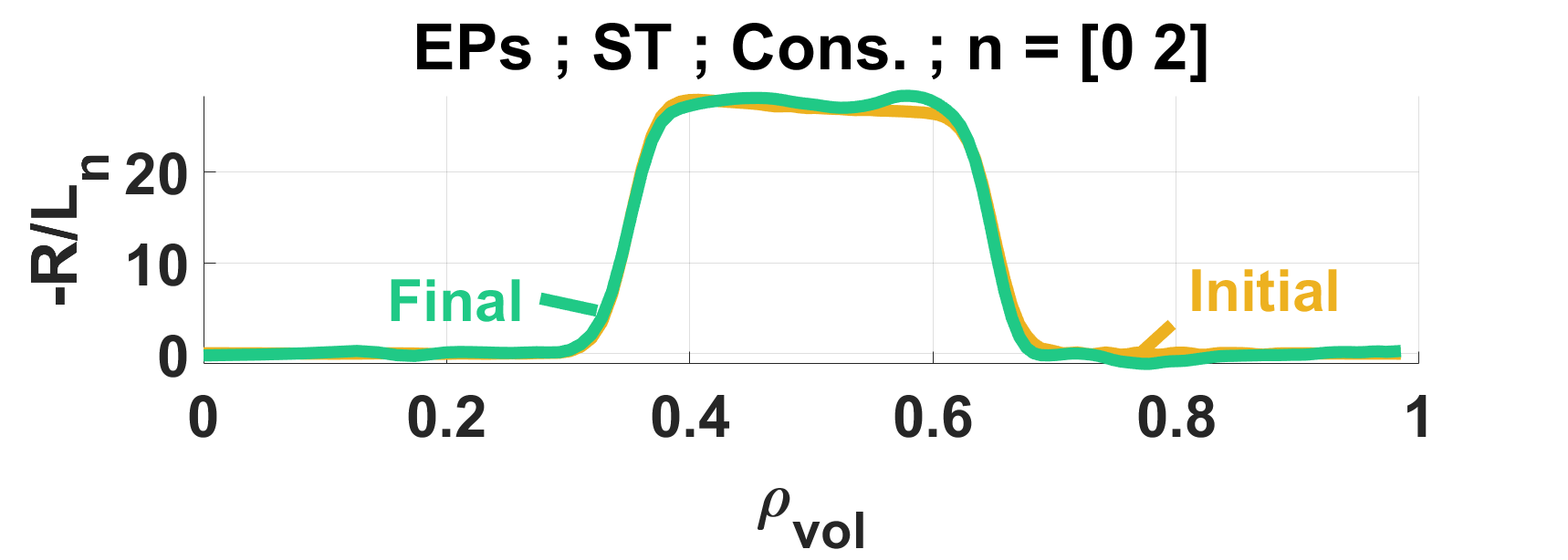}
\includegraphics[width=0.49\textwidth]{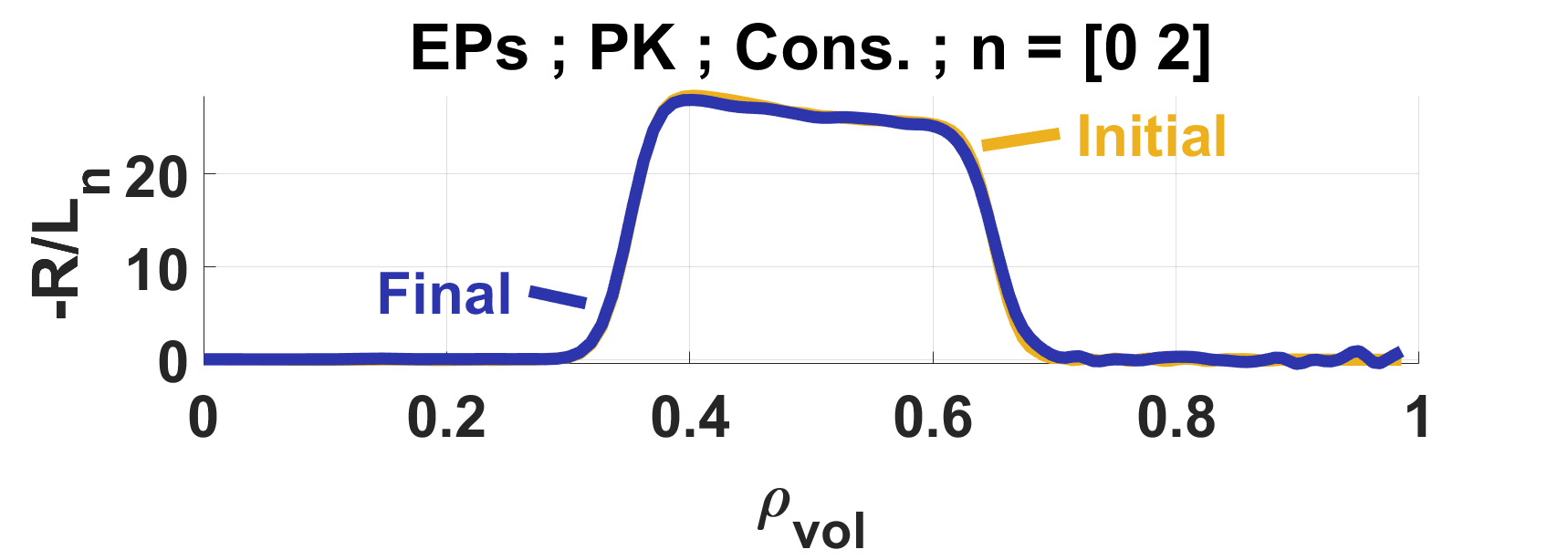}
\caption{\label{FIG:RLn_nonlin} \it 
The initial and final $R/L_n$ profiles for the consistent cases with standard (left) and peaked (right) bulk gradients.}
\end{center}
\end{figure}

In Figure \ref{FIG:RLn_nonlin} we show that the EP density profile relaxation is small and therefore is most likely not a major nonlinear saturation mechanism. 

\begin{figure}
\begin{center}
\includegraphics[width=1\textwidth]{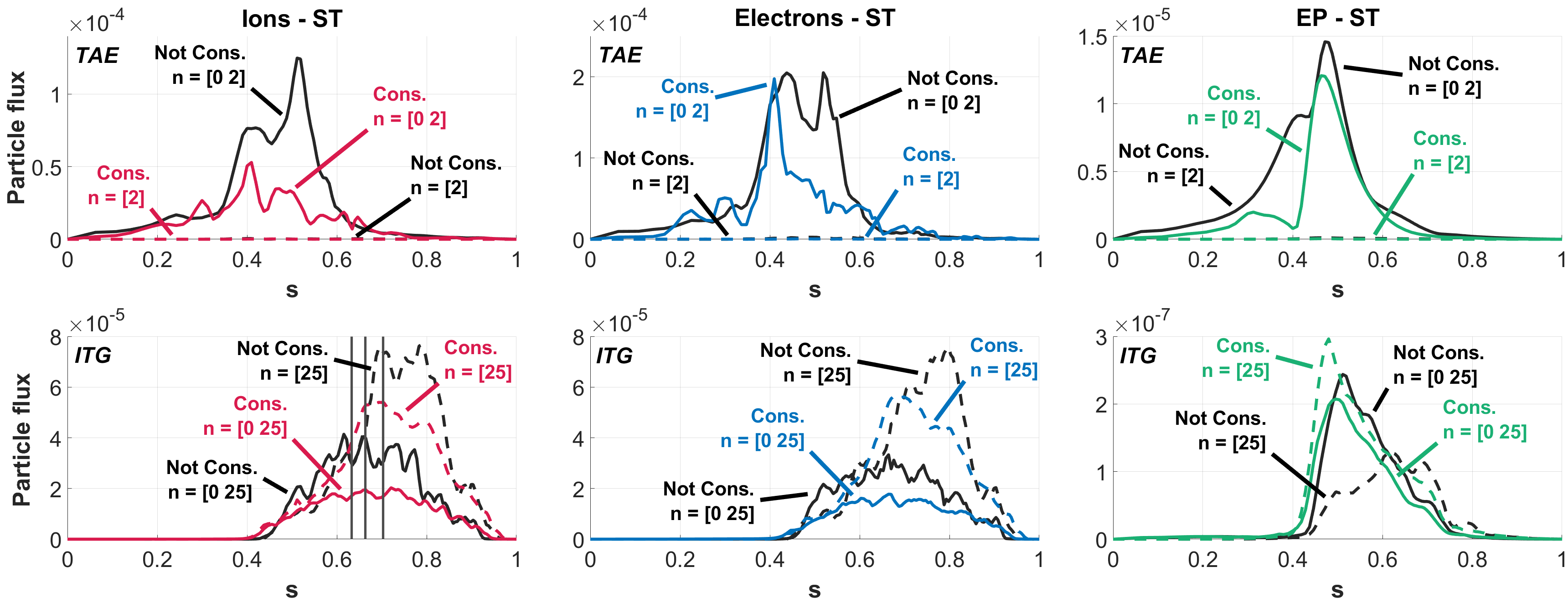}
\hspace{2pt}
\includegraphics[width=1\textwidth]{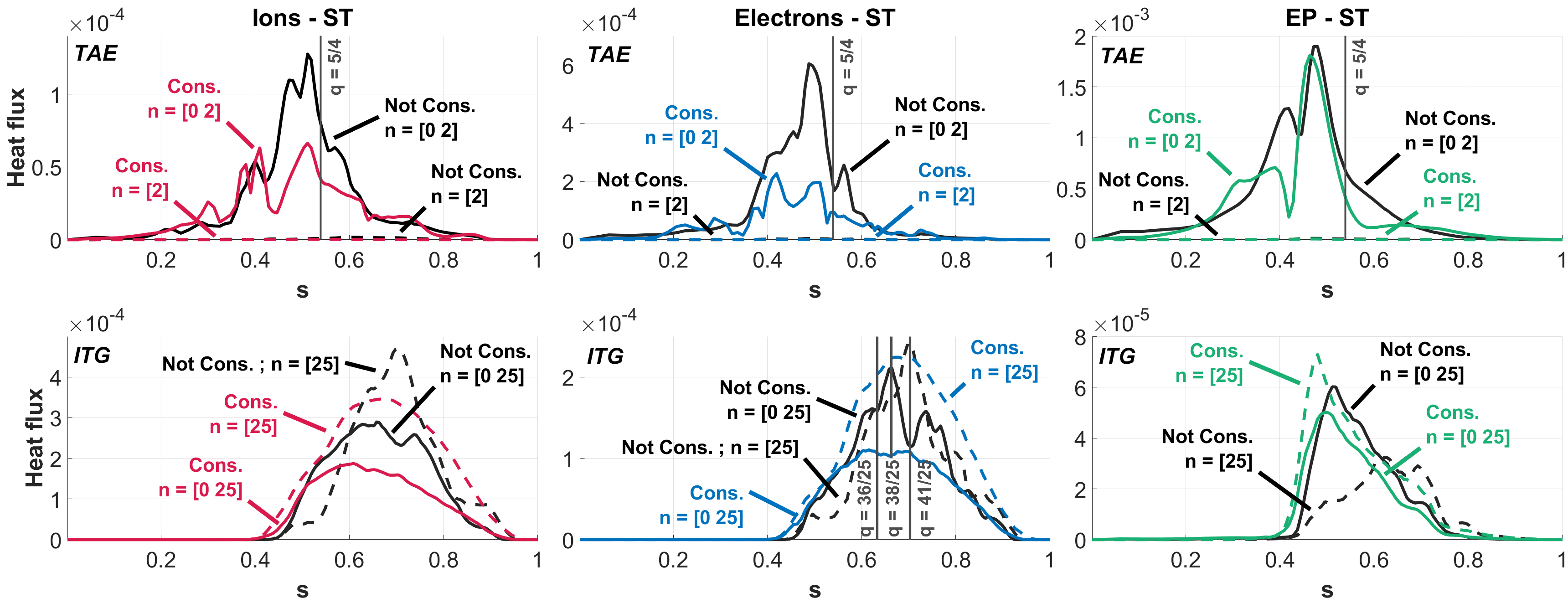}
\caption{\label{FIG:heat_part_flux} \it 
Radial profiles of particle and heat fluxes for the ST cases. The fluxes are averaged over a time window situated after the initial saturation phase. The peaked cases which exhibit a similar behavior, are omitted for brevity.  
Color coding: Black - Not Consistent (Not Cons.), and Colored - Consistent (Cons.), MHD equilibria. Solid lines - with ZS, and dashed lines - without ZS. Some of the mode rational surfaces where $q = m/n$ are plotted in gray, and TAE resonant surfaces, where $nq=m+1/2$ , are marked in black dashed vertical lines.}
\end{center}
\end{figure}

Figure \ref{FIG:heat_part_flux} shows the time-averaged radial profiles of the heat and particle fluxes (per species) for plasmas with standard bulk gradients (ST). The stabilizing effects of Shafranov shift are evident in the reduced fluxes of the thermal species for self-consistent MHD equilibria. For the ITG cases $n = [0 \ 25]$ including the $n = 0$ response had a stabilizing effect reducing the heat and particle fluxes of the thermal species. In general the EP flux carried by the ITG turbulence is small compared to the EP flux generated by the TAE. However, if we normalize the turbulent EP flux by the EP fraction, we find that it is comparable in magnitude to the fluxes of the thermal species. Thus showing that in our cases, turbulence indeed carries a non negligible EP flux, which in turn is smaller than the EP flux generated by the TAE.  

Including the zonal response in the TAE cases leads to the opposite behavior than for the ITG, i.e. a significant increase in heat and particle fluxes of all three species. In Figure \ref{FIG:nonliner_steady_fluct_fluxs} we divide the heat and particle fluxes of the consistent cases into steady and fluctuating parts. For the thermal species, both the steady and fluctuating parts increase, while the EP flux is predominantly steady in all cases. This increase in flux might be linked to a higher TAE activity seen in Figure \ref{FIG:nonlinear_TAE_timewindow} for the $n = 2 \in n =[0,2]$ TAE than the $n = 2 \in n =[2]$ TAE. 

A similar increase in fluxes when including the zonal response, was recently reported by Chen et. al. \cite{Chen_NF2025}. There it was attributed to a redistribution of the EP population in phase space. Here we propose an additional possible explanation. The nonlinearly excited AAE found in both the shearing rate and the fluxes in our plasmas, has a non-negligible $ m \neq 0$ components that can potentially drive a net flux. Further investigation of the EP phase space behavior is left for future works.

\begin{figure}
\begin{center}
\includegraphics[width=0.32\textwidth]{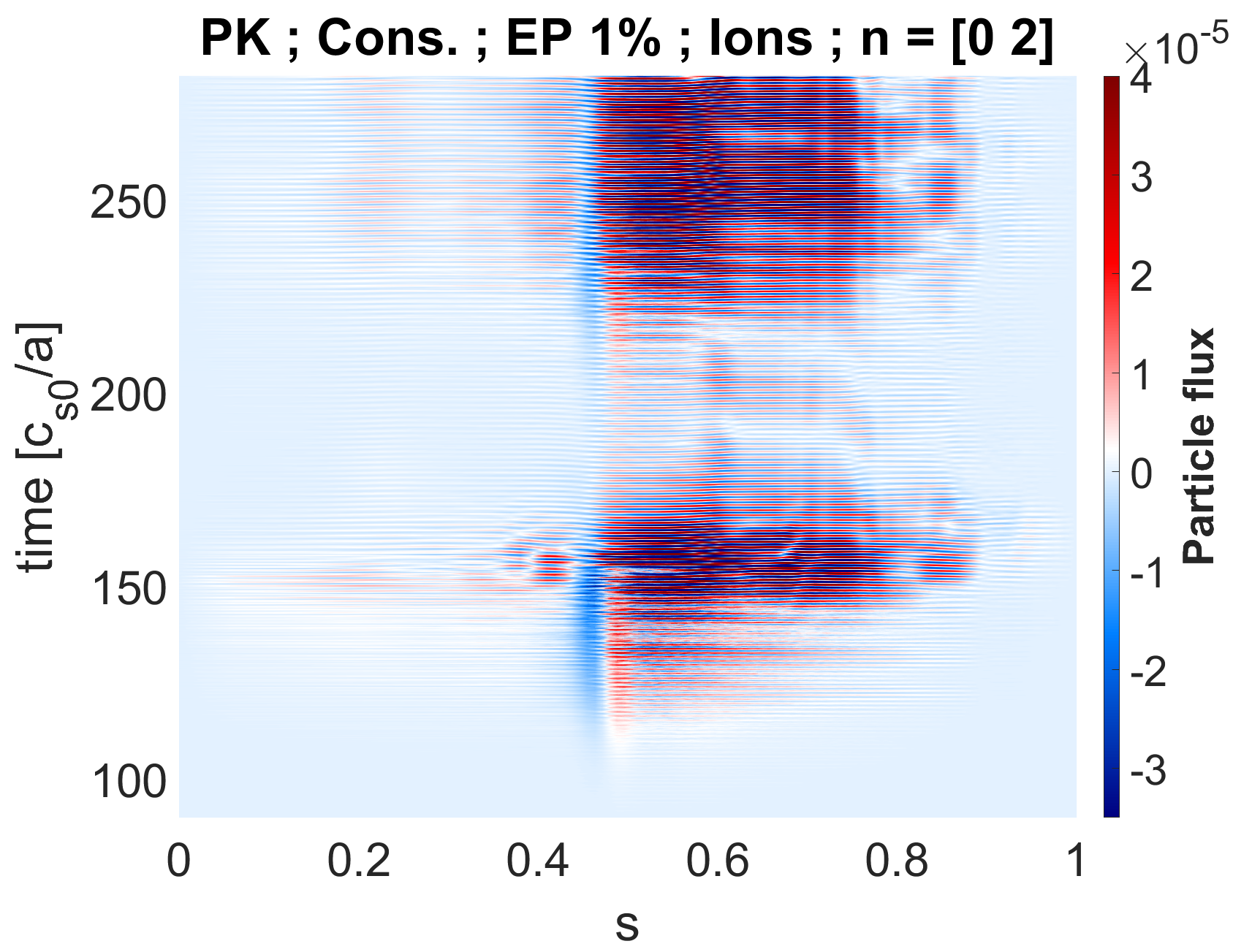}
\includegraphics[width=0.32\textwidth]{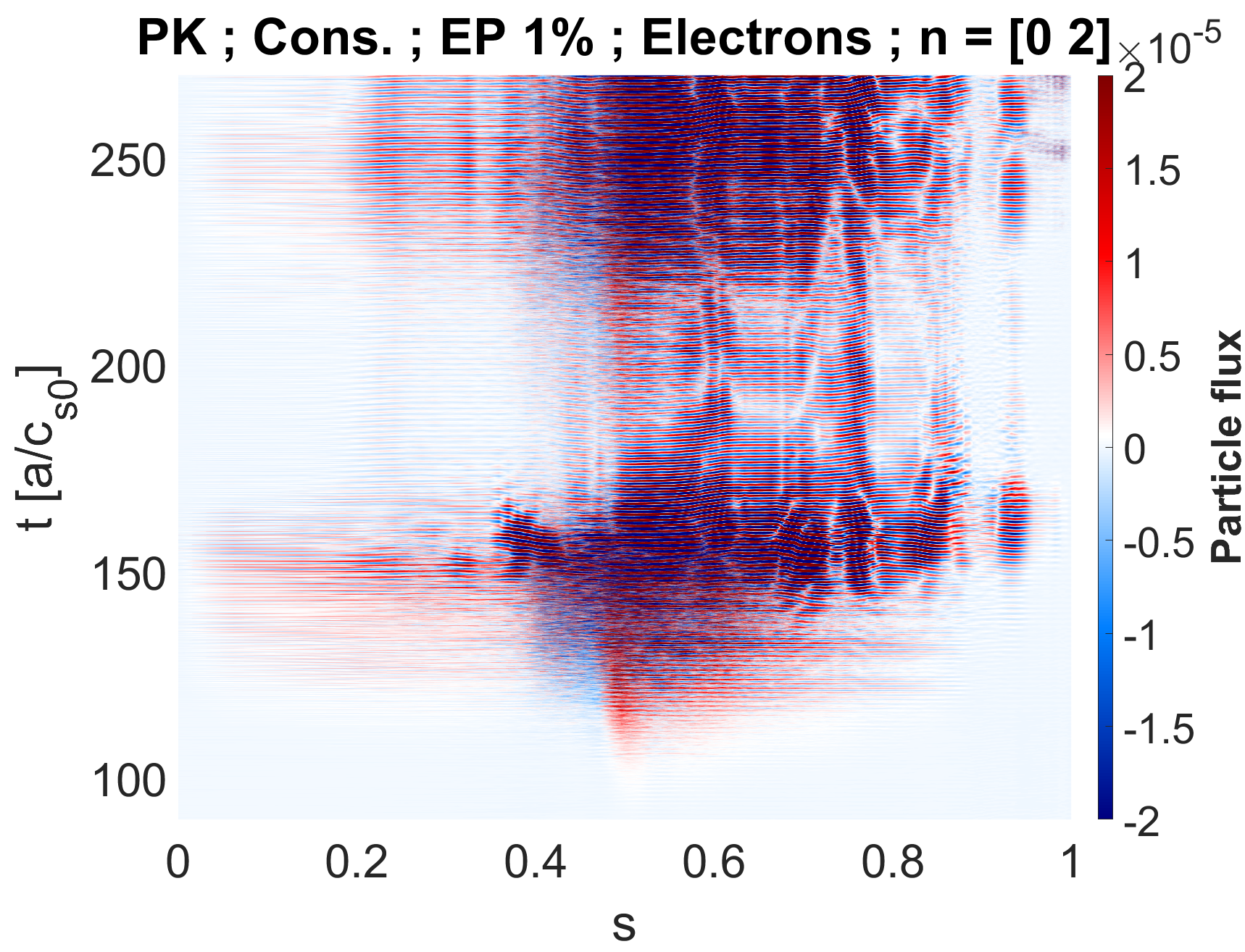}
\includegraphics[width=0.32\textwidth]{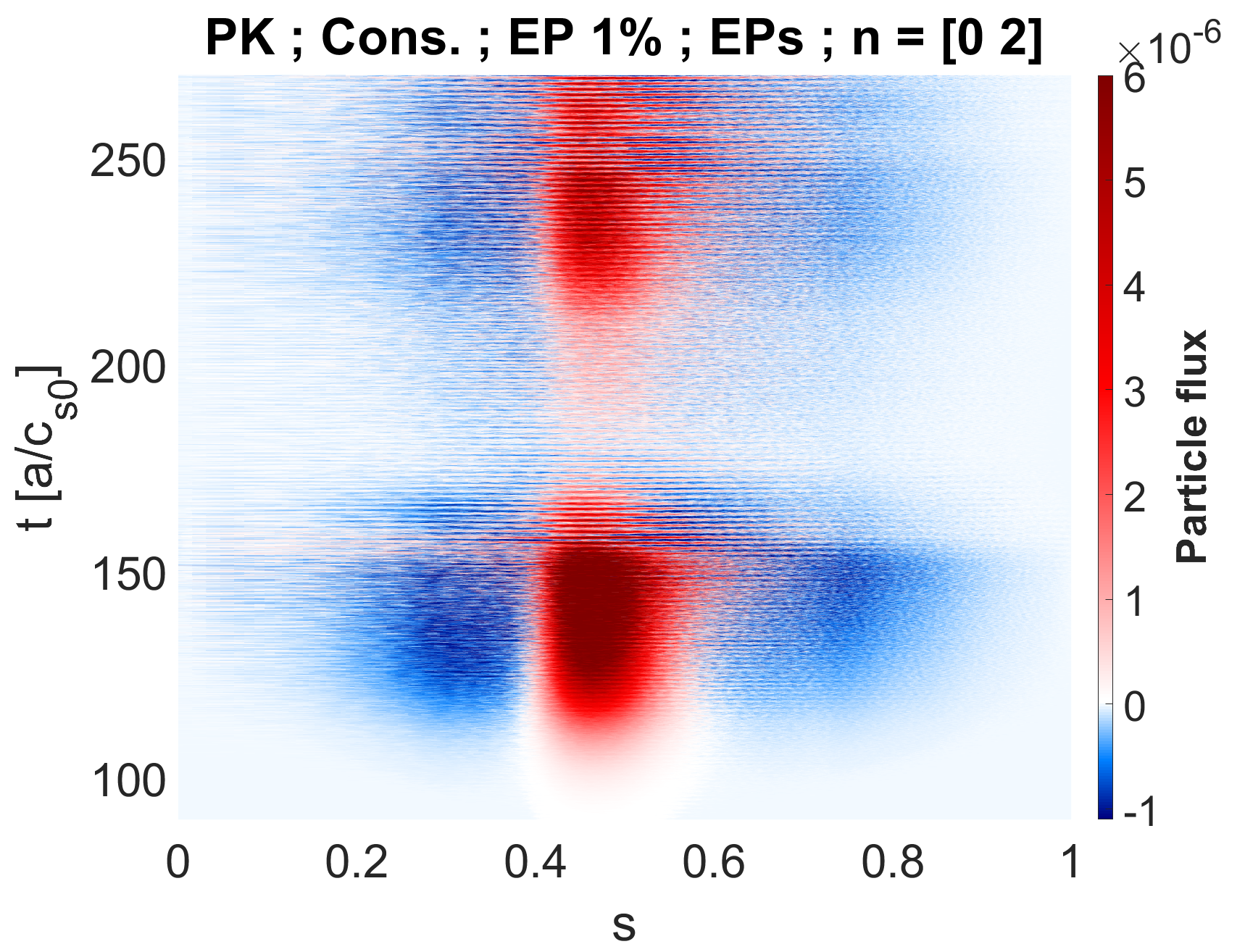}
\caption{\label{FIG:nonlinear_PK_Cons_TAE_part_flux} \it 
Contours of the particle flux vs. time and radius for each of the species (Ions, Electrons and EPs) in the case with peaked bulk gradients, self-consistent MHD equilibrium, with $1\%$ EP. }
\end{center}
\end{figure}

\begin{figure}
\begin{center}

\includegraphics[width=0.32\textwidth]{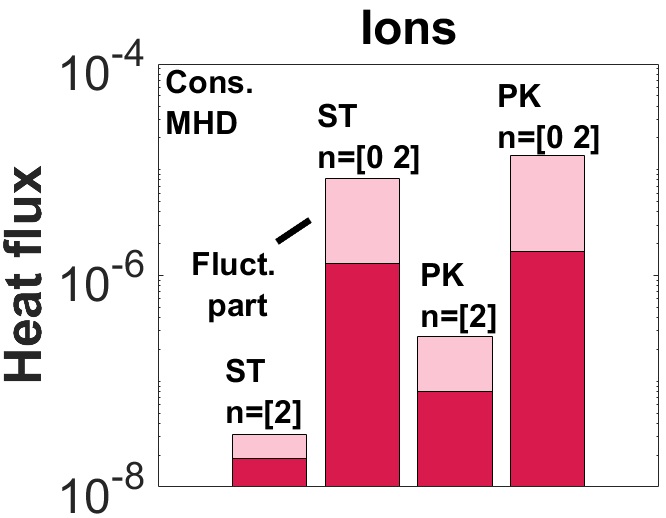}
\includegraphics[width=0.32\textwidth]{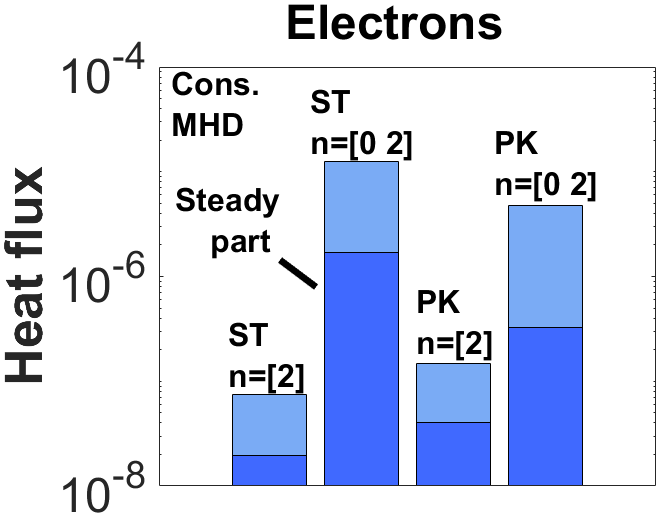}
\includegraphics[width=0.32\textwidth]{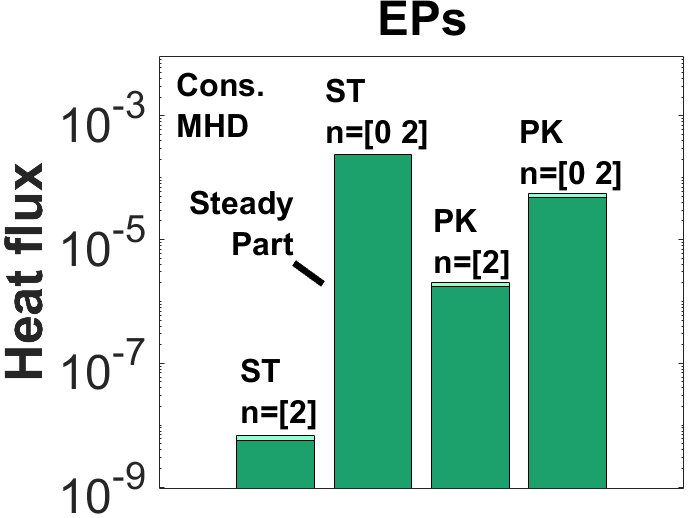}

\includegraphics[width=0.32\textwidth]{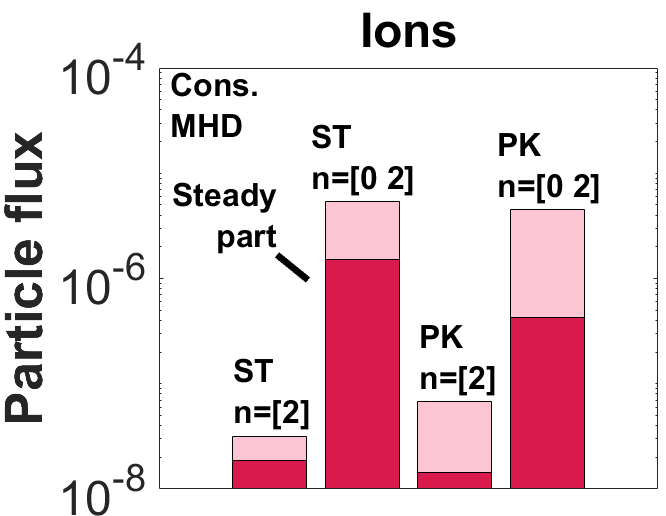}
\includegraphics[width=0.32\textwidth]{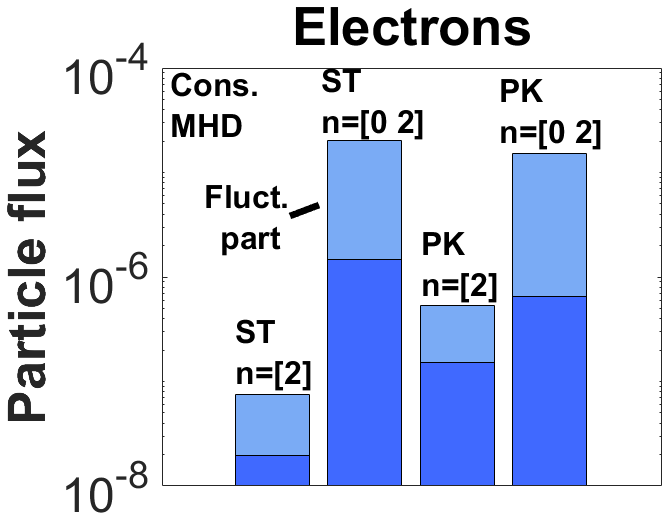}
\includegraphics[width=0.32\textwidth]{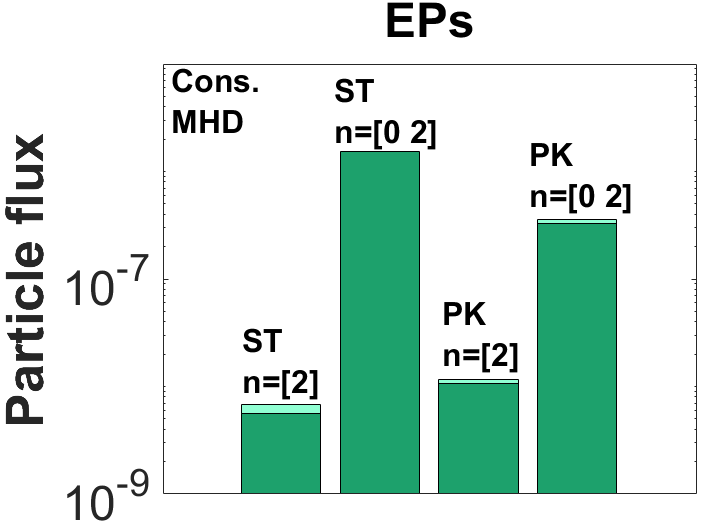}

\caption{\label{FIG:nonliner_steady_fluct_fluxs} \it 
Shows the steady and fluctuating parts of the heat and particles fluxes that are induced by a TAE. The time averaging is done in accordance with the nonlinear time windows. And the space averaging - across the entire width of the plasma. The steady parts are the lower pillar and have a darker color, and the fluctuating parts are colored in a lighter shade of the species color placed on top of them. We can notice the increase in fluxes associated with including the zonal response ($n = 0$)}.
\end{center}
\end{figure}

Figure \ref{FIG:nonlinear_PK_Cons_TAE_part_flux} shows an oscillation on top of the particle flux of ions and electrons in a self-consistent case with peaked gradients and which includes a zonal response. In its frequency spectrum we find two nonlinearly active modes, which match the TAE and the AAE frequencies. It is interesting to compare and contrast their behavior of the  $v_{E \times B}$ frequency spectrum (Fig. \ref{FIG:nonlinear_ExB_fft2D_timelaps}). For the fluxes, the location of the drive (see Fig. \ref{FIG:Combined_profiles_particles}) changes the dynamics of the system, with specific details depending on the species, Shafranov shift and whether or not we included the zonal response in the filter. The frequency spectrum of the particle fluxes changes between the saturation and early nonlinear phases of the systems, and is different between the TAE and ITG.

\begin{figure}
\begin{center}
\includegraphics[width=0.32\textwidth]{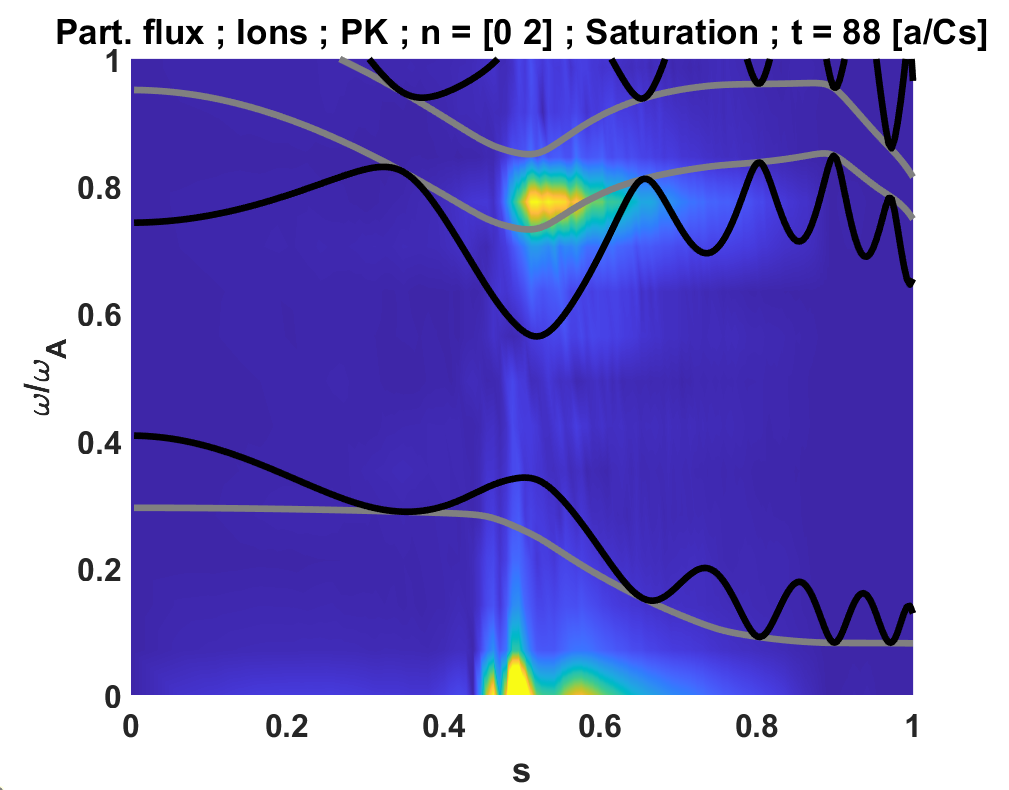}
\includegraphics[width=0.32\textwidth]{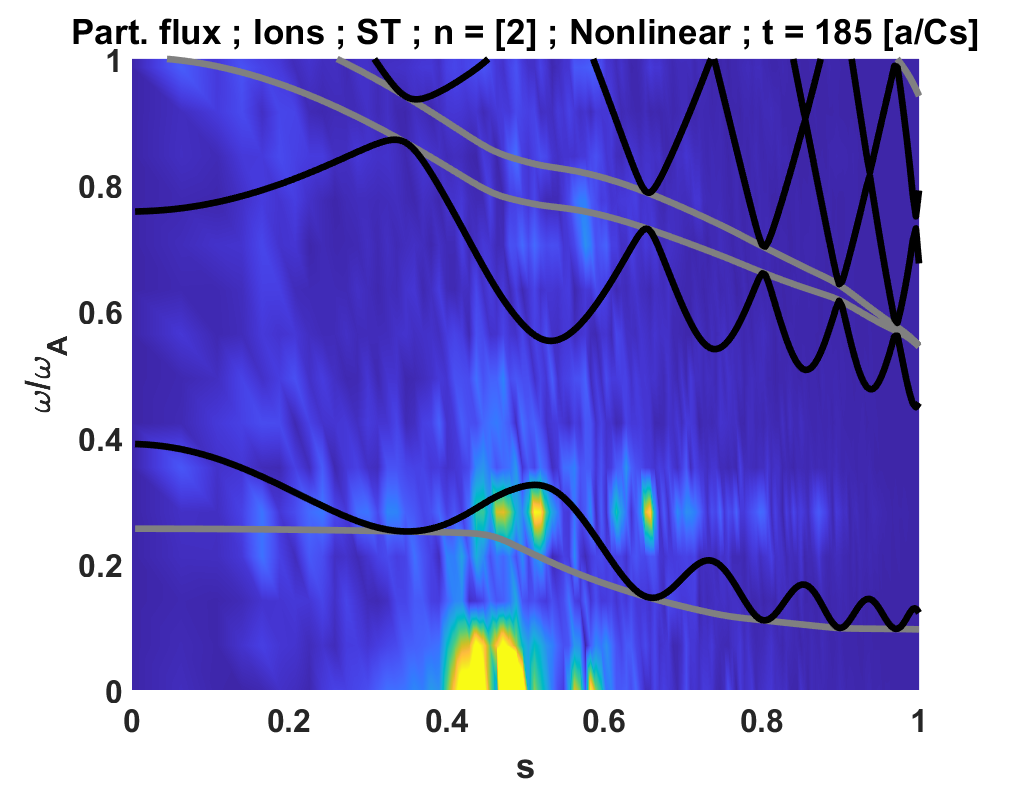}
\includegraphics[width=0.32\textwidth]{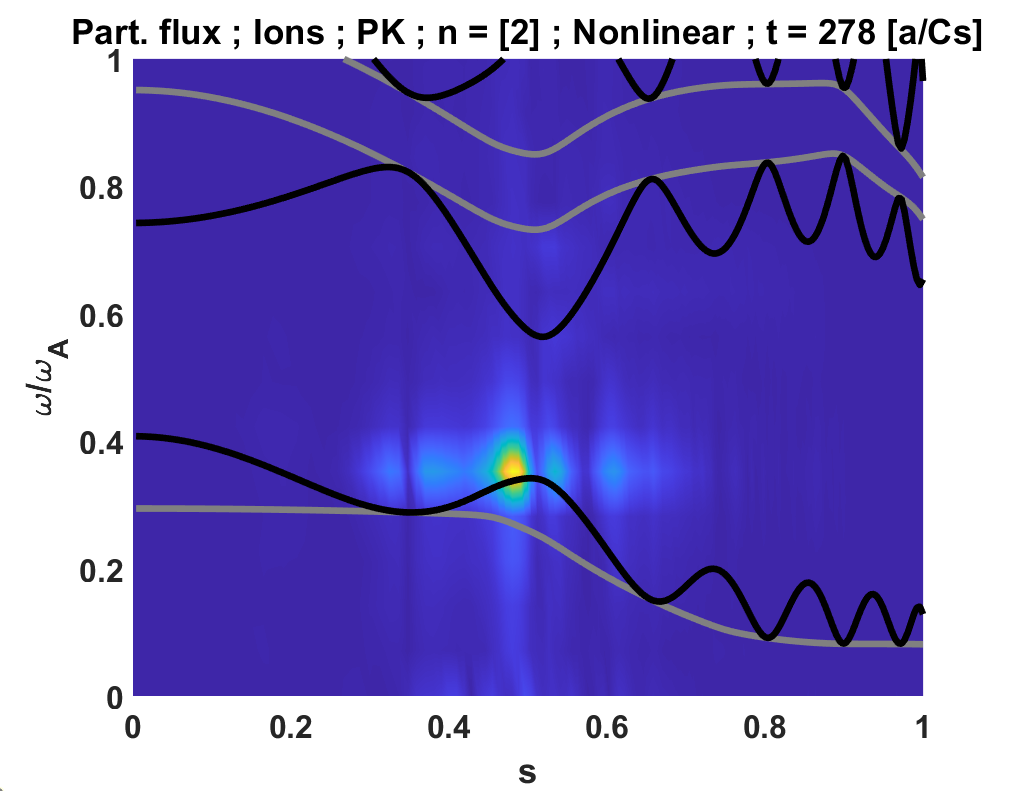}

\includegraphics[width=0.32\textwidth]{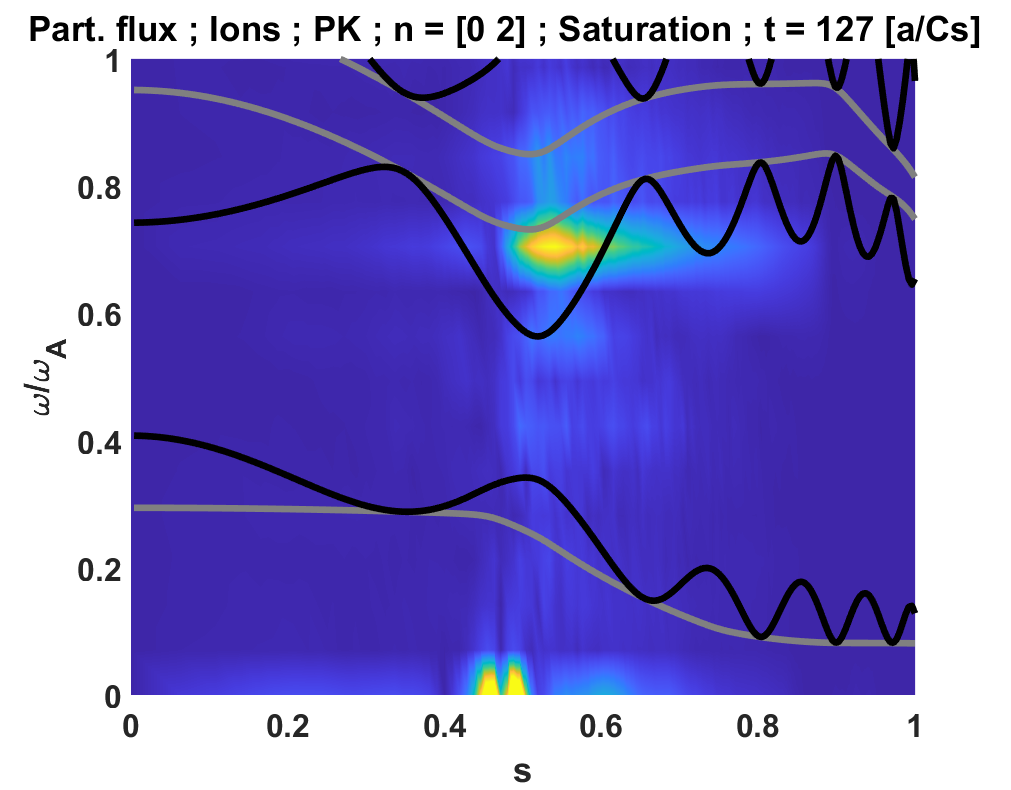}
\includegraphics[width=0.32\textwidth]{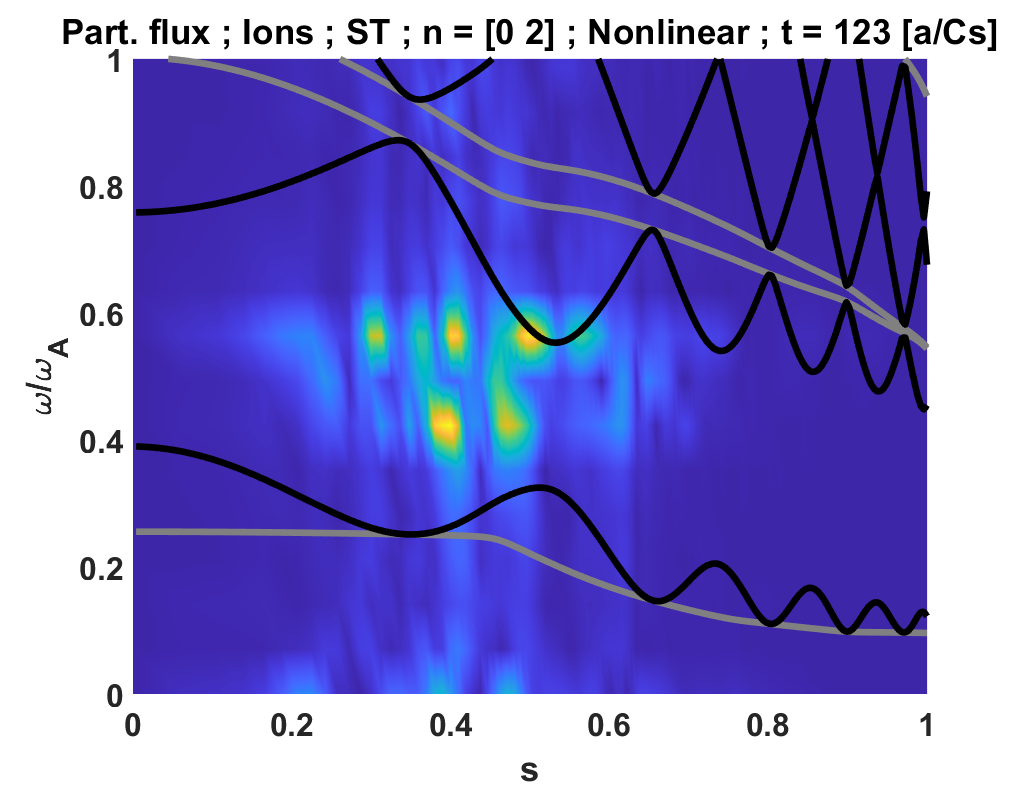}
\includegraphics[width=0.32\textwidth]{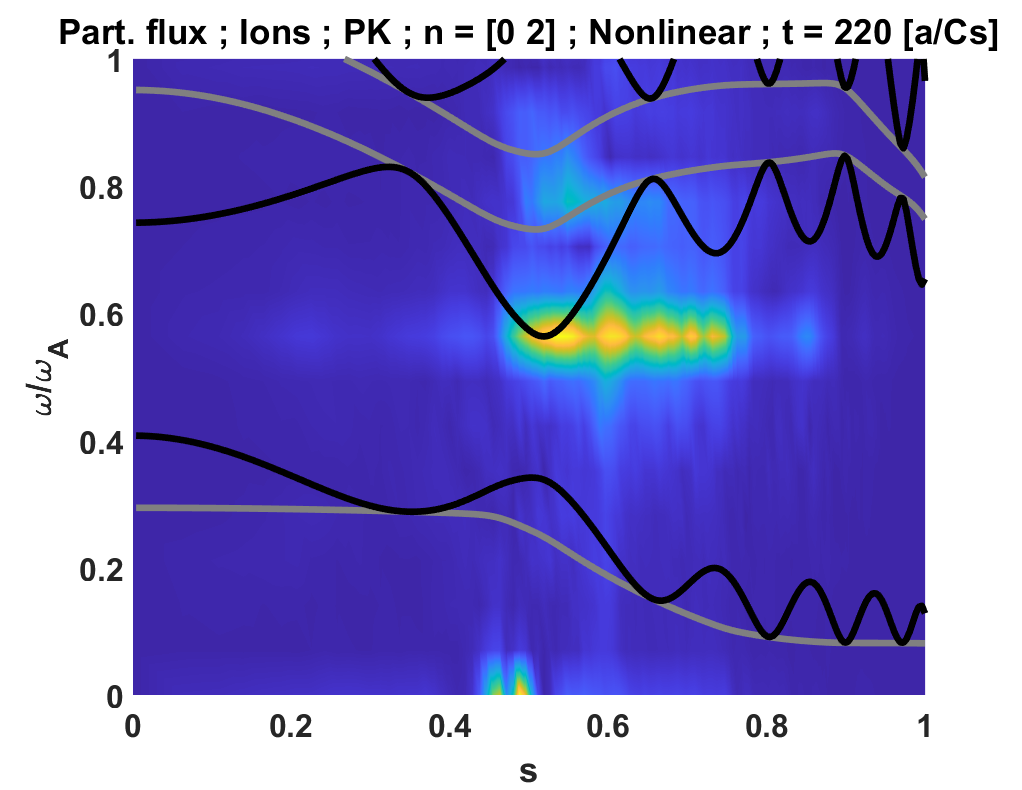}

\caption{\label{FIG:nonlinear_TAE_pflux_fft} \it 
Frequency spectrum in Alfv\'enic units of the particle flux produced by a TAE with an overlay of $n = 0$ Alfv\'en continuum in gray and an $n = 2$ Alfv\'en continuum in black. The "Saturation" and "Nonlinear" label indicates the system's stage at that timestamp. The Color map is normalized between $[0,1]$.
Left column shows an AAE chirping up and down during the saturation phase and the AAE in the bottom left tile, beats with the TAE in the bottom right tile.}
\end{center}
\end{figure}

\begin{figure}
\begin{center}
\includegraphics[width=0.325\textwidth]{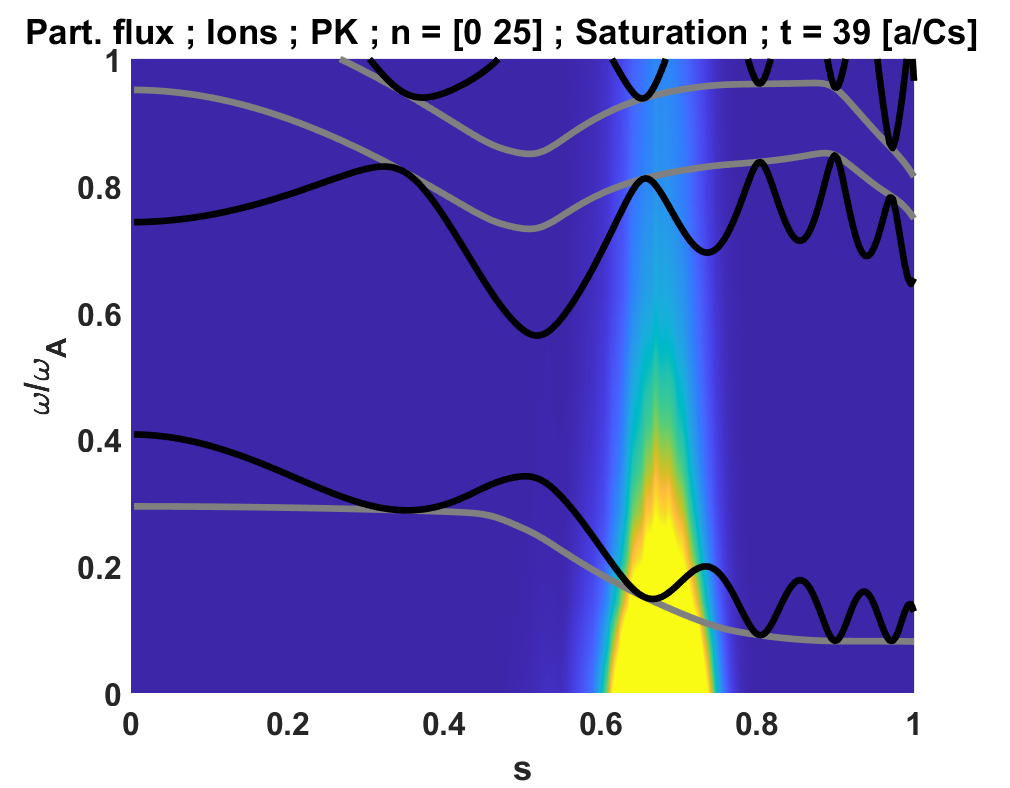}
\includegraphics[width=0.325\textwidth]{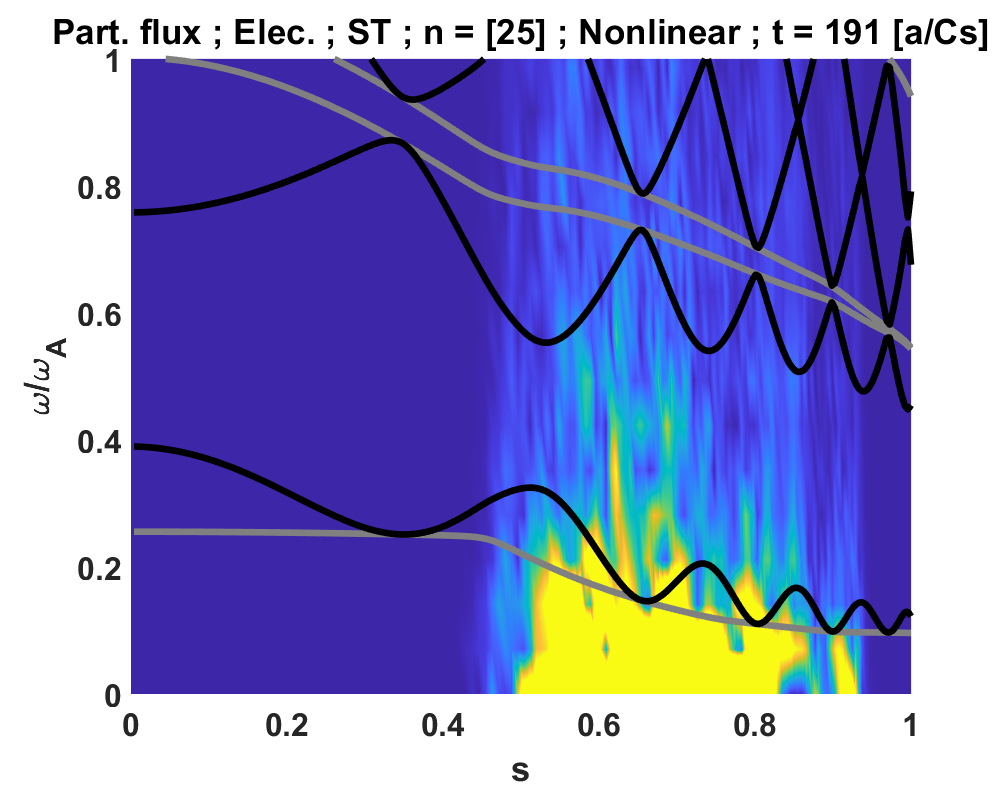}
\includegraphics[width=0.325\textwidth]{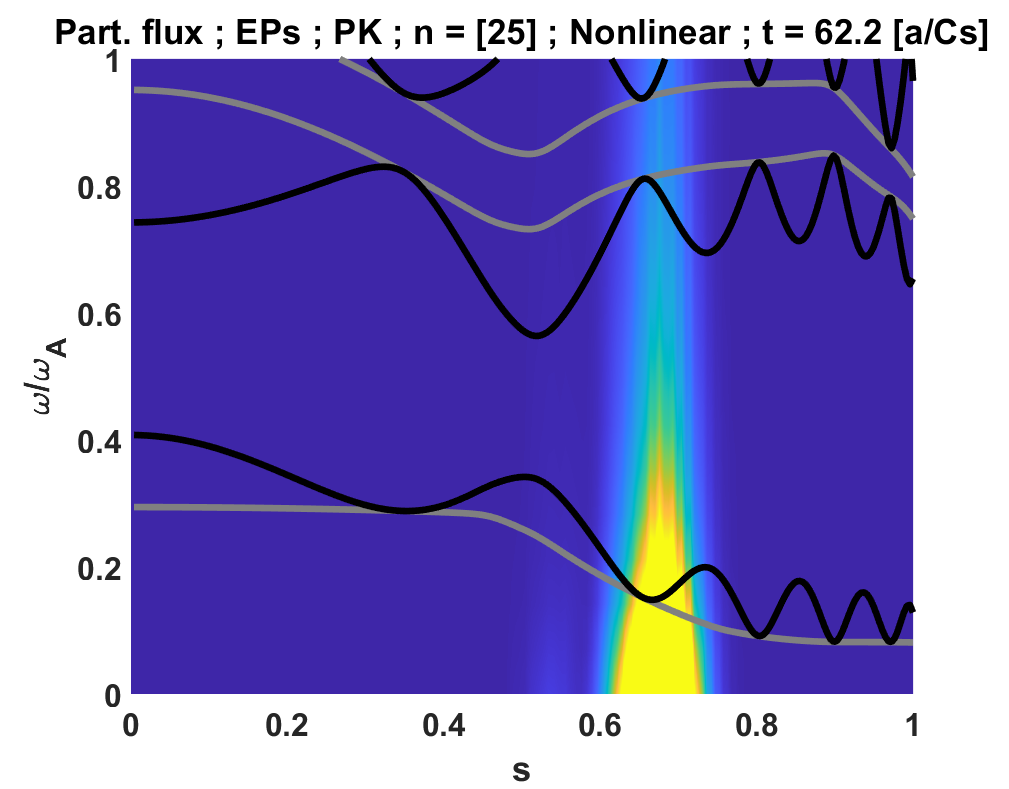}

\includegraphics[width=0.325\textwidth]{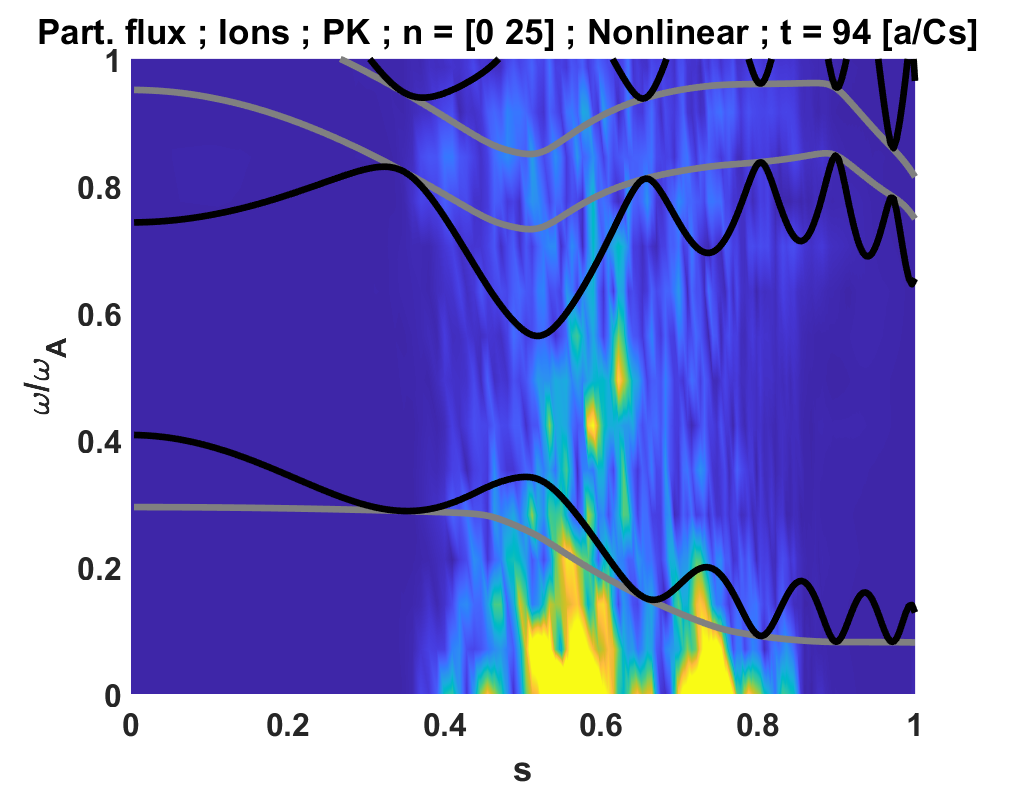}
\includegraphics[width=0.325\textwidth]{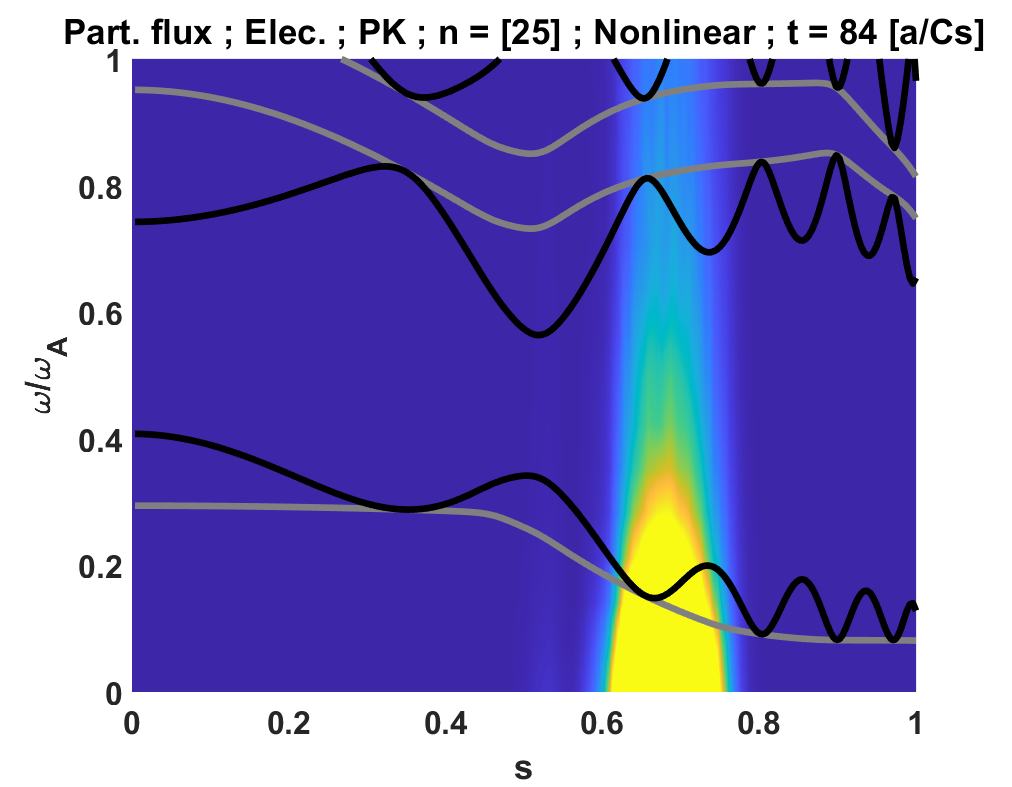}
\includegraphics[width=0.325\textwidth]{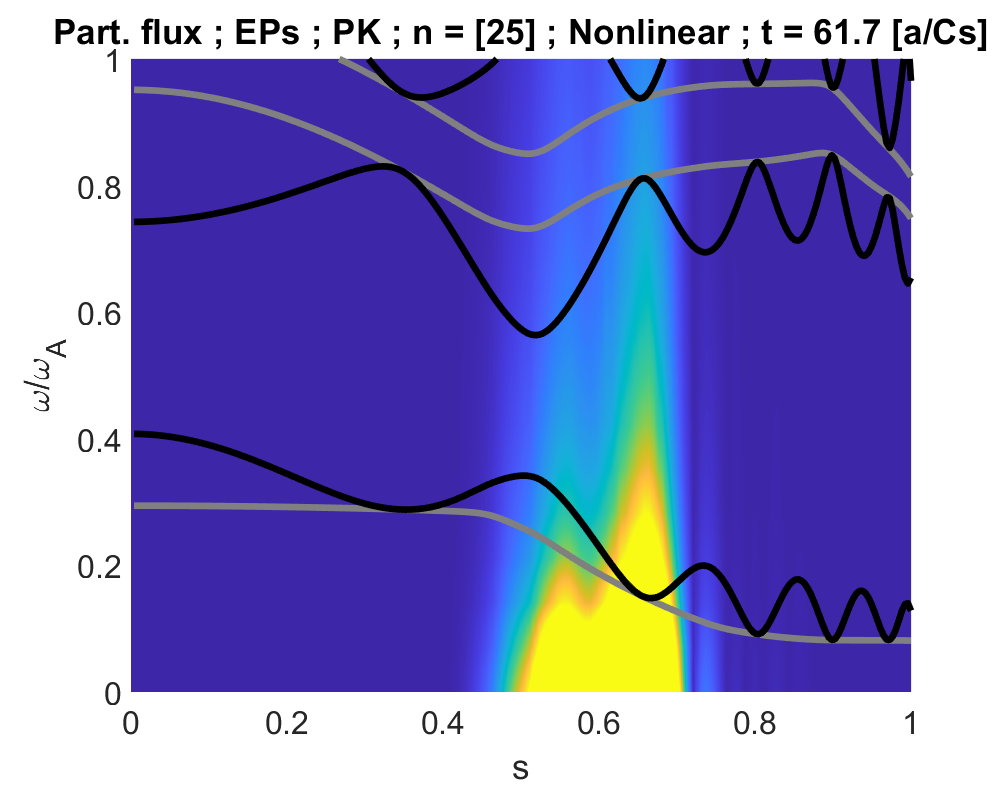}

\caption{\label{FIG:nonlinear_ITG_pflux_fft} \it 
Frequency spectrum in Alfv\'enic units of the particle flux produced by an ITG with an overlay of $n = 0$ Alfv\'en continuum in gray and an $n = 2$ Alfv\'en continuum in black. The "Saturation" and "Nonlinear" label indicates the system's stage at that timestamp. The Color map is normalized between $[0,1]$.}
\end{center}
\end{figure}

Figure \ref{FIG:nonlinear_TAE_pflux_fft} shows several typical behaviors of the TAE, summarized here:

\begin{itemize}
\begin{samepage}
    \item The saturation phase is characterized by a chirping up and down of the AAE mode. This behavior is accentuated in cases with peaked profiles (PK) due to the radial proximity between the minimum in the $n = 0$ continuum and the driving gradients.
    
    \item Cases with peaked profiles even without the $n = 0$ mode in the filter, show strong AAE activity during the saturation phase. This AAE activity does not emanate form the edge, but rather gets excited in the middle of the plasms. 
    
    \item All cases without the zonal response establish, during the nonlinear phase, a less coherent (weaker) TAE with a frequency partially intersecting the continuum. While including the $n = 0$ mode leads to systems with a more coherent (stronger) TAE. Which for the PK cases stays situated close to the top of the gap, and in the ST cases chirps up and down. 
    
    \item In cases with peaked profiles (PK), We see a beating between the TAE and AAE during the early nonlinear saturation. This also leads to the TAE chirping up, and staying there (unlike in the ST cases). 
\end{samepage}
\end{itemize}

\FloatBarrier

The frequency spectrum present in particle fluxes produced by the ITG microinstability, does not qualitatively change between the standard and peaked cases, and shows little AAE activity unlike the associated $E \times B$ flows. Figure \ref{FIG:nonlinear_ITG_pflux_fft} shows several typical behaviors of the ITG frequency spectrum:

\begin{itemize}
    \item During the quasi-linear phase the ITG radial mode structure is apparent in the radially distinct frequency bands and the envelope of mode. For the electromagnetic ITG these bands extend into the Alfv\'enic frequency spectrum.

    \item During the saturation phase the flux loses its radial coherence and spreads (radially) irregardless of whether or not we include the zonal response in the filter. However, the transient behavior and the (early) nonlinear state of the system will be different. Because when the zonal response is included, some of the spreading is achieved through the zonal meso-scales. Leading to simultaneously steeper gradients (Fig. \ref{FIG:RLT_nonlin}) and reduced heat and particle fluxes (Fig. \ref{FIG:heat_part_flux}). Excluding the zonal response can also lead to situations where for example the electron flux in the case with peaked profiles remains radially localized in the nonlinear stage perhaps leading to very strong corrugation in the temperature gradient (Fig. \ref{FIG:RLT_nonlin}).  

    \item Independent of the zonal response, the EP flux radial profile oscillates between two different positions. These positions correspond to the respective locations of the maximum $R/L_n$ of EPs and $R/L_T$ of bulk species. There is no signature at the TAE frequency. The radial oscillation is still going on at the end of our simulations, showing that a complete quasi-state has not yet been established.
\end{itemize}

\section{Conclusions}

In this work we explore the connections between MHD equilibria, in particular through the effects of Shafranov shift, Alfv\'en Eigenmodes, microturbulence and the nonlinear system response of a burning plasma containing Energetic Particles (EPs). We used ORB5, a PIC, global, electromagnetic, gyrokinetic code to explore the linear growth and nonlinear saturation of the instabilities driven by either gradients in Energetic Particles (EP) density, or by gradients in thermal species (ions and electrons) temperature and density. We adopted a Cyclone Base Case (CBC) parameter set and magnetic geometry - adjusted (or not) to bulk and EP pressure.

We mapped the linear dispersion relation and parts of the Alfvén continuum and found that accounting for consistent Shafranov shift completely stabilizes the KBMs appearing in the $n = [6 - 25]$ range for plasmas with $\beta = 0$ MHD equilibrium. notably, the KBMs are also very strongly stabilized (linearly) by the EP fraction. We find that Shafranov shift has a strong stabilizing effect on the TAE growth rate, leading to a less than linear dependence of the in TAE growth rate with EP fraction in the self consistent cases. The ITG spectrum is much less sensitive to the presence of EPs, especially for shorter wavelengths.

Taking into account the finite pressure in the MHD equilibrium plays a stabilizing role in the nonlinear saturation dynamics of the system and its evolution during the nonlinear stage. This picture is not complete without in addition considering the self-generated zonal flows and currents. Although both ITG and TAE instabilities generate zonal structures during the nonlinear saturation phase, they differ in their radial structure, relative intensity to the exciting mode (TAE or ITG), and the effects on the turbulent fluxes. Including the zonal response for the ITG mode lead to a smoother saturation, reduced heat and particle fluxes, and a reduced $R/L_T$ modulation in the nonlinear phase. In contrast to the ITG case, including the zonal response for the TAE leads to an increase in heat and particle fluxes and an increase in the mode amplitude. Both TAE and ITG cases show zonal currents leading to modifications of the radial $q$ profile of similar magnitude. 

We find that, during the nonlinear saturation phase all modes (TAE, KBM, and ITG) excite an Axisymmetric Alfv\'en Eigenmode (AAE) with a $[n,m] = [0,\pm1]$ mode structure and a frequency close to the $n = 0$ Alfv\'en continuum near its minimum (as predicted by theory). In this work we suggest that the AAE, which is dominantly $m \pm 1$ and appears to modulate the heat and particles fluxes of both TAE (strongly) and ITG (weakly), can carry finite radial fluxes and therefore might be responsible for the enhanced TAE fluxes. 

In future works we plan to explore the possible cross-talk between the ITG and TAE instabilities which may occur e.g. through the excitation of Zonal Structures. This will require simulations with all three modes present (e.g. $n=0$ (ZS), $n=2$ (TAE) and $n=25$ (ITG)). More work is also needed to study in detail the phase-space redistribution of EPs and how it relates to TAEs and AAEs.   

\section{Acknowledgment}
This work has been carried out within the framework of the EUROfusion Consortium, partially funded by the European Union via the Euratom Research and Training Program (Grant Agreement No 101052200 — EUROfusion). The Swiss contribution to this work has been funded by the Swiss State Secretariat for Education, Research and Innovation (SERI). Views and opinions expressed are however those of the author(s) only and do not necessarily reflect those of the European Union, the European Commission or SERI. Neither the European Union nor the European Commission nor SERI can be held responsible for them. This work is also supported by a grant from the Swiss National Supercomputing Center (CSCS) under project IDs s1232 $\&$ lp73. This work was supported in part by the Swiss National Science Foundation.


\end{document}